\documentclass[a4paper,11pt]{article}

\usepackage{jheppub} 

\usepackage[T1]{fontenc} 

 \usepackage{graphicx}

\usepackage{amsmath}
\usepackage{slashed}
\usepackage{amsfonts,amssymb}
\usepackage{graphicx}
\usepackage{pstricks}
\usepackage{subfigure}
\usepackage{bigfoot}
\usepackage{alphalph}

\author[a,b]{Marco Bochicchio}

\affiliation[a]{INFN sez. Roma 1, Piazzale A. Moro 2, Roma, I-00185, Italy}
\affiliation[b]{Scuola Normale Superiore (SNS), Piazza dei Cavalieri 7, Pisa, I-56100, Italy}
\emailAdd{marco.bochicchio@roma1.infn.it}
\abstract{Roughly speaking Morse-Smale-Floer homology associates the critical points of the action functional of a classical field theory over a manifold 
to its homology. We associate to the homology of two punctured Lagrangian submanifolds of $R^4$ intersecting at the cusps the critical points
of a quantum effective action of large-$N$ $SU(N)$ $YM$, thus realizing a quantum field-theoretical version of Lagrangian intersection Floer homology.
For this purpose we construct in $SU(N)$ $YM$ a trivial Topological Field Theory at $N=\infty$, defined by twistor Wilson loops
whose v.e.v. is 1 in the large-$N$ limit for any shape of the loops supported on certain punctured Lagrangian submanifolds. 
We derive a new version of Makeenko-Migdal loop equation for the topological twistor Wilson loops, the holomorphic loop equation, that
involves the change of variables in the $YM$ functional integral from the connection to the anti-selfdual part of the curvature
and the choice of a holomorphic gauge.  
Employing the holomorphic loop equation at $N=\infty$, and viewing Floer homology the other way around,
we associate to arcs asymptotic in both directions to the cusps of 
the Lagrangian submanifolds the critical points of an effective action implied by the holomorphic loop equation.
The critical points, being associated to the homology of the punctured Lagrangian submanifolds, consist of (magnetic) surface operators supported on the punctures,
thus realizing constructively by magnetic condensation a version of 't Hooft duality.
At the next-to-leading $\frac{1}{N}$ order a certain correlator of surface operators is non-topological and non-trivial, controls the mass gap of $YM$ theory, 
and is saturated by an infinite sum of pure poles of scalar and pseudoscalar glueballs with positive charge conjugation. 
Besides, it satisfies asymptotically for large momentum fundamental universal constraints arising from the asymptotic 
freedom and the renormalization group. We predict at large-$N$ the ratio of the masses of the two lower-mass scalar glueballs $\frac{m_{0^{++*}}}{m_{0^{++}}}=\sqrt 2=1.414 \cdots$, to be compared with the measure in lattice $SU(8)$ 
$YM$ by Meyer-Teper $\frac{m_{0^{++*}}}{m_{0^{++}}}=1.42(11)$, and with the value implied by the Particle Data Group (2014) $\frac{m_{f_0(2100)}}{m_{f_0(1500)}}=1.397(008)$.The construction extends
to massless Veneziano large-$N$ limit of $QCD$, for which we determine the lower edge of the conformal window $\frac{N_f}{N}=\frac{5}{2}$ and the corresponding quark-mass anomalous dimension $\gamma_m=-\frac{4}{5}$. }

\usepackage{braket}
\usepackage{amsmath}
\usepackage{amssymb}
\usepackage{graphicx}
\usepackage{hyperref}
\usepackage{setspace}
\usepackage{fancyhdr}
\newcommand{\gms}{g_{\overline{MS}}}

\newcommand{\Lambdams}{\Lambda_{\overline{MS}}}
\newcommand{\Lambdawb}{\Lambda_{\overline{W}}}

\newcommand{\plms}{\frac{p^2}{\Lambdams^2}}

\def\beq{\begin{equation}}
\def\eeq{\end{equation}}
\def\bea{\begin{eqnarray}}
\def\eea{\end{eqnarray}}
\def\bq{\begin{quote}}
\def\eq{\end{quote}}
\DeclareMathOperator{\Tr}{Tr}

\title{Yang-Mills mass gap, Floer homology, glueball spectrum, and conformal window in large-$N$ $QCD$}

\begin{document}
\maketitle
%
%
%


\begin{figure}[t] 
\centering
\includegraphics[width=.78\textwidth]{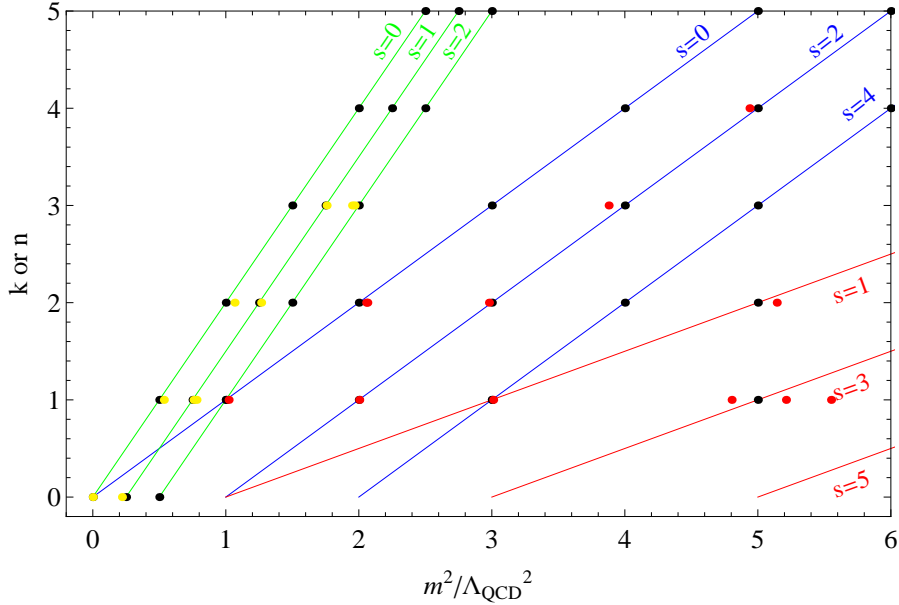}\\
\caption{Glueball and meson spectrum of massless $QCD$ in 't Hooft large-$N$ \cite{H1} limit, that in the pure-glue sector coincides with large-$N$ pure $YM$ \cite{H1}. The black points of the straight trajectory of spin $s = 0$ are the theoretical prediction of the  
$TFT$ Eq.(\ref{s}). $k$ is the internal quantum number for glueballs, that for $s = 0$ occurs in Eq.(\ref{s}). The red and yellow points represent respectively glueballs and mesons actually found in the lattice computations \cite{T0,T,L0,L}. The slope of the meson trajectories (green) is exactly the double of the slope of the even-spin glueball trajectories (blue), that in turn is the double of the slope of the odd-spin glueball trajectories (red). As an aside, we notice that the existence of slopes different by a factor of $2$ for the large-$N$ glueball trajectories challenges the standard picture that glueballs are excitations of closed string only. The plot is based on \cite{MBS} where a possible explanation is suggested.}
\label{fig:regge1}
\end{figure}

\section{Introduction and Conclusions} \label{s1}
The aim of this paper is fourfold. \par
Firstly, we show that in four-dimensional large-$N$ pure $SU(N)$ $YM$, that in the pure-glue sector coincides with $QCD$ in 't Hooft large-$N$ \cite{H1} limit, there exist some gauge-invariant observables, a special kind of  Wilson loops, that we call twistor Wilson loops for geometrical reasons explained later (see section \ref{s4}), whose vacuum expectation value (v.e.v.) is $1$ in the large-$N$ limit for any shape of the loops supported on a certain Lagrangian submanifold of Euclidean (complexified) space-time. They define a trivial Topological Field Theory ($TFT$) at $N=\infty$ underlying $YM$. We show that twistor Wilson loops can be localized at $N=\infty$, i.e. at the leading $\frac{1}{N}$ order, on a set of critical points of a certain quantum effective action, that turn out to be a special kind of surface operators, i.e. singular instantons, indexed by the integers $k=1,2,\cdots $ labeling the elements $e^{\frac{i 2 \pi k}{N}}$ of the center $Z_N$ of the gauge group. In fact, the localization on surface operators of $Z_N$ holonomy realizes in the $TFT$ by magnetic condensation a version of 't Hooft duality (see section \ref{s2}).  \par
Secondly, we show that a certain two-point connected correlator of these surface operators supported on the aforementioned Lagrangian submanifold analytically continued to Minkowski or to ultra-hyperbolic signature, at the next-to-leading $\frac{1}{N}$ order, controls the mass gap of $YM$ theory at large $N$, in fact
the joint spectrum of an infinite tower of scalar and pseudoscalar glueballs with positive charge conjugation. The spectrum of the glueballs that couple to our surface operators turns out to be:
\bea \label{s}
m_k^2= k \Lambda_{\overline W}^2
\eea
where $\Lambda_{\overline W}$ is the $YM$ renormalization-group ($RG$) invariant scale (see section \ref{s3}) in the scheme in which it coincides with the mass gap.
This statement about the exact linearity, as opposed to asymptotic linearity, of the large-$N$ joint scalar and pseudoscalar glueball spectrum of the masses squared is as strong as it sounds very unlikely, and easy to falsify by numerical lattice gauge theories computations. 
Therefore, despite the aim of this paper is to present Eq.(\ref{s}) as a deduction from first principles, we mention that Eq.(\ref{s}) agrees accurately with the first low-lying points of the glueball
spectrum computed by Meyer-Teper \cite{T0,T}, employing a supercomputer, for $SU(8)$ lattice $YM$ on the presently largest lattice ($16^3\times 24$) with the smallest value \footnote{This is the smallest value of $YM$ coupling constant $g_{YM}$ for $SU(8)$ ever reached to date in lattice computations for glueballs and perhaps in general. For a critical discussion see \cite{MBN}.} of $YM$ coupling constant ($\beta=\frac{2N}{g_{YM}^2}=45.5$): In loose words, in the lattice computation for $SU(8)$ that is presently closest \cite{MBN} to the continuum limit. Indeed, Meyer-Teper result \cite{T} for the ratios  $r_s=\frac{m_{0^{++*}}}{m_{0^{++}}}$, $r_{ps}=\frac{m_{0^{-+}}}{m_{0^{++}}}$, $r_s=r_{ps}=1.42(11)$,
agrees sharply \cite{MBN} with Eq.(\ref{s}), that implies  $r_s=r_{ps}=\sqrt{2}=1.4142 \cdots$ . 
Moreover, there is a numerical evidence \cite{MBS} that also the masses squared of glueballs of higher spin are integer valued in the large-$N$ limit in units of  $\Lambda_{\overline W}^2$ as summarized in the plot in Fig.(\ref{fig:regge1}), a fact that may be suggestive of the existence \cite{MBS} of a Topological String Theory dual to the $TFT$ of large-$N$ $YM$ (see appendix \ref{App}). \par
Thirdly, the physical identification of the spectrum in Eq.(\ref{s}) is based on our interpretation of a certain two-point correlator of surface operators \footnote{The overall normalization of our surface operators is chosen in order to match the normalization of the corresponding correlators in perturbation theory.} (see section \ref{s11}) on the Lagrangian submanifold, actually isomorphic to $R^2$ in the infinite volume limit, as the two-point correlator of the Euclidean local operator $\mathcal{O}_{ASD} \equiv \sum_{\alpha \beta} \Tr F^{-2}_{\alpha \beta} \equiv \Tr F^{-2}$, with  $F^-_{\alpha \beta}= F^{\alpha \beta}- \,^{*}\!F^{\alpha\beta}$ the anti-selfdual ($ASD$) part of the curvature of the gauge connection and $^{*}$ the Hodge dual 
$ ^{*}{F}_{\alpha \beta}=\frac{1}{2}\epsilon_{\alpha\beta\gamma\delta}F^{\gamma\delta} $ (see section \ref{s3}), analytically continued to ultra-hyperbolic signature:
\bea \label{corr}
\int \langle \mathcal{O}_{ASD}(x) \mathcal{O}_{ASD}(0) \rangle_{conn}\,e^{-ip\cdot x}d^4x 
=\frac{2} {\pi^2} \sum_{k=1}^{\infty} \frac{k^2 g_k^4\Lambda_{\overline{W}}^6}{p^2+k\Lambda_{\overline{W}}^2}  
\eea
and to Minkowski signature \footnote{In Minkowski signature the $ASD$ correlator is proportional to the difference of the scalar and pseudoscalar correlators because of the factor of $i$ in the Hodge dual  (see section \ref{s3}).
Nevertheless, according to Veltman conventions \cite{Velt2}, we maintain Euclidean notation for the momenta $p$. The analytic continuation is understood but not explicitly displayed.} ($ ^{*}{F}_{\alpha \beta}=\frac{i}{2}\epsilon_{\alpha\beta\gamma\delta}F^{\gamma\delta} $):
\bea \label{corrM}
\int \langle \mathcal{O}_{ASD}(x) \mathcal{O}_{ASD}(0) \rangle_{conn}\,e^{-ip\cdot x}d^4x =  16 \beta_0 < \frac{1}{N} \mathcal{O}_{ASD}(0)> \sum_{k=1}^{\infty}\frac{ g_k^4\Lambda_{\overline{W}}^2 }{p^2+k\Lambda_{\overline{W}}^2}
\eea
both restricted to momentum $p$ dual in Fourier sense to the aforementioned $R^2$. $g_k$ is 't Hooft (see section \ref{s3}) canonical running coupling constant $g(\frac{p^2}{\Lambda^2_{\overline{W}}})$ at the scale of the  $k$-th pole (in ultra-hyperbolic or in Minkowski space-time) $p^2= k \Lambda^2_{\overline{W}}$ in the scheme defined in Eq.(\ref{alfa}), i.e. $g_k=g(k)$. Since the lowest-mass glueball is believed to be a scalar (see section \ref{s3}) and the correlators in Eq.(\ref{corr}) and Eq.(\ref{corrM}) couple to scalars and pseudoscalars (see section \ref{s3}), Eq.(\ref{corr}) or Eq.(\ref{corrM}) suffice to imply the mass gap in the large-$N$ limit. 
We would like to make it clear that in this paper we take the point of view that functional integrals are defined by the rules by which they are computed, and therefore we refer to all the results of this paper as computations
rather than mathematical proofs. Nevertheless, we solve explicitly the $TFT$, in such a way that its existence in mathematical sense is a matter of definitions. \par
Moreover, we believe that the results of our computations are correct. Indeed, Eq.(\ref{corr}) and Eq.(\ref{corrM}) agree asymptotically \cite{MBN} for large momentum (see section \ref{s3}), up to physically-irrelevant contact terms, with the universal, i.e. the scheme-independent, leading and next-to-leading logarithms of the first-two coefficient functions of the $RG$-improved operator product expansion ($OPE$), according to fundamental principles of the $RG$ and the asymptotic freedom of $YM$ (see section \ref{s3}), in Euclidean (and ultra-hyperbolic) signature:
\bea \label{opeE}
\int\langle \mathcal{O}_{ASD}(x) \mathcal{O}_{ASD}(0) \rangle_{conn}\,e^{-ip\cdot x}d^4x 
\sim C^{(0)}_{ADS}(p^2) 
+ 0 <\frac{1}{N} \mathcal{O}_{ASD}(0)> \nonumber \\
\eea  
and in Minkowski signature: 
\bea \label{opeM}
\int\langle \mathcal{O}_{ASD}(x) \mathcal{O}_{ASD}(0) \rangle_{conn}\,e^{-ip\cdot x}d^4x 
 \sim p^4  0 
+  C^{(1)}_{ADS}(p^2) < \frac{1}{N} \mathcal{O}_{ASD}(0)> \nonumber \\
\eea
with the coefficients denoted by $0$ vanishing up to scheme-dependent terms on the order of $\frac{1}{\log^2\plms}$, and (see section \ref{s3}):
\begin{align}\label{OPE1}
C^{(0)}_{ASD}(p^2) = \frac{2 p^4}{\pi^2\beta_0}\Biggl[\frac{1}{\beta_0\log\frac{p^2}{\Lambdams^2}}\Biggl(1-\frac{\beta_1}{\beta_0^2}\frac{\log\log\frac{p^2}{\Lambdams^2}}{\log\frac{p^2}{\Lambdams^2}}\Biggr)+O\biggl(\frac{1}{\log^2\plms}\biggr)\Biggr]
\end{align} 
\,
\bea \label{OPE2}
C^{(1)}_{ADS}(p^2) = 16 \Biggl[\frac{1}{\beta_0 \log\frac{p^2}{\Lambdams^2}}\biggl(1-\frac{\beta_1}{\beta_0^2}\frac{\log\log\frac{p^2}{\Lambdams^2}}{\log\frac{p^2}{\Lambdams^2}}\biggr)
+O\biggl(\frac{1}{\log^2\frac{p^2}{\Lambdams^2}}\biggr) \Biggr]
\eea
\par
The vanishing of the universal part of either the second or the first coefficient function in the $OPE$ of the $ASD$ correlator, in Euclidean (and ultra-hyperbolic) signature and in Minkowski signature respectively, is due to 
peculiar partial cancellations (see section \ref{s3}) that occur taking the appropriate linear combination of the scalar and pseudoscalar correlators: They are implied both in the $TFT$ (see section \ref{s11}), and in massless $QCD$ on the basis of remarkable recent computations by Chetyrkin-Zoller \cite{Che3} and Zoller \cite{Zo} (see section \ref{s3}). \par
In particular, in the $TFT$ a powerful non-trivial check of the correct asymptotic behavior \cite{SM, MBN} of the $ASD$ correlators in Eq.(\ref{corr}) and Eq.(\ref{corrM}) occurs as the consequence of the conspiracy \cite{SM} between the fourth power $g^4(k)$ of the running coupling evaluated on shell, that arises by the anomalous dimension of our surface operators, and the linearity of the spectrum of the masses squared (see section \ref{s10}), that arises by the localization on surface operators of $Z_N$ holonomy (see section \ref{s2}). \par
We mention that in $n=1$ $SUSY$ $YM$ the aforementioned cancellations are complete due to space-time supersymmetry and the $ASD$ correlator actually vanishes \cite{Shif} in Minkowski space-time up to physically-irrelevant contact terms, as it has been proved by Shifman \cite{Shif}, by means of methods inspired by some of the techniques employed in this paper (i.e. the change to the $ASD$ variables, known also as Nicolai map (see section \ref{s5})). \par
The aforementioned agreement with fundamental principles of the $RG$ and the asymptotic freedom is presently a unique feature of Eq.(\ref{corr}) and Eq.(\ref{corrM}), not shared by any other proposal, in particular based on the $AdS$ String/ Large-$N$ Gauge Theories correspondence (see \cite{MBN,SM} and references therein for a critical discussion), for the scalar or pseudoscalar glueball propagators. \par
Fourthly, we extend the construction of the $TFT$ and the results for the $ASD$ correlator of surface operators to massless Veneziano limit \cite{Veneziano} of large-$N$ $QCD$, $N \rightarrow \infty, \frac{N_f}{N}=const$, with $N_f$ quarks in the fundamental representation (see section \ref{s12}). In the flavor-singlet sector, where the $ASD$ propagator of the $TFT$ lives, pure-glue operators mix with flavor-singlet quark operators in such a way that in general there is no clear distinction between glueballs and flavor-singlet mesons in Veneziano limit (see \cite{Veneziano} for a neat review and detailed estimates of the couplings 
of glueballs and mesons both in 't Hooft and Veneziano limits). \par
In particular the $ASD$ operator $\mathcal{O}_{ASD}$ in $QCD$ is a linear combination of the scalar $\mathcal{O}_{S}$ and pseudoscalar $\mathcal{O}_{P}$ pure-glue operators (see section \ref{s3}), and mixes with the divergence of the flavor-singlet axial current $\sum_{f} \partial_{\alpha} J_{f}^{5 \alpha}$ because of the chiral anomaly \cite{Che3,Zo}. 
But the mixing with the axial current affects the terms that involve the anomalous dimension in the Callan-Symansik equation only at order of $g^4$ \cite{Zo}, and therefore it does not affect the leading universal asymptotic behavior of the $ASD$ correlator but for the change of the coefficients of the beta function in Eq.(\ref{OPE1}) and Eq(\ref{OPE2}). Moreover, in the $TFT$ we compute only the correlator of $ASD$ surface operators, that field-theoretically are pure-glue operators. \par
Accordingly, we compute the $ASD$ correlator in massless Veneziano limit of $QCD$ in the local approximation (see section \ref{s12}) of the effective action in the $TFT$, from which we extract the critical value of $\frac{N_f}{N}=\frac{5}{2}$ at which the lower edge of the conformal window occurs,
and the corresponding value of the quark-mass anomalous dimension $\gamma_m=-\frac{\partial \log Z_m}{\partial \log M}=-\frac{4}{5}$ in the massless limit (see section \ref{s12}). \par
The lower edge of the conformal window occurs at the value of $\frac{N_f}{N}=\frac{5}{2}$ for which the kinetic term  $p^2$ in the $ASD$ correlator in the $TFT$ changes sign due to the quark contribution to the effective action, i.e. the kinetic part of the Hessian of the effective action at the critical points changes sign, signaling an instability of fluctuations around the magnetic condensate of surface operators, i.e. a phase transition from confinement to another phase, identified with a conformal Coulomb phase. We show that at this value the canonical beta function in our family of schemes develops an
infrared zero that occurs independently on the scheme (see section \ref{s12}). The argument mimics Seiberg argument for the lower edge of the conformal window in $n=1$ $SUSY$ gauge theories (see section \ref{s12}). In the last case the lower edge is determined as the point at which the anomalous dimension of operators in the holomorphic ring
reaches the unitarity limit, and correspondingly a zero of the $NSVZ$ beta-function \cite{NSVZ} occurs. \par
Remarkably, the canonical beta function in the $TFT$ exactly reproduces by first principles the universal part of the perturbative large-$N$ $QCD$ beta function in Veneziano limit (see section \ref{s12}).\par
Moreover, we compute the value of the ratio of the two lower-mass glueballs \footnote{Strictly speaking we cannot determine the parity of the glueball states, but only the joint spectrum of any parity, because the $ASD$ correlator
couples to scalar and pseudoscalars.} at the leading $\frac{1}{N}$ order, that turns out to be the same both in 't Hooft and in massless Veneziano large-$N$ limits: $\frac{m_{0^{++*}}}{m_{0^{++}}}=\sqrt 2=1.414 \cdots$, the only difference between the local parts of the effective action in the two limits being the change of the beta function that defines the $RG$-invariant scale $\Lambda_{\overline W}$, but the subtle difference being that in 't Hooft limit the effective action of the $TFT$ is large-$N$ exact (see section \ref{s2}), while in massless Veneziano limit only the local part of the effective action of the $TFT$ is in fact large-$N$ exact (see section \ref{s12}). In fact, the two-point correlators in Veneziano limit cannot be saturated only by single-particle states, because the widths are on the order of $1$, as opposed to 't Hooft limit.
Therefore, the fact that the spectrum is of pure poles also in Veneziano limit of the $TFT$ is an artifact of the local approximation for the effective action. \par
Here we outline the mathematical features of our approach. Our surface operators belong to a new $TFT$ underlying pure $YM$. For clarity we recall briefly topological field theories in the supersymmetric ($SUSY$) setting, because of the loose connection with our non-$SUSY$ $TFT$.
The basic idea of $SUSY$ topological field theories \cite{W1} 
is to view certain functional integrals as cohomology classes associated to a nilpotent differential $Q^2=0$ defined by the $BRS$ charge $Q$ obtained by a
twist of the supersymmetry, with the property $\int Q \alpha=0$ for any $\alpha$. 
Thus these physical $SUSY$ theories contain a very special topological subsector defined by closed forms $Q C=Q S_{SUSY}= 0$ modulo total differentials $Q \alpha$, the cohomology class $[C]$ of $C$.
Moreover, the $SUSY$ topological field theory is often solvable, since the cohomology classes $[C]$ are localized on critical points by means of deformations $Q \alpha$ trivial in cohomology:
\begin{equation}
[C]= \int C e^{-S_{SUSY}} =\lim_{t\rightarrow \infty} \int C e^{-S_{SUSY}-t Q\alpha}
\end{equation}
because the saddle-point approximation for large $t$, being $t$ independent for the class $[C]$, is in fact exact \cite{DE,A,Bis1,W}. The aforementioned localization extends to the whole cohomology ring generated by the closed forms. Hence a small \emph{subset} of the observables of the theory is localized on critical points. For a review of cohomological localization see \cite{Szabo}. 
The canonical example from the physics point of view is the gluino condensate \cite{NSVZ}, for which cohomological localization on critical points (see section 3.2 in \cite{GGI}), in this case on gauge connections with vanishing $ASD$ curvature, leads to the exact $NSVZ$ beta function \cite{NSVZ} of $n=1$ $SUSY$ $YM$. \par
The canonical example from the mathematical point of view leads to a twisting of the supersymmetry of $n=1$ $SUSY$ $YM$ and to a field theoretical interpretation \cite{W1} of Donaldson invariants by localization
of the functional integral in the topological sector of $n=1$ $SUSY$ $YM$ on the very same configurations with vanishing $ASD$ curvature. \par
This cannot work in pure $YM$, because of the lack of the differential of the would-be cohomology, i.e. the lack of $SUSY$. \par
However, in the finite dimensional setting, Morse-Smale theory \cite{Morse} (and references therein) associates to the critical points of a function on a compact manifold the homology groups of the manifold. 
We recall briefly Morse theory. Given a non-degenerate Morse function $f$ on a manifold,
Morse inequality bounds from below the number $N_f$ of critical points of $f$ by the sum of the dimension of the homology 
groups of the manifold: $N_f  \geq \sum_i \dim H_i(M)$. 
Moreover, Morse-Smale homology allows us to reconstruct the homology groups directly from the knowledge of the critical points. \par
The basic idea of Smale is to connect two critical points, whose Morse index differs by $1$, by arcs associated to the
gradient flow of $f$ on the manifold. Then to the arcs it is associated a complex, that turns out to be isomorphic to
the usual singular homology of the manifold. Morse index is the number of negative eigenvalues of the Hessian of $f$ at the critical points. Hence Morse-Smale homology involves necessarily both the stable and the unstable manifold of the critical points, since the arcs of Smale flow connect necessarily the unstable with the stable manifolds. \par
Besides, in the infinite dimensional setting, Floer theory \cite{Floer} (and references therein) associates to the critical points of a \emph{classical} field theory on a manifold certain Floer homology groups, that are topological invariants of the underlying manifold. \par
The basic idea of Floer homology is to extend Smale construction to infinite dimension, i.e. to a field theory over a manifold, for which the function $f$ is now the \emph{classical} action functional of the theory, by allowing Morse index 
to be infinite. Floer preserves the construction of the gradient flow associated to the functional $f$, by requiring that the flow connects two critical points
whose relative Morse index is finite and equal to $1$, despite each Morse index may actually be infinite. 
Floer shows that in certain cases this construction defines new homology groups $HF(M)$ that are topological invariants of the manifold, that may, or may not in the infinite dimensional setting,
be isomorphic to the ordinary singular homology $H(M)$ of $M$. 

\section{A synopsis of the main argument and plan of the paper} \label{s2}

To say it in a nutshell, we replace cohomology in function space (i.e. $SUSY$) with homology in submanifolds of space-time (see section \ref{s1}), in order to localize on critical points of a \emph{quantum} effective action in large-$N$ $SU(N)$ $YM$. \par
Indeed, our construction associates critical points of a quantum effective action of large-$N$ $YM$ to the intersection homology of certain Lagrangian submanifolds of (complexified) space-time
(see Fig.\ref{fig:1} and Fig.\ref{fig:2}): The critical points occur at the cusps that lie at the intersection of two Lagrangian submanifolds. Conceptually, our construction is a quantum field-theoretical version of Lagrangian intersection Floer homology, that instead, going in the opposite direction, associates the intersection homology of the Lagrangian submanifolds to the critical points of certain functionals \cite{Wiki,Floer}. \par
Thus, roughly speaking, our strategy consists in constructing a very special topological subsector of pure $YM$, that represents by gauge-invariant observables numerical topological invariants defined on homology classes $[L_{zz}]$ in the loop algebra of closed curves $L_{zz}$ based over a point $z$ on a complex \emph{submanifold} of space-time $R^4$ (or of its complexification $C^4$). Then we associate, by means of new field theoretical methods, to the aforementioned homology classes the critical points of a \emph{quantum} effective action of $YM$ \emph{restricted} \footnote{This restriction is the counterpart in homology of the restriction to the cohomology ring in cohomological localization.} to the subalgebra of gauge-invariant observables in the topological subsector. Therefore, our construction can be regarded as Morse-Smale-Floer homology (see section \ref{s1}) seen the other way around: From the homology classes to the critical points. \par
The natural candidates to represent homology classes of closed curves in gauge theories are the Wilson 
loops. However, in pure $YM$ in general they represent the loop algebra over the manifold (see \cite{MBT} for a short summary of the properties of the representation by Wilson loop operators of the loop algebra at large-$N$), but this representation
does not lift to homology classes. Indeed, the v.e.v. of Wilson loops in gauge theories is not in general homotopy invariant, as opposed to the homology groups. This is due to the fact that Wilson loops in general capture detailed information about the physics of pure $YM$, that certainly is not a topological field theory. \par
Therefore, we define in section \ref{s4} special Wilson loops $\Psi(B_{\lambda}; L_{ww})= P \exp i \int_{L_{ww}} (B_{\lambda})_z dz+ (B_{\lambda})_{\bar z} d \bar z $ constructed by means of a modified gauge connection $B_{\lambda}$, that we call twistor Wilson loops for geometrical reasons explained in section \ref{s4}, whose v.e.v. is a numerical homotopy and homology invariant. The twistor connection $B_{\lambda}$ depends on a complex non-vanishing fixed parameter $\lambda$. $B_{\lambda}$ is a functional of the ordinary $YM$ connection $A$. Thus we introduce new gauge-invariant observables, but we do not change the action of $YM$ theory, that defines the v.e.v. through the functional integral. In fact, twistor Wilson loops in the adjoint representation 
are completely trivial homotopy 
and homology invariants, but only in the large-$N$ limit of $SU(N)$ $YM$ theory: Their v.e.v. $<\frac{1}{\mathcal N}Tr\Psi(B_{\lambda}; L_{ww})>$ is $1$ in the large-$\mathcal N$ limit ($\mathcal{N}=N\hat N$, see section \ref{s4}) for any shape of loops supported on certain Lagrangian submanifolds immersed in  $R^4$ or in its complexification $C^4$. Yet, the very existence of twistor Wilson loops is non-trivial. \par
By a completely different argument, Witten argued \footnote{Talk at the Simons Center workshop \emph{Mathematical Foundations of Quantum Field Theories}, Jan 2012. See in this respect \cite{MBT,MB1}.} that every theory with a mass gap should contain a possibly trivial topological field theory in the infrared. But what makes the existence of twistor Wilson loops a powerful tool is their triviality at all scales, in particular in the ultraviolet. Indeed, it is precisely the large-$N$ ultraviolet triviality of twistor Wilson loops, that implies the absence of cusp anomalies (see below), that allows us to localize them on critical points: A feature 
much more specific than the 
triviality of the topological theory in the infrared. 
Besides, the adjoint twistor Wilson loops factorize in the large-$N$ limit into twistor Wilson loops in the fundamental and conjugate representation, that turn out to be homology invariants valued in the center $Z_N$ of the gauge group $SU(N)$. Twistor Wilson loops in the fundamental representation are the building blocks of our new $TFT$ underlying the large-$N$ limit of $YM$.
The main conceptual point of this paper is that twistor Wilson loops satisfy a new kind of loop equation, the holomorphic loop equation \cite{3MB} derived in section \ref{s6}:
\bea \label{hol0}
&&<\frac{1}{\mathcal N}Tr(\Psi(B_{\lambda}; L^{(1)}_{zz})\frac{\delta \Gamma}{\delta \mu_{\lambda}(z, \bar z)}\Psi(B_{\lambda}; L^{(2)}_{zz}))> \nonumber \\
&&=\frac{1}{ \pi } \int_{L_{zz}} \frac{ dw}{z-w} <\frac{1}{\mathcal N}Tr\Psi(B_{\lambda}; L^{(1)}_{zw})> <\frac{1}{\mathcal N}Tr\Psi(B_{\lambda}; L^{(2)}_{wz})>
 \eea
that involves a change of variables in the functional integral that defines $YM$ theory. One of these new variables is the curvature $-i F_{z \bar z}(B_{\lambda})=\mu_{\lambda}$ of the connection $B_{\lambda}$
that occurs in twistor Wilson loops. $\mu_{\lambda}$ turns out to be a non-Hermitian field of $ASD$ type: $\mu_{\lambda} = \frac{1}{2} F^-_{01}+\frac{1}{4} \lambda^{-1}(F^-_{02}+i F^-_{03}) - \frac{1}{4} \lambda (F^-_{02}-i F^-_{03})$.\par
In fact, the aforementioned change of variables is the map, defined in section \ref{s5}, from the $3$ independent components of the connection $A_{\alpha}$ in any fixed gauge to the $3$ independent components of the $ASD$
field $F^-_{\alpha \beta}$, composed with the change to $\mu_{\lambda}$ in a holomorphic gauge $(B_{\lambda})_{\bar z} =0$ (see section \ref{s6}). This map is implemented
employing the resolution of identity in the functional integral in the original variables \cite{3MB}:
\bea \label{id}
1=\int \delta(F_{\alpha\beta}^-(A)-\mu_{\alpha\beta}^-) \delta\mu_{\alpha\beta}^-
\eea
that allows us to compute the Jacobian of the change to the $ASD$ variables (see section \ref{s5}), by integrating exactly in any gauge on the gauge connection $A$ in the functional integral in the original variables, because of the delta functional (see section \ref{s5}).
Since the change to the $ASD$ variables is already a non-standard tool in quantum $YM$ theories, with A. Pilloni we have proved \cite{BP} the identity of the one-loop perturbative one-particle irreducible ($1PI$) effective action of $YM$, $QCD$ and $n=1$ $SUSY$ $YM$, and more generally of any gauge theory that extends pure $YM$, in the original and in the $ASD$ variables.
Therefore, we think that the change to the $ASD$ variables is sufficiently tested at perturbative level to be employed in our non-perturbative approach. \par
Coming back to our $TFT$, we choose as support of twistor Wilson loops a Lagrangian surface with the topology of a high-genus Riemann surface immersed in space-time with a canonical basis of $a$ cycles and certain $b$ cycles
Fig.\ref{fig:1}. We make this choice in order to get a large homology group
and potentially a large number of critical points in our $TFT$, according to Morse-Smale-Floer theory. We specify later in this section the concrete realization of the Lagrangian surface. For our initial considerations only topology matters.
Then we use the freedom to deform the loops of our trivial $TFT$. We deform the Lagrangian surface in such a way that the $b$ cycles degenerate to nodal points Fig.\ref{fig:2}. 
Each node is then the intersection point of a pair of cusps
belonging to the deformed $a$ cycles of the Lagrangian nodal surface Fig.\ref{fig:2}. The first reason for introducing $a$ cycles intersecting at cusps is that loop equations carry quantum information precisely at the intersection points.
The second reason will become apparent momentarily. Thus pairs of deformed $a$ cycles form a loop $L$ of Makeenko-Migdal ($MM$) type \cite{MM,MM1}, with the shape of the symbol $\infty$, that decomposes into two petals $L^{(1)}$ and $L^{(2)}$. 
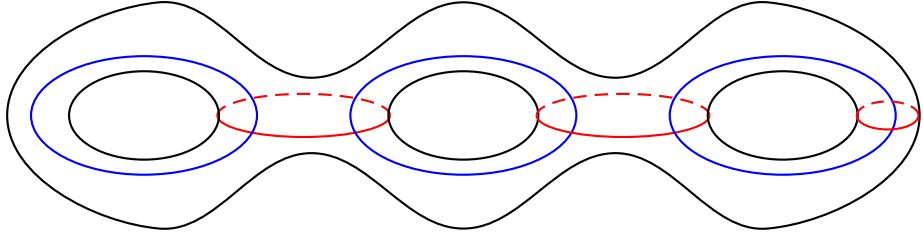
\begin{figure}[t]
\centering
\begin{pspicture}(12,4)(0,0)
 %
 \psccurve(0,2)(2,0.5)(4,1.5)(6,0.5)(8,1.5)(10,0.5)(12,2)(10,3.5)(8,2.5)(6,3.5)(4,2.5)(2,3.5)
 \psellipticarc[linecolor=red](3.9,2)(1.15,.3){180}{0}
 \psellipticarc[linecolor=red,linestyle=dashed](3.9,2)(1.15,.3){0}{180}
 \psellipse(1.8,2)(1,0.6)
 \psellipse[linecolor=blue](1.8,2)(1.5,0.8)
 \psellipse(6,2)(1,0.6)
 \psellipse[linecolor=blue](6,2)(1.5,0.8)
 \psellipticarc[linecolor=red](8.1,2)(1.15,.3){180}{0}
 \psellipticarc[linecolor=red,linestyle=dashed](8.1,2)(1.15,.3){0}{180}
 \psellipse[linecolor=blue](10.2,2)(1.5,0.8)
 \psellipse(10.2,2)(1,0.6)
 \psellipticarc[linecolor=red](11.58,2)(.42,.2){180}{0}
 \psellipticarc[linecolor=red,linestyle=dashed](11.58,2)(.42,.2){0}{180}

\end{pspicture}
\caption{A high genus Lagrangian surface immersed in $C^4$, which twistor Wilson loops are supported on, with a basis of $a$ cycles in blue and certain $b$ cycles in red. A real structure is introduced choosing the surface to be the topological double of a sphere with the $b$ cycles as boundaries.}
\label{fig:1}
\end{figure}
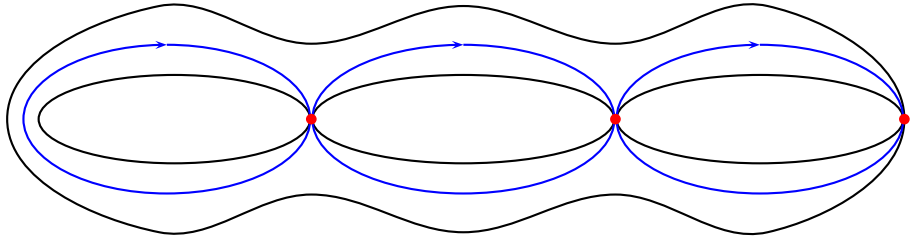
\begin{figure}[t]
\centering
\begin{pspicture}(12,4)(0,0)
 %
 \psccurve(0,2)(2,0.5)(4,1)(6,0.5)(8,1)(10,0.5)(11.8,2)(10,3.5)(8,3)(6,3.5)(4,3)(2,3.5)
 \psellipse(2.2,2)(1.8,0.6)
 \psellipse(6,2)(2,0.6)
 \psellipse(9.9,2)(1.9,0.6)
 \psellipticarc[linecolor=blue]{<-}(2.1,2)(1.9,1){90}{90}
 \psellipticarc[linecolor=blue]{<-}(6,2)(2,1){90}{90}
 \psellipticarc[linecolor=blue]{<-}(9.9,2)(1.9,1){90}{90}
 \psdot[linecolor=red,fillcolor=red,dotstyle=*,dotsize=4pt](4,2)
 \psdot[linecolor=red,fillcolor=red,dotstyle=*,dotsize=4pt](8,2)
 \psdot[linecolor=red,fillcolor=red,dotstyle=*,dotsize=4pt](11.8,2)
\end{pspicture}
\caption{ All the the $b$ cycles but the one on the extreme left degenerate to nodal points. Correspondingly, pairs of $a$ cycles form a Makeenko-Migdal loop with the shape of the symbol $\infty$ intersecting at the nodal points.
The orientation of the $a$ cycles implies that the $RHS$ of the holomorphic loop equation vanishes at each cusp of the two mirror surfaces intersecting at the nodal points.
As a consequence the holomorphic loop equation reduces to a critical equation for a quantum effective action at the nodal points of the Lagrangian submanifold. The construction can be viewed as Lagrangian intersection Floer homology seen the other way around: from the intersection homology of the resulting two mirror Lagrangian submanifolds to the critical points.}
\label{fig:2}
\end{figure}

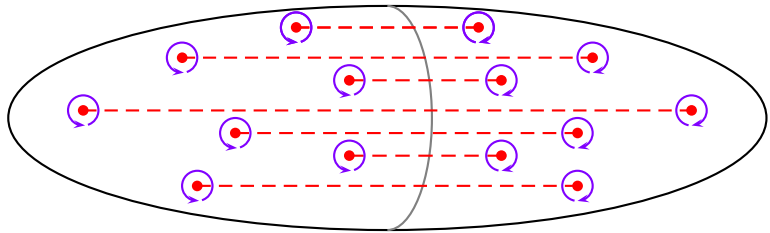
\begin{figure}[t]
\centering
\begin{pspicture}(12,4)(0,0)
 %
 \newrgbcolor{mpurple}{.5 0 1}
 \psellipse(6,2)(5,1.5)
 \psellipticarc[linecolor=gray](6,2)(0.6,1.5){-90}{90}
 \psdots[linecolor=red,fillcolor=red,dotstyle=*,dotsize=4pt](3.3,2.8)(5.5,1.5)(3.5,1.1)(4.8,3.2)(4,1.8)(5.5,2.5)(2,2.1)
 \psdots[linecolor=red,fillcolor=red,dotstyle=*,dotsize=4pt](8.7,2.8)(7.5,1.5)(8.5,1.1)(7.2,3.2)(8.5,1.8)(7.5,2.5)(10,2.1)
 \psline[linecolor=red,linestyle=dashed](4.8,3.2)(7.2,3.2)
 \psline[linecolor=red,linestyle=dashed](5.5,2.5)(7.5,2.5)
 \psline[linecolor=red,linestyle=dashed](5.5,1.5)(7.5,1.5)
 \psline[linecolor=red,linestyle=dashed](3.5,1.1)(8.5,1.1)
 \psline[linecolor=red,linestyle=dashed](4,1.8)(8.5,1.8)
 \psline[linecolor=red,linestyle=dashed](2,2.1)(10,2.1)
 \psline[linecolor=red,linestyle=dashed](4.8,3.2)(7.2,3.2) 
 \psline[linecolor=red,linestyle=dashed](3.3,2.8)(8.7,2.8)
 \psarc[linecolor=mpurple]{->}(4.8,3.2){.2}{-72}{-83}
 \psarc[linecolor=mpurple]{->}(5.5,2.5){.2}{-72}{-83}
 \psarc[linecolor=mpurple]{->}(5.5,1.5){.2}{-72}{-83}
 \psarc[linecolor=mpurple]{->}(3.5,1.1){.2}{-72}{-83}
 \psarc[linecolor=mpurple]{->}(4,1.8){.2}{-72}{-83}
 \psarc[linecolor=mpurple]{->}(2,2.1){.2}{-72}{-83}
 \psarc[linecolor=mpurple]{->}(4.8,3.2){.2}{-72}{-83}
 \psarc[linecolor=mpurple]{->}(3.3,2.8){.2}{-72}{-83}
 \psarc[linecolor=mpurple]{<-}(7.2,3.2){.2}{270}{259}
 \psarc[linecolor=mpurple]{<-}(7.5,2.5){.2}{270}{259}
 \psarc[linecolor=mpurple]{<-}(7.5,1.5){.2}{270}{259}
 \psarc[linecolor=mpurple]{<-}(8.5,1.1){.2}{270}{259}
 \psarc[linecolor=mpurple]{<-}(8.5,1.8){.2}{270}{259}
 \psarc[linecolor=mpurple]{<-}(10,2.1){.2}{270}{259}
 \psarc[linecolor=mpurple]{<-}(7.2,3.2){.2}{270}{259}
 \psarc[linecolor=mpurple]{<-}(8.7,2.8){.2}{270}{259} 
\end{pspicture}
\caption{The normalizing surface of the nodal surface with a real structure is a punctured sphere with punctures pairwise identified, that can be thought of as two punctured disks glued at their boundaries. The pairwise identification of the puncture implies that surface operators supported on the punctures have the same holonomy for opposite orientations of the cycles around pairwise identified punctures. Assuming the global gauge group to be unbroken, surface operators with $Z_N$ holonomy around the punctures must occur at
the critical points. }
\label{fig:3}
\end{figure}

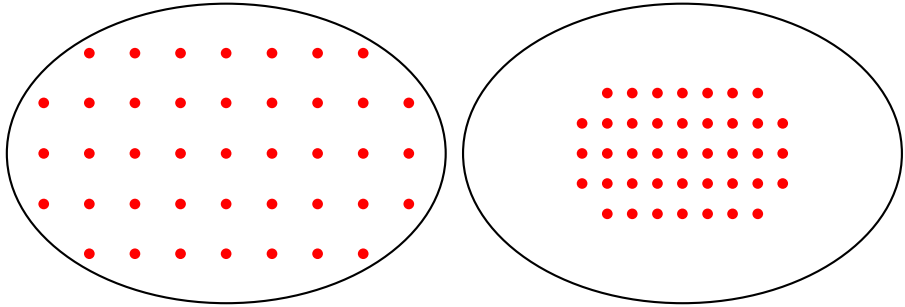
\begin{figure}[t]
\centering
\begin{pspicture}(12,4)(0,0)
\psellipse(3.00,2.00)(2.90,2.00)
\psellipse(9.00,2.00)(2.90,2.00)
\psdots[linecolor=red,fillcolor=red,dotstyle=*,dotsize=4pt](1.20,0.67)(1.80,0.67)(2.40,0.67)(3.00,0.67)(3.60,0.67)(4.20,0.67)(4.80,0.67)
\psdots[linecolor=red,fillcolor=red,dotstyle=*,dotsize=4pt](0.60,1.33)(1.20,1.33)(1.80,1.33)(2.40,1.33)(3.00,1.33)(3.60,1.33)(4.20,1.33)(4.80,1.33)(5.40,1.33)
\psdots[linecolor=red,fillcolor=red,dotstyle=*,dotsize=4pt](0.60,2.00)(1.20,2.00)(1.80,2.00)(2.40,2.00)(3.00,2.00)(3.60,2.00)(4.20,2.00)(4.80,2.00)(5.40,2.00)
\psdots[linecolor=red,fillcolor=red,dotstyle=*,dotsize=4pt](0.60,2.67)(1.20,2.67)(1.80,2.67)(2.40,2.67)(3.00,2.67)(3.60,2.67)(4.20,2.67)(4.80,2.67)(5.40,2.67)
\psdots[linecolor=red,fillcolor=red,dotstyle=*,dotsize=4pt](1.20,3.33)(1.80,3.33)(2.40,3.33)(3.00,3.33)(3.60,3.33)(4.20,3.33)(4.80,3.33)
\psdots[linecolor=red,fillcolor=red,dotstyle=*,dotsize=4pt](8.01,1.2)(8.34,1.2)(8.67,1.2)(9.,1.2)(9.33,1.2)(9.66,1.2)(9.99,1.2)
\psdots[linecolor=red,fillcolor=red,dotstyle=*,dotsize=4pt](7.68,1.6)(8.01,1.6)(8.34,1.6)(8.67,1.6)(9.,1.6)(9.33,1.6)(9.66,1.6)(9.99,1.6)(10.32,1.6)
\psdots[linecolor=red,fillcolor=red,dotstyle=*,dotsize=4pt](7.68,2.)(8.01,2.)(8.34,2.)(8.67,2.)(9.,2.)(9.33,2.)(9.66,2.)(9.99,2.)(10.32,2.)
\psdots[linecolor=red,fillcolor=red,dotstyle=*,dotsize=4pt](7.68,2.4)(8.01,2.4)(8.34,2.4)(8.67,2.4)(9.,2.4)(9.33,2.4)(9.66,2.4)(9.99,2.4)(10.32,2.4)
\psdots[linecolor=red,fillcolor=red,dotstyle=*,dotsize=4pt](8.01,2.8)(8.34,2.8)(8.67,2.8)(9.,2.8)(9.33,2.8)(9.66,2.8)(9.99,2.8)
\end{pspicture}
\caption{ The normalizing surface is the union at the boundaries of two topological disks with
$N_2$ punctures pairwise identified. Two weights, $a_{UV}$ and $a_{IR}$, that
are the lattice spacing on the two disks, are introduced. The continuum limit can be defined by $a_{UV,IR} \rightarrow 0$ with $N_2 \rightarrow \infty $ and $N_2 a^2_{UV,IR}=const$. A scaling limit can be defined by $N_2 \rightarrow \infty$ with $a_{UV,IR} = const $.}
\label{fig:4}
\end{figure}
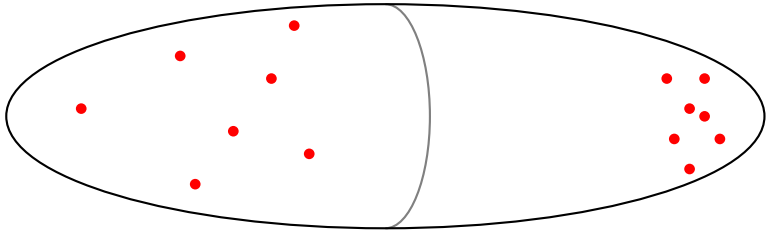
\begin{figure}[t]
\centering
\begin{pspicture}(12,4)(0,0)
 %
 \psellipticarc[linecolor=gray](6,2)(0.6,1.5){-90}{90}
 \newrgbcolor{mpurple}{.5 0 1}
 \psellipse(6,2)(5,1.5)
 \psdots[linecolor=red,fillcolor=red,dotstyle=*,dotsize=4pt](10.4,1.7)(10.2,2.5)(10.2,2)(10,1.3)(9.8,1.7)(9.7,2.5)(10,2.1)
 \psdots[linecolor=red,fillcolor=red,dotstyle=*,dotsize=4pt](3.3,2.8)(5,1.5)(3.5,1.1)(4.8,3.2)(4,1.8)(4.5,2.5)(2,2.1)
\end{pspicture}
\caption{The Wilsonian scheme for the Wilsonian coupling $g_W$ corresponds to the continuum limit with the Wilsonian normalization of the local part of the effective action, with the same choice of cutoffs and subtraction scales at the $UV$ and at the $IR$. The canonical scheme for the canonical coupling $g$ corresponds to the continuum limit for the local part of the effective action in canonical form, with the subtraction scale at the $UV$ on the order of the cutoff and the subtraction scale at the $IR$ close to the scale of Landau infrared singularity of the Wilsonian coupling.}
\label{fig:5}
\end{figure}
We need a gauge-invariant regularization of the right-hand side ($RHS$) of the holomorphic loop equation \cite{3MB}, that is obtained introducing a real structure on our Riemann surface (see \cite{real} and references therein), that therefore is chosen to be the topological double of a sphere with boundary circles, all of which but one degenerate to cusps Fig.\ref{fig:2}. Thus the cusps are now real points of the Riemann surface. For real points the result of the $i \epsilon$ regularization of the Cauchy kernel \footnote{Alternatively a real Cauchy kernel is obtained by analytic continuation to Minkowski space-time \cite{3MB}.} is the sum of
two distributions, the principal part of the real Cauchy kernel and
a one-dimensional delta function:
\bea \label{prin}
\frac{1}{w_+ - z_+  +i\epsilon}=  P\frac{1}{w_+ - z_+ }- i \pi \delta(w_+ -z_+)
\eea
Hence for real points $z=z_+$, the holomorphic loop equation reduces to:
\bea
&&<Tr(\Psi(B_{\lambda}; L^{(1)}_{z_+z_+})\frac{\delta \Gamma}{\delta \mu_{\lambda}(z_+, z_+)}\Psi(B_{\lambda}; L^{(2)}_{z_+z_+}))> \nonumber \\
&&= i \int_{L_{z_+z_+}}  dw_+ \delta(w_+ -z_+) <Tr\Psi(B_{\lambda}; L^{(1)}_{z_+w_+})> <Tr\Psi(B_{\lambda}; L^{(2)}_{w_+z_+})> 
\eea
because by gauge invariance the principal part does not contribute \cite{3MB}, being supported on open loops for which, by (non-)gauge invariance \cite{Mak2}, $0=<Tr\Psi(B_{\lambda}; L_{w_+z_+})>$.
As a consequence the $RHS$ of the holomorphic loop equation is supported on closed loops only, as it must be by gauge invariance, and it is non-vanishing in general, as it is the $RHS$ of the $MM$ loop equation \cite{MM,MM1}: It represents the quantum contribution, that is the obstruction to localize the loop equation on the critical points defined by the
equation of motion that occurs in the $LHS$. 
But the remarkable fact about the holomorphic loop equation, as opposed to the $MM$ equation, is that for \emph{nodal} points the $RHS$ of the holomorphic loop equation vanishes, provided the $a$ cycles on the Riemann surface are oriented in the same way. Indeed, in this case pairwise intersecting cycles are asymptotic to each cusp in both directions Fig.\ref{fig:2}. It follows that the contribution in the $RHS$ of the holomorphic loop equation of each 
backtracking cusp intersecting at a node is exactly $0$:
\bea
\int dw_+(s)
\delta(z_+(s_{cusp}) -w_+(s))&=&\frac{1}{2}\frac{\dot w_+(s^+_{cusp})}{ |\dot w_+(s^+_{cusp})|}+ \frac{1}{2}
\frac{\dot w_+(s^-_{cusp})}{|\dot w_+(s^-_{cusp})|} =0
\eea
because of the opposite orientation of the arcs asymptotic to the cusps.
Thus the holomorphic  loop equation associates to each node of the Lagrangian surface with a real structure a critical equation for the quantum effective action $\Gamma$:
\bea \label{101}
<Tr(\Psi(B_{\lambda}; L^{(1)}_{z_+z_+})\frac{\delta \Gamma}{\delta \mu_{\lambda}(z_+, z_+)}\Psi(B_{\lambda}; L^{(2)}_{z_+z_+}))>=0
\eea
restricted to the loop algebra generated by twistor Wilson loops, that is our homological counterpart of the cohomology ring in cohomological localization (see section \ref{s1}).
We can rewrite Eq.(\ref{101}) as:
\bea \label{102}
<\Psi(B_{\lambda}; L^{(1)}_{z_+z_+})|\frac{\delta \Gamma}{\delta \mu_{\lambda}(z_+, z_+)}|\Psi(B_{\lambda}; L^{(2)}_{z_+z_+}))>=0
\eea
that states that all the matrix elements of the critical equation vanish between states created by operators in the subalgebra generated by (cusped) twistor Wilson loops. 
Hence restricted to this subalgebra:
\bea \label{7}
\frac{\delta \Gamma}{\delta \mu_{\lambda}(z_+, z_+)}=0
\eea
This is Morse-Smale-Floer homology seen the other way around, from the non-trivial homology to the critical points. In particular, in our inverse construction, the arcs interpolating critical points associated to Smale gradient flow
(see section \ref{s1}) are replaced 
by arcs belonging to the arc complex \cite{3MB} of the double of our punctured Riemann surface, asymptotic in both directions to the pairwise identified punctures Fig.\ref{fig:2}. Moreover, despite the $RHS$ of the holomorphic loop equation vanishes also for simple backtracking cusps, and thus this vanishing does not require necessarily nodes, Eq.(\ref{102}) needs cusps that are double points in order to imply the vanishing of matrix elements between states created by the vacuum by independent petals of the twistor Wilson loops operators.
For a single cusp one of these petals would degenerate to the identity operator. Thus the localization on critical points is associated really to the intersection homology.  \par
We observe that this localization does not hold
for $MM$ loop equation \cite{MM,MM1} and ordinary Wilson loops $\Psi(x,y;A)=P \exp i\int_{L_{xy}} A_{\alpha} dx_{\alpha}$:
\bea \label{loop}
&&\int_{L_{xx}} dx_{\alpha}<\frac{N}{2 g^2} Tr(\frac{\delta S_{YM}}{\delta A_{\alpha}(x)}\Psi(x,x;A))> \nonumber \\
&&=i \int_{L_{xx}} dx_{\alpha} \int_{L_{xx}} dy_{\alpha} \delta^{(4)}(x-y) <Tr \Psi(x,y;A)>< Tr \Psi(y,x;A)> 
\eea
Indeed, while deforming an ordinary Wilson loop by a backtracking cusp 
leaves the loop invariant at classical level because of the zig-zag symmetry \cite{Gr} of Wilson loops, at quantum level it introduces an additional logarithmic divergence known as cusp anomaly \cite{Po2,Kr}.
The cusp anomaly is reflected in an extra divergence \cite{Gr} in $MM$ loop equation at the cusp, as opposed to the holomorphic loop equation, in addition to the usual perimeter divergence:
\bea
&&\int_{L_{xx}} dx_{\alpha}<\frac{N}{2 g^2}Tr(\frac{\delta S_{YM}}{\delta A_{\alpha}(x)}\Psi( x,x;A))> \sim  i(P \Lambda^3+\frac{ \cos \Omega_{cusp}}{ \sin \Omega_{cusp}} (\pi -\Omega_{cusp})\Lambda^2)  \nonumber \\
&&\times <Tr \Psi(x,x;A)> <Tr \Psi(x,x;A)> 
\eea
where $P$ is the perimeter of the loop and $\Omega_{cusp}$ the cusp angle at the cusp. For our conventions $\Omega_{cusp}=\pi$ for no cusp,
while $\Omega_{cusp}=0$ for a backtracking cusp. 
Thus in no way the quantum contribution in the $RHS$ of the $MM$ equation vanishes for ordinary Wilson loops, as opposed to the holomorphic loop equation for (cusped) twistor Wilson loops.
It is therefore clear that our localization depends crucially on the large-$N$ triviality of our $TFT$ at all scales and in particular in the ultraviolet, that implies the absence of cusp anomalies. \par
Correspondingly to the localization, the effective action $\Gamma$ does not get quantum corrections at $N=\infty$ and therefore it must contain information, as we check a posteriori by direct computation in the following sections, about the $YM$ beta function through the Jacobian of the change 
to the $ASD$ variables, and about the mass gap through the Jacobian to the holomorphic gauge. We prove for consistency in this section, after some preparation, that $\Gamma$ does not get quantum corrections at $N=\infty$ around the critical points, as an independent check of the localization of the holomorphic loop equation (see section \ref{s11} for the direct computation). \par
Indeed, in a certain technical sense Eq.(\ref{7}), that for the moment is restricted to the subalgebra generated by adjoint twistor Wilson loops in Eq.(\ref{101}), can be extended in the large-$N$ limit to a set dense in the whole algebra.
In fact, the algebra generated by adjoint twistor Wilson loops for any fixed $\lambda$ contains only the local algebra generated by $\mu_{\lambda}$ and $\bar \mu_{\lambda}$ because of the large-$N$ factorization into the fundamental and conjugate representation,
but we will show for consistency that fluctuations around the critical points of Eq.(\ref{7}) in the \emph{whole} local algebra generated by $\mu^-_{\alpha \beta}$, our new independent variables in force of Eq.(\ref{id}), are suppressed in the large-$N$  limit,
provided that a certain basis dense in function space is chosen to compute the functional integral in the $TFT$. This justifies interpreting Eq.(\ref{101}) strongly, as Eq.(\ref{7}) extended to the whole local algebra. \par
We describe now how the choice of this basis arises. Since our considerations are of topological nature, we can deform our nodal surface arbitrarily, to the extent that we do not change its topology. Topologically to the nodal surface it is associated its normalization, that is a punctured sphere with certain pairwise identifications of the punctures Fig.\ref{fig:3}. 
The normalizing surface can also be seen as two topological disks glued at the boundary, with certain identifications of the fibers of the vector bundles involved over the punctures Fig.\ref{fig:3}.
The local picture obtained restricting to each punctured disk Fig.\ref{fig:4} is the most convenient for the following field-theoretical considerations, and it is our actual choice in the functional integral. \par
In the $TFT$ the $YM$ functional integral on a punctured disk is fixed specifying as boundary condition that the $YM$ connection carries a polar singularity at the punctures (see section \ref{s7} and \ref{s8}). Correspondingly, the $ASD$ curvature is a $\delta^{(2)}$ distribution supported on the punctures. But while in the $TFT$ the number of punctures depends on our choice and it is arbitrary, field theoretically we have the freedom to construct the vacuum of the $TFT$ in such a way that it is translational invariant, and therefore there must be a critical point for each point of space-time. 
Since in the $TFT$ critical points arise by double points, translational invariance cannot be achieved globally because of the opposite orientations of the cycles along which the holonomy of twistor Wilson loops is computed Fig.\ref{fig:3}, but only locally on each Lagrangian disk. However, it turns out that the critical points at the punctures on the two disks are charge conjugate, in such a way that both the action and the fluctuations are the same, being the theory
invariant for charge conjugation. \par
Besides, we introduce a lattice of punctures of not-necessarily-equal uniform spacings $a_{UV}$ and $a_{IR}$, as in lattice field theory, and then we take either the continuum limit $a_{UV,IR} \rightarrow 0$ keeping fixed the areas $a_{UV,IR}^2 N_2 = const$ for largely separated  ultraviolet ($UV$) and infrared ($IR$) divisors, or a scaling limit (see section \ref{s3})  $a_{UV,IR} =const$ for largely separated divisors, with the number of punctures $N_2$ going to infinity Fig.\ref{fig:4}. Locally, to the extent that the areas of the $UV$ and $IR$ Lagrangian disks are much larger than the $YM$ $RG$-invariant scale, translational invariance is achieved for all the physical purposes. 
In both the continuum limit and the scaling limit the lattice spacing plays the role of the (inverse of the) $UV$ cutoff $a^{-1}=\frac{\Lambda}{2 \pi}$, while an intermediate infrared subtraction scale $M$, not necessarily equal at the $UV$ and $IR$ divisors, is introduced in the renormalized
theory. Taking the continuum limit or the large-area scaling limit must be compatible with the renormalization of $YM$ implied by ordinary perturbation theory, as in lattice gauge theories: The universal scheme-independent part of the beta function implied by the effective action of the $TFT$ must coincide with the universal part of the perturbative beta function, that is fixed by its first and second coefficient. Most remarkably, this turns out to be the case \cite{3MB}, as it is checked by direct computation in section \ref{s9}. \par
In fact, as in $n=1$ $SUSY$ $YM$ theory, two different renormalization schemes can be defined for the gauge coupling of the $TFT$ underlying large-$N$ $YM$, the Wilsonian and the canonical scheme (see section \ref{s9}). In the Wilsonian scheme the gauge coupling $g_W(\frac{\Lambda^2}{\Lambdawb^2})$ turns out to be one-loop exact, as in $n=1$ $SUSY$ $YM$ (for a compact exposition of the supersymmetric case see section 3.2 in \cite{GGI}). The Wilsonian scheme is implemented in the $TFT$ for the Wilsonian normalization of the action (see section \ref{s9}) if the same choice is made for both the cutoff and the subtraction scales on the two disks. If different subtraction scales are instead chosen, one large at the $UV$ divisor and one small at the $IR$ divisor, a canonical coupling in the $TFT$ can be defined (see section \ref{s9}), that correctly reproduces in the $TFT$ the universal part of the perturbative beta function for 't Hooft coupling (see section \ref{s3}) $g(\frac{\Lambda^2}{\Lambdawb^2})$:
\begin{align} \label{alfa}
&\frac{\partial g}{\partial \log \Lambda}
=\frac{-\beta_0 g^3+\frac{1}{(4\pi)^2}g^3 \frac{\partial \log Z}{\partial \log \Lambda}}{1-\frac{4}{(4\pi)^2}g^2}=-\beta_0 g^3-\beta_1 g^5+\cdots \\
&\frac{\partial g_W}{\partial \log \Lambda}=-\beta_0 g_W^3 \\ \label{beta0}
&\frac{\partial\log Z}{\partial\log \Lambda} =\frac{2\gamma_0 g_W^2}{1+c' g_W^2}=2\gamma_0 g^2 +\cdots  
\end{align}
with $\gamma_0=\frac{1}{(4\pi)^2}\frac{5}{3}$ and $c'$ a scheme-dependent constant. $c'$ can be fixed imposing a physical condition on the scheme that occurs in the $ASD$ correlators (see section \ref{s11}). It is easy to check that the canonical beta function reproduces \cite{3MB} the correct universal one- and two-loop coefficients of the perturbative \cite{AF} $YM$ beta function $\beta_0=\frac{1}{(4\pi)^2}\frac{11}{3}$ and  $\beta_1=\frac{1}{(4\pi)^4}\frac{34}{3}$, by noticing that for small $g_W^2$,$\, g_W^2 \sim g^2 $ within the leading logarithmic accuracy, and expanding in powers of $g^2$ (see section \ref{s9}). Remarkably, the occurrence of the correct canonical beta function in the $TFT$ \cite{3MB} is somehow linked to the entanglement of the $UV$ and $IR$ degrees of freedom due to the pairwise identification of the fibers over the punctures on the normalization of the nodal surface, a direct consequence of the homological nature of our localization. Indeed, the canonical coupling of the $TFT$ gets also contributions from the infrared divisor, as opposed to the Wilsonian coupling, as somehow expected by the analogy with the $n=1$ $SUSY$ case (see \cite{Sh} p.89). In fact, the homology interpretation associated to the nodal points allow us a more precise and rigid realization, with respect to the original computation of the beta function in  \cite{3MB}, of the Wilsonian and canonical $RG$ flow in the $TFT$. \par
 Analytically, in the functional integral we associate to the lattice of punctures a special basis in the resolution of identity that defines the change to the $ASD$ variables \cite{3MB} in Eq.(\ref{id}).
Indeed, we assume the further resolution (see section \ref{s7}) \cite{3MB}:
\bea \label{la}
\hat \mu^{-}_{\alpha \beta}(z, \bar z)=\sum_p \hat \mu^{-}_{\alpha \beta}(p)\delta^{(2)} (z-z_p)
\eea
parametrized by the lattice field $\hat \mu^{-}_{\alpha \beta}(p)$, that is dense in function space in the sense of distributions in the gauge theory on non-commutative space-time
 $R^2 \times  R^2_{\theta}$ in which twistor loops are in fact defined (see section \ref{s4}). The non-commutativity $\theta$ of space-time arises here just as a device to define the large-$N$ limit, since it 
is known that the large-$\theta$ limit and the large-$N$ limit on commutative space-time coincide, by the modern version \cite{Mak2,Szabo2} of the non-commutative \cite{Twc,Twl1} Eguchi-Kawai reduction \cite{EK}, both at perturbative level \cite{Seiberg} and at level of loop equations \cite{loop}. This special choice of a dense basis of cylindrical functions in the resolution of identity has crucial properties that we summarize as follows, first on the mathematics side and then on the physics side. \par
Firstly, it allows us to actually expand the $TFT$ around its critical points only, by restricting to the fluctuations of the lattice field $\hat \mu^{-}_{\alpha \beta}(p)$,
since critical points of topological origin occur only at the singular fibers over the punctures: Any other choice of basis, say the ordinary Fourier transform, would imply that fluctuations are computed outside the singular support of the critical points, and therefore it would be meaningless in the framework of the $TFT$. \par
Secondly, it implies that the functional integral of the $TFT$, despite being over a dense set in function space in the $ASD$ curvature, is in fact restricted to $SD$ connections flat away from the punctures, because of 
Eq.(\ref{la}) and Eq.(\ref{id}). The $ASD$ equations reduce in $YM$ theory on non-commutative space-time (see section \ref{s6})  to (an infinite dimensional version of) Hitchin equations \cite{H2} (see section \ref{s7} and \ref{s8}) for a flat (away from the punctures) non-Hermitian two-dimensional connection, precisely our twistor connection $B_{\lambda}$.
Therefore, as in Floer theory \cite{Floer}, our construction involves the fundamental group of the underlying Riemann surface rather than the homology group, and a field theoretical representation of it by twistor Wilson loops $\Psi(B_{\lambda};L_{ww})$ evaluated on the flat connection $B_{\lambda}$.
Besides, as in Floer theory, to the representation of the fundamental group it is associated a homology theory in the moduli space of the flat connection \cite{Floer} over the Riemann surface, that reflects
the topology of the underlying manifold via its fundamental group. \par
Thirdly, the choice of this basis guarantees us that the critical points of the effective action carry homological information about the Lagrangian surface, since they are associated, in the spirit of Floer theory, to a representation of the fundamental group of the Lagrangian surface, whose Abelianization is the middle homology group of the surface. In fact, we can determine a priori what the critical points are: Under the assumption that the global gauge group of $YM$, and of the $TFT$ as well, be unbroken,
the critical points of the $TFT$ must be the twistor connections $B_{\lambda}$ whose holonomy in the fundamental representation is valued in the center $Z_N$ of $SU(N)$. In section \ref{s9} we check by direct computation that the twistor connections with $Z_N$ holonomy are in fact degenerate minima of the effective action of the $TFT$. 
The point-like singularities on the Lagrangian disks lift to surface-like singularities \cite{MB2,MB3} known as surface operators \cite{W2} because of the immersion in $R^4$. 
They are singular instantons, i.e. singular connections satisfying $ASD$ equations away from the punctures, carrying, at the critical points of the effective action, magnetic charge valued in $Z_N$ on each point of the lattice of punctures. Therefore, since the theory must be translational invariant on the local wedge at the $UV$ or at the $IR$, the translational invariant critical points belong to sectors labelled by the integer $k$ that determines the element $e^{\frac {i 2 \pi k}{N}}$ of $Z_N$. \par
Moreover, the degeneracy of the eigenvalues of the $ASD$ curvature for surface operators of $e^{\frac {i 2 \pi k}{N}}$ holonomy is $k$ and $N-k$, and the glueball masses squared, in the spectral formula Eq.(\ref{s}) derived in section \ref{s10}, turn out to be precisely
proportional to these degeneracies, thus making the statement of the integral nature (in units of $\Lambda^2_{\overline W}$ ) of the glueball masses squared that couple to our  surface operators a very rigid prediction of the $TFT$.  \par
Finally, the most remarkable self-consistency check of our localization by the holomorphic loop equation Eq.(\ref{7}) evaluated in our special basis arises by the following property of the $ASD$ equations around the local singularities: The $3$ lattice $ASD$ fields $\hat \mu^{-}_{\alpha \beta}(p)$ at each puncture $p$
must commute \cite{S5,S6} (see also \cite{W3} p.6 and p.31 for a proof of this statement in the physics style) in order for the $ASD$ equations $ F_{\alpha\beta}^-=\sum_p \hat \mu^{-}_{\alpha \beta}(p)\delta^{(2)} (z-z_p)$  in the resolution of identity to admit a solution. \par
This is most relevant to compute two-point connected correlators of surface operators at the next-to-leading $\frac{1}{N}$ order, by computing non-topological and non-trivial fluctuations around the critical points of the $TFT$, that is trivial at the leading $\frac{1}{N}$ order. Indeed, as a consequence of the aforementioned commutativity there exists a gauge, on the complement of the set of measure zero in the functional integral for which no solution of the $ASD$ equations in the resolution of identity Eq.(\ref{la}) and Eq.(\ref{id}) exists, in which the local magnetic fluctuations of the $ASD$ field around the critical points can be diagonalized simultaneously at each puncture, implying the local Abelianization of the $ASD$ lattice field. \par
Hence by the standard large-$N$ argument in the large-$N$ solution of vector-like models \cite{Zinn,Mak1} and of one-matrix models \cite{Par}, the quantum corrections to the effective action in Eq.(\ref{7}) are suppressed in the large-$N$ limit, since only the $3(N-1)$ eigenvalues fluctuate around each singularity, as opposed to the $3(N^2-1)$ matrix elements of the $ASD$ field around \emph{smooth} points, that \emph{are not} critical points. This argument applies also to fluctuations outside the subalgebra generated by twistor Wilson loops, provided they are computed around the critical points in the topological sector, i.e. around the punctures. This furnishes the promised extension of Eq.(\ref{7}) outside the subalgebra generated by twistor Wilson loops, that for the adjoint representation
contains only the holomorphic/anti-holomorphic sector generated by $\mu_{\lambda}$ and $\bar \mu_{\lambda}$ in the fundamental representation. \par
Thus, on the mathematics side, the special choice of variables obtained restricting to the $TFT$ makes the non-topological fluctuations around the $TFT$ weakly coupled at large-$N$ and explicitly computable by saddle-point methods in the large-$N$ limit. 
In loose words the fluctuations around the condensate of surface operators are self-consistently suppressed in the large-$N$ limit, in such a way that the aforementioned formal localization is in fact checked by direct computation. \par
This is perhaps the deepest result of this paper, because it allows us to get a real control of the fluctuations around the condensate of critical points in the large-$N$ limit, and because of its importance we rephrase it in another way,
that does not make use of restricting a priori to the fluctuations of the lattice field.  \par
The fluctuations around a few isolated critical points cannot be controlled in the large-$N$ limit because in general there are $3(N^2-1)$ smooth fluctuating fields away from the punctures, say by taking the usual Fourier transform.
Thus in this case the $TFT$ is still strongly coupled. But in the $TFT$ we have the freedom to deform in order to localize on the configurations that are more convenient to perform an actual computation.
Thus we \emph{choose} to localize on a lattice of critical points, simply introducing a lattice of nodes. For a lattice of critical points the fluctuations of the $ASD$ field can be decomposed into the sum of the distribution-valued fluctuations supported on the lattice and smooth fluctuations, that, to avoid overcounting, must have Fourier modes orthogonal to the momenta of the lattice field:
\bea
\delta \hat  \mu^{-}_{\alpha \beta}(z) = \sum_p \delta \hat \mu^{-}_{\alpha \beta}(p)\delta^{(2)} (z-z_p) + \text{orthogonal smooth fluctuations}
\eea
In this equation the fluctuations of the lattice field are supported on the two-dimensional commutative $R^2$ and the smooth fluctuations are consistently taken on the same $R^2$, by means of the modern version of the large-$N$ non-commutative Eguchi-Kawai reduction, i.e. the fact that the gauge group of gauge theories on non-commutative space-time contains the translations, that allows us to reabsorb the space-time degrees of freedom in the non-commutative directions into color degrees of freedom by a gauge transformation (see section \ref{s4}). \par
But in the continuum limit of the lattice field this orthogonal complement of smooth fluctuations converges to zero, because the lattice field becomes dense everywhere in Fourier space in the sense of the distributions (see section \ref{s7}). Therefore, in the continuum limit of the critical points only the fluctuations of the lattice field survive. As a consequence, after gauge-fixing to the gauge in which the $3$ components of the lattice $ASD$ field are diagonal at the same time, that is an allowed gauge for the lattice field on the complement of the set of measure zero in the functional integral for which the resolution of identity in Eq.(\ref{la}) and Eq.(\ref{id}) has no solution, the counting of fluctuating fields in the large-$N$ limit becomes on the order of $O(N)$, entirely similar to the counting of eigenvalues in the one-matrix model \cite{Par} or in vector-like models \cite{Zinn,Mak1}, and therefore the saddle-point approximation reliably describes the large-$N$ theory for the special observables that are in the subalgebra generated by twistor Wilson loops. \par
On the physics side, the choice of variables associated to the $TFT$ realizes a new version of 't Hooft electric/magnetic duality \cite{Super1} (for a neat review see \cite{DW}):
If $YM$ theory has a mass gap, either the electric charge condenses (Higgs phase, broken global gauge group) or the magnetic charge condenses (confinement phase, unbroken global gauge group). 
In the $TFT$ the condensation of the magnetic singularities follows by the asymptotic freedom \cite{AF} of the effective action in Eq.(\ref{7}), that we establish by checking that the beta function of the $TFT$ reproduces the universal, i.e. the scheme independent, first and second coefficients \cite{AF} of the perturbative $YM$ beta function.  
In fact, introducing the density $\rho_k$ of surface operators in the sector of magnetic charge $k$, $\rho_k=\sum_p \delta^{(2)} (z-z_p)$, we find at the critical points, from the asymptotic freedom of the effective action renormalized at a scale on the order of the classical action density itself (see section \ref{s10}) \footnote{This is precisely the prescription that occurs in Veneziano-Yankielowicz effective action of $n=1$ $SUSY$ $YM$ (see section \ref{s9}).}
\bea \label{14}
\rho_k^2= const \frac{N}{\hat N k (N-k)}  \Lambda_{\overline W}^4
\eea
with $k=1,2,\cdots$,
in the large-$N$ limit, that is equivalent to the condition that all the critical points labelled by the elements of $Z_N$, but for $k=0$, are in fact degenerate minima of the local part of the effective action (see section \ref{s10}).
Indeed, the condensation of the magnetic $Z_N$ is believed to be a necessary and sufficient condition for confinement \cite{DW}, according to the aforementioned 't Hooft duality alternatives. \par
At this point we can forget the holomorphic loop equation and simply regard the $TFT$ as a convenient choice of variables, that furnishes the effective action $\Gamma$ for special observables in special variables and in a special basis: Purely field theoretically the $TFT$ is a way of testing concretely long-standing conjectures about confinement and mass gap in $YM$, since the functional integral of the $TFT$ is precisely the measure induced by
the $YM$ functional integral on a condensate of surface operators in the $ASD$ variables in the holomorphic gauge. \par
Indeed, we check by direct computation in section \ref{s10} that the effective action is a non-degenerate Morse functional, that has a stable critical manifold 
in the holomorphic/anti-holomorphic sector analytically continued to Minkowski or ultra-hyperbolic signature, obtained factorizing at large-$N$ twistor Wilson loops in the adjoint representation into the fundamental and the conjugate representation:  i.e. the $TFT$ of $YM$ has a mass gap in this sector. \par
In fact, in section \ref{s11} we compute at each critical point the holomorphic/anti-holomorphic two-point correlator of lowest dimension constructed by the local single-trace gauge invariant $ASD$ surface operator, on the (analytically-continued) Lagrangian surface of the $TFT$, that corresponds to $g^2 TrF^{-2}$ in a canonical scheme in $YM$ perturbation theory. Besides, according to Eq.(\ref{corr}) and Eq.(\ref{corrM}) the fluctuations around each minimum contribute one glueball propagator for each term in the spectral sum that saturates the correlator of surface operators. We anticipate that the factor of $g_k^4$ that occurs in Eq.(\ref{corr}) and Eq.(\ref{corrM}) in the residue of each propagator, that is the square of the on-shell renomalization factor associated to the anomalous dimension of the surface operator that 
corresponds to $TrF^{-2}$, and that it is fundamental to reproduce (see section \ref{s3}) the correct large-momentum asymptotics of the propagator, arises 
by the canonical normalization $g^2$ of the surface operator that corresponds to $g^2 TrF^{-2}$ in the two-point correlator. The correlators of our surface operators are related \cite{GGI} to the ground state of the large-$N$ one-loop integrable sector 
of Ferretti-Heise-Zarembo \cite{F}, that involves scalar composite operators constructed by the $ASD$ curvature.   \par
To summarize, what makes our computation possible in the large-$N$ limit is the aforementioned Abelianization of the local distribution-valued  
lattice $ASD$ field and the scaling with $\frac{1}{\hat N}$ of the density $\rho_k$ in Eq.(\ref{14}), because it implies that the loop expansion of the non-local effective action $\Gamma$ in powers of $\rho_k$ reduces to a purely local one at the relevant $\frac{1}{ \hat N}$ order, quadratic in $\rho_k$.   \par
The plan of the paper is as follows. \par
Before presenting the derivation of our results we clarify in section \ref{s3} which are the constraints that any solution of the mass-gap problem in large-$N$ $SU(N)$ $YM$ or in any large-$N$ confining asymptotically-free gauge theory has to satisfy. We think that this clarification is needed, since there is some confusion in the physics literature, but essentially no mathematics literature, on the subject.   \par
In section \ref{s4} we introduce twistor Wilson loops in non-commutative gauge theories. \par
In section \ref{s5} we define the change to the $ASD$ variables and we remark that the Jacobian of the change of variables and its zero modes contribute to the beta function. \par
In section \ref{s6} we derive the holomorphic loop equation.  \par
In section \ref{s7} we restrict the $YM$ functional integral to a dense basis of surface operators. \par
In section \ref{s8}  we classify the moduli of surface operators, that are associated to the zero modes of the Jacobian. \par
In section \ref{s9} we obtain the
effective action of the $TFT$ and we use it to evaluate the Wilsonian and canonical beta function of the $TFT$ at the leading large-$N$ order. \par
In section \ref{s10} we show that the mass gap of the $TFT$ underlying large-$N$ $YM$ arises by the Jacobian to the holomorphic gauge, that the kinetic term arises by the Jacobian to the $ASD$ variables, and we compute explicitly the Wilsonian effective action to the next-to-leading $\frac{1}{N}$ order.  \par
In section \ref{s11} we compute the two-point correlator of the surface operator that corresponds to $g^2 Tr(F^{-2})$ at the next-to-leading $\frac{1}{N}$ order.\par
In section \ref{s12} we extend our approach to large-$N$ $QCD$ with $N_f$ quarks in the fundamental representation in massless Veneziano limit, and we determine the beta function, the lower edge of the conformal window and the 
quark-mass anomalous dimension. We stress that our technique requires that the quark masses be vanishing in 
Veneziano limit, and therefore our results hold only in the massless limit $m \rightarrow 0$. \par
In appendix \ref{App}, to make contact with reality, we discuss heuristically the actual experimental glueball spectrum in relation to our large-$N$ computation.

\section{Constraints on any solution of the problem of Yang-Mills mass gap} \label{s3}

In gauge theories the only observables that have a physical meaning are the gauge-invariant ones. For example, the gluon propagator is not gauge invariant, and therefore it is not interesting for the problem of the mass gap. Since the lowest-mass state above the vacuum in pure $YM$ is believed to be a scalar \footnote{There is an overwhelming numerical evidence from lattice gauge theory computations that sustains this belief, see \cite{T0,T} and references therein.}, the scalar gauge-invariant local correlators are relevant for the mass-gap problem. 
By general arguments the Euclidean two-point correlator of a local gauge-invariant single-trace scalar operator $\mathcal{O}(x)$ in $SU(N)$ $YM$ admits the Kallen-Lehmann representation:
\begin{equation}
G^{(2)}(p)=\int \langle \mathcal{O}(x)\mathcal{O}(0) \rangle_{conn}e^{ip\cdot x} d^4p
=\int_0^{+\infty}\frac{\rho(m^2)}{p^2+m^2}dm^2
\end{equation} 
$YM$ theory has a mass gap if and only if the closure of the support of the spectral distribution $\rho(m^2)$ does not contain zero. 
For $N$ finite $\rho(m^2)=R_{gap} \delta(m^2-m_{gap}^2)+\rho_{cont}(m^2)$ is believed to contain a term that accounts for the mass gap and a term that accounts for the possibly continuous spectrum due to the interaction and
the multi-particle states. \par
The mass-gap problem is very difficult in pure $YM$, or in any confining asymptotically-free gauge theory with no mass scale in perturbation theory \footnote{These theories include $n=1$ $SUSY$ $YM$ and $QCD$ with massless quarks. In the last case the theory is believed to have no mass gap since the pion is massless because of the spontaneous breaking of the chiral symmetry, but a mass gap would still occur in the pure-glue sector in the large-$N$ limit.} as well, because the renormalization group ($RG$) together with the asymptotic freedom
($AF$) require that any mass scale of the theory that has a physical meaning, such as the mass gap,
must depend on the canonical coupling constant $g_{YM}$
only through the $RG$ invariant scale $\Lambda_{YM}$, in such a way that in some renormalization scheme, say in the $\overline{MS}$ scheme \cite{schema}:
\bea \label{12}
m_{gap}&=& const  \Lambda_{YM} \nonumber \\
\Lambda_{YM} &=& \Lambda \exp(-\frac{1}{2\beta_0 g_{YM}^2}) (\beta_0 g_{YM}^2)^{-\frac{\beta_1}{2 \beta_0^2}}(1+ \cdots)
\eea
where the $\overline{MS}$ scheme is defined by the relation:
\bea \label{sc}
\log(\frac{\Lambda}{\Lambda_{\overline{MS}}})^2=2 \int_{g_{YM}(\Lambda_{\overline{MS}})}^{g_{YM}(\Lambda)}\frac{dg_{YM}}{\beta(g_{YM})}=\frac{1}{\beta_0 g_{YM}^2(\Lambda)}+\frac{\beta_1}{\beta_0^2}\log g_{YM}^2(\Lambda)+ C +\cdots \nonumber \\
\eea
with $C=\frac{\beta_1}{\beta_0^2} \log\beta_0$, in order to cancel \cite{schema} the term proportional to $\frac{1}{\log^2\frac{\Lambda^2}{\Lambda_{\overline {MS}}^2}}$ in the solution for $g_{YM}$.
The constant $const$ can be reabsorbed in a redefinition of the scheme, in such a way that there is a scheme in which the mass gap is  $\Lambda_{YM}$. Therefore, apart from being a $RG$-invariant scale, the mass gap is an
arbitrary parameter of the theory, whose value in physical units is determined only experimentally. However, fixed the mass gap, any other physical quantity of the theory is determined uniquely. Alternatively, fixed any other physical quantity, the mass gap is uniquely determined. \par
Physically, the continuum limit is defined removing the cutoff $\Lambda \rightarrow \infty$ sending at the same time $g_{YM} \rightarrow 0$, in such a way that $\Lambda_{YM}$ is kept constant.
Alternatively but equivalently, in lattice gauge theory computations it is often defined a scaling limit in which the (lattice) cutoff is kept constant, but the infinite-volume limit is taken, while $g_{YM} \rightarrow 0$.
In the scaling limit the mass gap becomes exponentially small, but the physics is extracted looking at correlators at distances much larger than the correlation length given by the inverse of the mass gap,
in such a way that the (finite) lattice spacing becomes invisible. \par
In both the continuum and the scaling limit the dependence of $\Lambda_{YM}$ on $g_{YM}$ is equivalent to the knowledge of the exact beta function of the theory in some scheme.
In Eq.(\ref{12}) the result implied by the two-loop beta function is explicitly displayed, while the dots refer to the scheme-dependent higher-loop contributions irrelevant in the $UV$.
Eq.(\ref{12}) in turn implies that an amazing asymptotic accuracy, as $g_{YM}$ vanishes when the cutoff $\Lambda$ diverges, is needed to solve the mass-gap problem
and that the mass gap is zero to every order of perturbation theory, since the Taylor expansion of Eq.(\ref{12}) around $g_{YM}=0$ is identically zero. Besides, Eq.(\ref{12}) requires by consistency to sum to all orders the perturbative expansion associated to correlators involving the mass gap, since perturbative corrections being polynomial in $g_{YM}^2$ are much larger for small $g_{YM}^2$ than the dimensionless function of the coupling in Eq.(\ref{12}). \par
Therefore, a finest asymptotic accuracy of non-perturbative type is needed to get control over the mass gap. \par
This rules out any perturbative method, since the mass gap
is identically zero to every order of perturbation theory. \par
This rules out also strong coupling methods, because the mass gap has nothing to do with the coupling being large,
since Eq.(\ref{12}) implies that the existence of the mass gap in the continuum limit, i.e. for arbitrarily-large cutoff $\Lambda$, or in the scaling limit, i.e. at fixed cutoff but for exponentially-small mass gap, requires an estimate uniformly in a neighborhood of zero coupling asymptotic to Eq.(\ref{12}). 
The same holds if we substitute in Eq.(\ref{12}) to the cutoff the renormalization scale $M$ and to the bare coupling the renormalized coupling at the scale $M$,
since the physics of the mass gap must not depend on how large the renormalization scale $M$ is chosen. \par
In fact, strong coupling methods do not allow us to remove the cutoff,
since if the coupling is large the cutoff scale that occurs in Eq.(\ref{12}) must be finite and cannot be large for the mass gap to stay bounded. But then, in absence of a uniform estimate that extends to any positive arbitrarily-small neighborhood of zero coupling, the continuum limit cannot be taken, and the proof is lacking that the would-be mass gap survives the continuum limit and it is not an artifact of the finite cutoff introduced in the theory at strong (or fixed) coupling. \par
Another unfit feature of any strong coupling approach (see \cite{MBN} for a critical discussion of a strong-coupling approach to the mass gap popular in the physics literature, based on the $AdS$ String/ Large-$N$ Gauge Theories correspondence) is that it implicitly assumes
that the strong coupling expansion in a neighborhood of $g_{YM}=\infty$, or of $g=\infty$ for 't Hooft coupling (see below), is connected to the $RG$ flow of the $AF$ theory and that it computes meaningful numbers. 
This statement has no theoretical foundation and on the contrary there is opposite numerical evidence from finite-coupling phase transitions in lattice $YM$ theory, that imply that the only phase of pure $YM$ at zero temperature
that admits a continuum limit is the asymptotically-free one.\par
Therefore, in section \ref{s2} we have suggested that the proper framework to address the mass-gap problem is to limit ourselves to very special observables, and to weak-coupling semi-classical techniques able to capture the non-perturbative dependence on the coupling in Eq.(\ref{12}) and to resum, to all orders of perturbation theory and to the leading large-$N$ order, the v.e.v. of adjoint twistor Wilson loops, because of their large-$N$ triviality. In fact, the problem of the mass gap, as formulated in full generality in \cite{AJ} for any compact gauge group and any correlation function, is in our opinion presently hopeless, especially for those compact gauge groups that have trivial center,
since the techniques of this paper crucially depend on the fact that $Z_N$ is the center of $SU(N)$. \par
Hence following the theoretical-physics wisdom, we have suggested in section \ref{s2} a simpler problem, the mass gap in the large-$N$ limit of $SU(N)$ $YM$, whose basic features we briefly recall in the following.
In 't Hooft large-$N$ limit \cite{H1} ($N\rightarrow \infty , g^2\equiv g^2_{YM} N = const$) $r$-point connected correlators of single-trace local operators scale as $N^{2-r}$. Therefore, after a suitable normalization, only one-point condensates survive at leading $\frac{1}{N}$ order, and two-point connected correlators survive at next-to-leading order. Hence the interaction vanishes in the large-$N$ limit at next-to-leading $\frac{1}{N}$ order, since it is associated to the three- and multi-points connected correlators. 
Because of confinement and the mass gap and the vanishing of the interaction, it is believed that the two-point connected correlators at next-to-leading order are an infinite sum of propagators of massive free fields, i.e. the spectral distribution in Eq.(\ref{14}) is saturated by massive free one-particle states only, the glueballs \cite{M1,P1} (see also \cite{MBN,MB1,3MB}). For example, for the correlator of a scalar operator 
$\mathcal{O}$ the relation would hold:
\bea \label{15}
G_{\mathcal{O}}^{(2)}(p)= \sum_k\frac{R_{k}}{p^2+m_{k}^2}
\eea
Proving Eq.(\ref{15}) for positive $R_k$ and $m_{k}$, say for the correlator of the action density, would imply the mass gap in the large-$N$ limit. \emph{Computing} Eq.(\ref{14}) in some non-perturbative scheme
may not be ludicrously ambitious, since it represents the propagator as a sum of free fields. Besides, we can restrict  Eq.(\ref{14}) to momentum dual in the Fourier sense to certain submanifolds of space-time, or of its complexification by means of analytic continuation, as we suggested in section \ref{s1}, without
loosing information about the mass spectrum and the residues of the poles of the propagators. 
In fact, in this paper we study the $ASD$ two-point correlator.
The $ASD$ correlator reduces in Euclidean (and ultra-hyperbolic) signature to the sum of the scalar $\mathcal{O}_S=\sum_{\alpha \beta} \Tr F_{\alpha \beta}F^{\alpha \beta}$ and pseudoscalar correlators $\mathcal{O}_P=\sum_{\alpha \beta} \Tr (F^{\alpha \beta}\, ^{*}F_{\alpha \beta})$:$\langle \mathcal{O}_{ASD}(x)\mathcal{O}_{ASD}(0) \rangle_{conn}= 4 \langle \mathcal{O}_S(x)\mathcal{O}_S(0) \rangle_{conn}+ 4 \langle \mathcal{O}_P(x)\mathcal{O}_P(0) \rangle_{conn}$ and in Minkowskian signature to their difference.
Thus computing:
\bea
G_{\mathcal{O}_{ASD}}^{(2)}(p^2)= \sum_k\frac{R_{k}}{p^2+m_{k}^2}
\eea
would suffice for the mass-gap problem as well. \par
The structure of two-point correlators of local gauge-invariant operators is severely constrained \cite{M1} by the $RG$-improved \cite{MBN} version of perturbation theory \cite{M1}, by the operator product expansion ($OPE$) \cite{M1} and by the low-energy theorems of $NSVZ$ (see \cite{MBN} for a short review). 
Any proposed solution has to satisfy these constraints as well, that we summarize as follows. \par
Firstly, we proved \cite{MBN} by standard $RG$ methods based on the Callan-Symanzik equation plus the Kallen-Lehmann representation plus the assumption that the one-particle spectrum is a discrete diverging sequence with asymptotic distribution $ \rho_s(m^2)$, that the two-point large-$N$ correlators of any operator of spin $s$ and given anomalous dimension
must satisfy the following asymptotic theorem in any confining asymptotically-free gauge theory massless in perturbation theory, such as large-$N$ $YM$. \par
The connected two-point Euclidean correlator of a local single-trace gauge-invariant operator $\mathcal{O}^{(s)}$, of integer spin $s$ and naive mass dimension $D$ and with anomalous dimension $\gamma_{\mathcal{O}^{(s)}}(g)$,
must factorize asymptotically for large momentum, and at the leading order in the large-$N$ limit, over the following poles and residues:
\bea \label{eq:1}
\int \langle \mathcal{O}^{(s)}(x) \mathcal{O}^{(s)}(0) \rangle_{conn}\,e^{-ip\cdot x}d^4x
\sim \sum_{n=1}^{\infty}  P^{(s)} \big(\frac{p_{\alpha}}{m^{(s)}_n}\big) \frac{m^{(s)2D-4}_n Z_n^{(s)2}  \rho_s^{-1}(m^{(s)2}_n)}{p^2+m^{(s)2}_n  } \nonumber \\
\eea
where $ P^{(s)} \big( \frac{p_{\alpha}}{m^{(s)}_n} \big)$ is a dimensionless polynomial in the four momentum $p_{\alpha}$ that projects on the free propagator of spin $s$ and mass $m^{(s)}_n$ and:
\bea \label{g}
\gamma_{\mathcal{O}^{(s)}}(g)= - \frac{\partial \log Z^{(s)}}{\partial \log \mu}=-\gamma_{0} g^2 + O(g^4)
\eea 
with $Z_n^{(s)}$ the associated renormalization factor computed on shell, i.e. for $p^2=m^{(s)2}_n$:
\bea \label{z}
Z_n^{(s)}\equiv Z^{(s)}(m^{ (s)}_n)= \exp{\int_{g (\mu)}^{g (m^{(s)}_n )} \frac{\gamma_{\mathcal{O}^{(s)}} (g)} {\beta(g)}dg}
\eea
The sum in the $RHS$ of Eq.(\ref{eq:1}) is in fact badly divergent, but the divergence is a contact term, i.e. a polynomial of finite degree in momentum. Thus the infinite sum in the $RHS$ of Eq.(\ref{eq:1}) makes sense only after subtracting the contact terms (see remark below Eq.(\ref{eq:2})). Fourier transforming Eq.(\ref{eq:1}) in the coordinate representation
projects away for $x\neq 0$ the contact terms and avoids convergence problems:
\bea \label{eq:0}
\langle \mathcal{O}^{(s)}(x) \mathcal{O}^{(s)}(0) \rangle_{conn} 
\sim \sum_{n=1}^{\infty} \frac{1}{(2 \pi)^4} \int  P^{(s)} \big(\frac{p_{\alpha}}{m^{(s)}_n}\big) \frac{m^{(s)2D-4}_n Z_n^{(s)2} \rho_s^{-1}(m^{(s)2}_n)}{p^2+m^{(s)2}_n  } \,e^{ip\cdot x}d^4p \nonumber \\
\eea
The proof of the asymptotic theorem reduces to showing that Eq.(\ref{eq:1})
matches asymptotically for large momentum, within the universal leading and next-to-leading logarithmic accuracy,
the $RG$-improved perturbative result implied by the Callan-Symanzik equation:
\bea \label{CS}
&& \int \langle \mathcal{O}^{(s)}(x) \mathcal{O}^{(s)}(0) \rangle_{conn}\,e^{-ip\cdot x}d^4x  \nonumber \\
&& \sim P^{(s)}\big(\frac{p_{\alpha}}{p}\big) \, p^{2D-4}    \Biggl[\frac{1}{\beta_0\log (\frac{p^2}{\Lambda^2_{QCD} } )}\biggl(1-\frac{\beta_1}{\beta_0^2}\frac{\log\log (\frac{p^2}{\Lambda^2_{QCD} } ) }{\log (\frac{p^2}{\Lambda^2_{QCD} } )}    + O(\frac{1}{\log (\frac{p^2}{\Lambda^2_{QCD} } )} ) \biggr)\Biggr]^{\frac{\gamma_0}{\beta_0}-1}
\eea
up to contact terms (i.e. distributions supported at coinciding points), and that this matching fixes uniquely the universal asymptotic behavior of the residues in Eq.(\ref{eq:1}).
More precisely, the asymptotic behavior of the residues is fixed by the asymptotic theorem within the universal, i.e. the scheme-independent, leading and next-to-leading logarithmic accuracy.
This implies that the renormalization factors are fixed asymptotically for large $n$ to be:
\begin{equation}\label{z1}
Z_n^{(s)2}\sim 
\Biggl[\frac{1}{\beta_0\log \frac{ m^{ (s) 2}_n }{ \Lambda^2_{QCD} }} \biggl(1-\frac{\beta_1}{\beta_0^2}\frac{\log\log \frac{ m^{ (s) 2}_n }{ \Lambda^2_{QCD} }}{\log \frac{ m^{ (s) 2}_n }{ \Lambda^2_{QCD} }}    + O(\frac{1}{\log \frac{ m^{ (s) 2}_n }{ \Lambda^2_{QCD} } } ) \biggr)\Biggr]^{\frac{\gamma_0}{\beta_0}}
\end{equation}
where $\beta_0, \beta_1,\gamma_0$ are the first and second coefficients of the beta function and the first coefficient of the anomalous dimension respectively and $\Lambda_{QCD}$
is the $QCD$ $RG$-invariant scale in some scheme. 
Eq.(\ref{eq:1}) for the propagator can be rewritten equivalently as:
\bea \label{eq:2}
\int \langle \mathcal{O}^{(s)}(x) \mathcal{O}^{(s)}(0) \rangle_{conn}\,e^{-ip\cdot x}d^4x  
\sim    P^{(s)} \big(\frac{p_{\alpha}}{p} \big)  \, p^{2D-4}   \sum_{n=1}^{\infty} \frac{Z_n^{(s)2}   \rho_s^{-1}(m^{(s)2}_n)  }{p^2+m^{(s)2}_n  }
\eea
up to contact terms, where now the sum in the $RHS$ is convergent for $\gamma'=\frac{\gamma_0}{\beta_0} > 1$. Otherwise, it is divergent but the divergence is again a contact term. $P^{(s)} \big(\frac{p_{\alpha}}{p} \big)$ is the projector obtained substituting $-p^2$  to $m_n^2$ in  $P^{(s)} \big(\frac{p_{\alpha}}{m_n} \big)$\footnote{We use Veltman conventions for Euclidean and Minkowski propagators of spin $s$ \cite{Velt2}.}. \par
With S. Muscinelli we have checked \cite{SM} the aforementioned asymptotic estimates for the scalar, pseudoscalar and $ASD$ Euclidean glueball propagator (they correspond to the case $\gamma_0=2\beta_0)$ by means
of a very remarkable three-loop perturbative computation by Chetyrkin et al. \cite{Che1,Che2}, that sums up about $800$ Feynman diagrams \footnote{The earlier two-loop computation was performed in \cite{Kat}.}, by comparing the perturbative computation re-expressed in terms of the two-loop running coupling with the actual asymptotic estimates, finding perfect agreement \cite{SM}. \par
Secondly, the asymptotic estimates extend to the coefficient of the glueball condensate in the $OPE$ \cite{MBN} and imply Eq.(\ref{OPE2}) up to overall normalization.\par
In fact, peculiar cancellations occur between the coefficients of the $OPE$ of the scalar and pseudoscalar correlators combined in the $ASD$ correlator. We report here the results of the perturbative computation  \cite{Che1,Che2,Che3,Zo}, in the notation of \cite{SM} which we refer to for further details:
\bea
\int\langle \mathcal{O}_{S}(x) \mathcal{O}_{S}(0) \rangle_{conn}\,e^{-ip\cdot x}d^4x 
 \sim \, C^{(0)}_{S}(p^2) 
+  C^{(1)}_{S}(p^2) <\mathcal{O}_{S}(0)> \nonumber \\
\eea 
with
\begin{align} \label{contact}
 C^{(0)}_{S}(p^2)  
&=-\bigl(1-\frac{1}{N^2}\bigr)\gms^4(\mu)\frac{p^4}{4\pi^2}\log\frac{p^2}{\mu^2}
\biggl[1+\gms^2(\mu)\biggl(f_0-\beta_0\log\frac{p^2}{\mu^2}\biggr)\nonumber\\
&+\gms^4(\mu)\biggl(f_1+f_2\log\frac{p^2}{\mu^2}+\beta_0^2\log^2\frac{p^2}{\mu^2}\biggr)\biggr] \nonumber\\
C^{(1)}_{S}(p^2)&= - 4\gms^4(\mu)\beta_0\log\frac{p^2}{\mu^2}
\biggr[1- \gms^2(\mu) \beta_0 \log\frac{p^2}{\mu^2}+ \gms^2(\mu) g_0\biggr] + \text{infinite contact terms}
\end{align}
and
\bea
\int\langle \mathcal{O}_{P}(x) \mathcal{O}_{P}(0) \rangle_{conn}\,e^{-ip\cdot x}d^4x 
 \sim \, C^{(0)}_{P}(p^2) 
+  C^{(1)}_{P}(p^2) <\mathcal{O}_{S}(0)> \nonumber \\
\eea 
with
\begin{align}
C^{(0)}_{P}(p^2)&=
-\bigl(1-\frac{1}{N^2}\bigr)\gms^4(\mu)\frac{p^4}{4\pi^2} \log\frac{p^2}{\mu^2}
\biggl[1+\gms^2(\mu)\biggl(\tilde{f}_0-\beta_0\log\frac{p^2}{\mu^2}\biggr)\nonumber\\
&+\gms^4(\mu)\biggl(\tilde{f}_1+\tilde{f}_2\log\frac{p^2}{\mu^2}+\beta_0^2\log^2\frac{p^2}{\mu^2}\biggr)\biggr]
\nonumber \\
C^{(1)}_{P}(p^2)&= 4\gms^4(\mu)\beta_0\log\frac{p^2}{\mu^2}
\biggr[1- \gms^2(\mu) \beta_0 \log\frac{p^2}{\mu^2}+ \gms^2(\mu) \tilde{g}_0\biggr] + \text{contact terms}
\end{align}
where the coefficients $f,g,\tilde f, \tilde g$ are not explicitly displayed because they are irrelevant for the scheme-independent leading logarithmic behavior.
Therefore, the leading perturbative logarithms cancel in $C^{(1)}_{S}(p^2)+C^{(1)}_{P}(p^2)$ and in $C^{(0)}_{S}(p^2)-C^{(0)}_{P}(p^2)$, implying the aforementioned partial cancellations in the $RG$-improved $ASD$ correlator in
Euclidean (or ultra-hyperbolic) and Minkowskian signature respectively, that are summarized in Eq.(\ref{opeE}) and in Eq.(\ref{opeM}). The $RG$-improved result for the second coefficient in the $OPE$ can be obtained directly by the perturbative expansion, employing the same elementary technique
that in \cite{SM} has been applied to the first coefficient. Alternatively, the general $RG$ estimates in \cite{MBN} can be applied. \par
Moreover, we have checked in \cite{SM} that the asymptotic behavior of the $ASD$ Euclidean correlator in the $TFT$ matches the $RG$-improved result in large-$N$ $QCD$ and \cite{MBN} that the $ASD$ correlator in the $TFT$ in Minkowski matches the $RG$-improved  result in large-$N$ $QCD$ as well. \par
Thirdly, the $NSVZ$ theorem is discussed in section \ref{s11}. \par
Thus Eq.(\ref{corr}) and Eq.(\ref{corrM}) are compatible with everything that we know presently about large-$N$ $QCD$, both in the infrared numerically and experimentally (see appendix \ref{App}), and asymptotically by first principles in the ultraviolet \cite{MBN,SM}.

\section{Twistor Wilson loops in the $TFT$, $YM$ on non-commutative space-time, and Morita equivalence} \label{s4}

The definition of twistor Wilson loops involves $YM$ theory on non-commutative space-time.  Non-commutativity in our framework is just a technical tool to define the large-$N$ limit. 
Indeed, the limit of infinite non-commutativity $\theta$ is equivalent to the large-$N$ limit of $YM$ theory on commutative space-time \cite{Twc,Seiberg,loop,Szabo2} \footnote{Once localization on critical points is obtained, the glueball spectrum is computed employing the effective action in large-$N$ $YM$ theory on commutative $R^4$ around the critical points, using Morita equivalence recalled below.}. 
Twistor Wilson loops  $\Psi(\hat B_{\lambda};L_{ww})$  are defined in $YM$ theory with gauge group $U(N)$ on $R^2 \times R^2 _{\theta}$, with complex coordinates $(z=x_0+i x_1, \bar z=x_0-i x_1, \hat u=\hat x_2+i \hat x_3, \hat {\bar u}=\hat x_2-i \hat x_3)$ satisfying $ [\hat \partial_u, \hat \partial_{\bar u}]=\theta^{-1} 1$,
where $R^2 _{\theta}$ is the non-commutative plane with non-commutative parameter $\theta$. Their v.e.v. is $1$ in the large-$N$ limit for any shape of the loops:
\bea
\lim_{\theta \rightarrow \infty} <\frac{1}{ \cal N} Tr_{\cal N} \Psi(\hat B_{\lambda};L_{ww})  >=1 
\eea
This property is proved in the second part of this section. 
Twistor Wilson loops are defined as follows:
\bea
Tr_{\cal N} \Psi(\hat B_{\lambda};L_{ww})=Tr_{ \cal{N}} P \exp i \int_{L_{ww}}(\hat A_z+\lambda \hat D_u) dz+(\hat A_{\bar z}+ \lambda^{-1} \hat D_{\bar u}) d \bar z 
\eea
where $\hat D_u=\hat \partial_u+i \hat A_u$ is the covariant derivative along the non-commutative direction $\hat u$, and $\lambda$ is a complex non-vanishing parameter. The plane $(z, \bar z)$ is commutative. The loop $L_{ww}$ starts and ends at the marked point $w$ and lies in the commutative plane. Thus we regard the twistor connection $B_{\lambda}$, whose holonomy the twistor Wilson loop computes, as a non-Hermitian connection
in the commutative plane valued in the tensor product of the Lie algebra $u(N)$ of $U(N)$ and of the infinite-dimensional operator algebra that generates the Fock 
representation of the non-commutative plane $(\hat u, \hat {\bar u})$. The trace $Tr_{\cal N}$ is defined on the tensor product of the Lie algebra and the Fock space. $B_{\lambda}$ is indeed a connection in the
commutative plane, since the non-commutative covariant derivative transforms as a Higgs field of the commutative plane.\par
However, we will see below that there exists a four-dimensional description, obtained by means of a gauge transformation, for which the support of twistor Wilson loops is in fact a planar Lagrangian surface immersed in (the complexification of) four-dimensional space-time.\par 
We briefly recall the definition of $YM$ theory on non-commutative space-time. It admits two equivalent representations: One involves the Moyal product and the other one its representation as an operator algebra (see for details section 4 in \cite{GGI}). In this paper we employ only the second one. Non-commutative Euclidean space is defined by:
\bea \label{1}
[\hat x^{\alpha}, \hat x^{\beta}]=i \theta^{\alpha \beta} 1
\eea
Defining: $\hat \partial^i(\hat x^j)= \delta^{ij}1$, the non-commutative derivatives $\hat \partial^i$ can be represented via Eq.(\ref{1}) and satisfy:
\bea
[\hat \partial_{\alpha}, \hat \partial_{\beta}]=i \theta^{-1}_{\alpha \beta} 1
\eea
For brevity, in this paper we refer to gauge theories on non-commutative or commutative space-time, as non-commutative or commutative gauge theories respectively.
In non-commutative gauge theories translations can be absorbed by gauge transformations.
Correspondingly, there is a gauge in which  ordinary Wilson lines involve a constant gauge connection \cite{Mak2,Mak1}:
\bea \label{line}
 \Psi(\hat A; L_{yz})= P \exp  \int_{L_{yz}}( \hat \partial_{\alpha} + i  \hat A_{\alpha}) dx_{\alpha}
\eea
and a gauge in which the space-time dependence of the connection can be restored by a unitary gauge transformation
$\hat U(x)= e^{ x_{\alpha} \hat \partial_{\alpha}} $,
where $x_{\alpha}$ is a commutative space-time coordinate.
The operator-valued connection transforms under this gauge transformation in the usual way:
$\hat A_{\alpha}^ {\hat U} = \hat U(x) \hat A_{\alpha} \hat U(x)^{-1} + i \partial_{\alpha} \hat U(x) \hat U(x)^{-1}$,
where the partial derivative is the usual partial derivation with respect to the commutative 
parameter $x_{\alpha}$. Correspondingly, the Wilson line in Eq.(\ref{line}) transforms as:
\bea
&& \hat U(z)\Psi( \hat A; L_{yz}) \hat U(y)^{-1}  =  P \exp i \int_{L_{yz}}( \hat U(x) \hat A_{\alpha} \hat U(x)^{-1} -i U(x)\hat \partial_{\alpha}U(x)^{-1} + i \partial_{\alpha} \hat U(x) \hat U(x)^{-1} )dx_{\alpha}  \nonumber \\
&&=  P \exp i \int_{L_{yz}} \hat U(x) \hat A_{\alpha} \hat U(x)^{-1} dx_{\alpha}  
\eea
where we ignore central terms that vanish for large $\theta$.
As a consequence the action of the $U(N)$ non-commutative gauge theory can be written in the operator representation in terms of a constant $YM$ connection:
\bea \label{nctr}
\frac{N}{2 g^2}  (2 \pi)^{\frac{d}{2}} Pf(\theta )   tr_N \hat Tr(i [\hat \partial_{\alpha}+i \hat A_{\alpha},\hat \partial_{\beta}+i \hat A_{\beta}] + \theta^{-1}_{\alpha \beta} 1)^2
\eea
where the non-commutative gauge connection is valued in the tensor product of the Lie algebra $u(N)$ in the fundamental representation and of the operator-algebra on Fock space. 
Hence in non-commmutative $YM$  there must be a relation between the $UV$ cutoff of the regularized theory and the color degrees of freedom, since they contain space-time information.
This leads to the modern version of what we refer to as non-commutative \cite{Twc,Twl1} Eguchi-Kawai ($EK$) reduction \cite{EK}:
\bea \label{NC}
\frac{N}{2 g^2}  \hat N (\frac{2 \pi}{\Lambda})^d tr_N Tr_{\hat N} (i [\hat \partial_{\alpha}+i \hat A_{\alpha},\hat \partial_{\beta}+i \hat A_{\beta}]+ \theta^{-1}_{\alpha \beta} 1)^2
\eea
where the trace $Tr_{\hat N}$ is taken now over a subspace of dimension $\hat N$ in the large $\hat N, \theta, \Lambda $ limit with:
\bea \label{cutoff}
\hat N (\frac{2 \pi}{\Lambda})^d=(2 \pi)^{\frac{d}{2}} Pf(\theta )
\eea
where $Pf$ is the Pfaffian, i.e. the square root of the determinant.
The factor of $(\frac{2 \pi}{\Lambda})^d$ in the normalized non-commutative action is essential to reproduce the large-$N$  loop equation 
of the commutative gauge theory Eq.(\ref{loop}), since its effect is equivalent to the insertion of $\delta^{(d)}(0)$ in the $RHS$ of the non-commutative loop equation, that is missing
because of the functional integration over constant gauge connections only, in the non-commutative gauge theory Eq.(\ref{NC}). As a consequence the distribution  $\delta^{(d)}(x-y)$ in the $RHS$ of $MM$ loop equation on commutative space-time is reproduced provided the trace of open $x \neq y$ Wilson lines vanishes \cite{Mak1,Mak2}:$
 < \frac{1}{{\cal N}} Tr_{\cal N}\Psi(\hat A; L_{xy})>=0$.\par
 $YM$ can also be defined on a non-commutative torus of sides $\hat L$, that provides via Morita equivalence described in the following, a stronger version of the large-$\theta$/ large-$N$ equivalence. For deep technical reasons, essentially because we are interested in the spectrum of fluctuations of $YM$ on \emph{commutative} space-time, we employ in our computations this stronger version of the equivalence. Provided the non-commutativity $2 \pi \theta= \frac{\hat M}{\hat N} \hat L^{2}$ is rational in natural units, $YM$ on a non-commutative torus of sides $\hat L$ enjoys Morita equivalence with $YM$ with the same 't Hooft coupling on a commutative torus of sides $\frac{\hat L}{\hat N}$, with gauge group $U(\hat N N)$, and twisted boundary conditions. More explicitly, in the rational case any $U(N)$ connection of $YM$ theory on $R^2 \times T^2_{\theta}$, with coordinates $(y, \hat x)$ and periodic boundary conditions on the non-commutative torus $T^2_{\theta}$, admits the expansion \cite{Gu,AG}:
\bea
A(\hat x,y)=\sum_{l \in  Z^2} a_l(y) e^{-2 \pi i l \cdot \hat x/ \hat L}
\eea
The corresponding Morita equivalent  $U( N \hat N)$ connection (up to perhaps a sign redefinition of Fourier coefficients) reads \cite{Gu,AG}:
\bea \label{NC1}
A'(x,y)=\sum_{l \in  Z^2} a_l(y) U^{ l_1} V^{l_2} \omega^{-\hat M l_1 l_2/2} e^{-2 \pi i l \cdot x/ \hat L}
\eea
where $(U, V)$ are the clock and shift matrices of $SU(\hat N)$: $UV=\omega^{\hat M} VU$, $P^{\hat N}=Q^{\hat N}=1$, with $\omega= e^{2 \pi i / \hat N}$.  The traceless part of $A'$ is a connection on the twisted 't Hooft bundle $SU(N \hat N)/Z_{N \hat N}$ on the commutative torus of sides $L/\hat N$ \cite{Gu,AG}:
\bea
A'(x_j+L/\hat N)=\Gamma_j A'(x_j)  \Gamma_j^{-1}
\eea
with: $\Gamma_1= 1_N \times V^{r}$, $\Gamma_2=1_N \times  U^{-r} $,
where $r \hat M=1 \text{mod} \hat N$  \cite{Gu,AG}. 
Therefore, in the Morita equivalent theory, because of the twisted boundary conditions, there is a relation between color and momentum $2 \pi \frac{l}{\hat L}$ of Fourier modes, as implied by Eq.(\ref{NC1}).
Projecting on a finite number $\hat N$ of Fourier modes, the theory is regularized in the $UV$ with a cutoff given by Eq.(\ref{cutoff}) and computed in section \ref{s10}.
Moreover, the theory contains the $l=0$ untwisted sector $SU(N) \times 1_{\hat N}$ that is diagonally embedded in $U(N \times \hat N)$.\par
The $l=0$ untwisted sector plays a key role, because it carries zero momentum on the commutative torus and thus fields inside it may admit a translationally-invariant condensate (of surface operators, see section \ref{s7}).
In fact, we employ the untwisted sector of the Morita equivalent $U(\hat N N)$ twisted theory to construct the critical points of the $TFT$ in the large-$\theta$ limit of non-commutative $YM$.\par
But once the critical points in the $U(\hat N N)$ theory are obtained, fluctuations are computed in section \ref{s10}
in a carefully-defined thermodynamic limit $\hat L \rightarrow \infty$. Thus the non-commutative theory is employed only to construct the large-$\theta$ vacua of the $TFT$ and their $RG$ flow (that coincides with the $RG$ flow of the large-$N$ commutative theory), but fluctuations are safely computed in ordinary $YM$ around the aforementioned vacua (see section \ref{s10}).\par
Twistor Wilson loops are supported on null Lagrangian surfaces immersed in the complexification of Euclidean space-time. 
To show this it is convenient to gauge away the non-commutative derivatives that occur in the definition of twistor Wilson loops.
This is obtained performing a local gauge transformation with values in the complexification of the gauge group. Although this is not a symmetry
of the theory, the trace of  twistor Wilson loops is left invariant, because of the cyclicity property of the trace. Let be
$\hat S(z, \bar z)= e^{i \lambda z \hat \partial_{ u}+ i {\lambda}^{-1} \bar z \hat \partial_{\bar u}} $
a gauge transformation in the complexification of the gauge group, where $(z, \bar z)$ are commutative coordinates.
The operator-valued gauge connection $\hat B_{\lambda}$ transforms under this gauge transformation in the usual way:
$\hat B_{\lambda}^ {\hat S} = \hat S(z, \bar z) \hat B_{\lambda} \hat S(z, \bar z)^{-1} + i d \hat S(z, \bar z) \hat S(z, \bar z)^{-1}$
where the differential $d=\partial+\bar \partial$ is the usual differential with respect to the commutative 
parameters $(z, \bar z)$. 
Correspondingly, a twistor Wilson line transforms as: 
\bea \label{supp}
&& \hat S(v, \bar v)\Psi( \hat B_{\lambda}; L_{wv}) \hat S(w, \bar w)^{-1} \nonumber \\
&& = P \exp i \int_{L_{wv}} \hat S(z, \bar z)   (\hat A_z+i \lambda \hat A_u) \hat S(z, \bar z)^{-1}  dz +  \hat S(z, \bar z)   (\hat A_{\bar z}+ i \lambda^{-1} \hat A_{\bar u})  \hat S(z,\bar z)^{-1}  d \bar z \nonumber \\
\eea
where we have disregarded central terms that vanish for large $\theta$. Therefore, a twistor loop lies effectively on the submanifold of four-dimensional complexified commutative space-time defined by:
$( z,  \bar z, u,  {\bar u})=(z, \bar z, i \lambda  z, i \lambda^{-1} \bar z)$.
This is a Lagrangian submanifold of the complexified Euclidean space-time with respect to the Kahler form $dz \wedge d \bar z+ du \wedge d \bar u$, that lifts to a Lagrangian submanifold
of twistor space, provided $\lambda$ is either real or a unitary phase. The two cases correspond to Lagrangian submanifolds of antipodal and circle type
respectively. \par
The proof of triviality of twistor Wilson loops to all orders of perturbation theory in the limit $\theta \rightarrow \infty$ follows now almost immediately. 
Indeed, at any order in perturbation theory
a generic contribution to an ordinary Wilson loop of a commutative gauge theory contains a correlator of gauge fields,
i.e. a Schwinger function, with tensor indices contracted with a product of monomials in
$\dot x_{\alpha}(s)$ at generic insertion points on the loop, labeled by $s$:
\bea 
\int ds_1 ds_2 \cdots G_{{\alpha_1}{\alpha_2} \cdots}(x_{\beta}(s_1)-x_{\beta}(s_2), \cdots) \dot x_{\alpha_1}(s_1) \dot x_{\alpha_2}(s_2) \cdots
\eea
Because of the $O(4)$ invariance of the commutative theory, $\dot x_{\alpha_1}(s_1)$ is contracted either with another  $\dot x_{\alpha_2}(s_2)$ or with an $ x_{\alpha_2}(s_2)$  to form  polynomials in  $\dot x_{\alpha}(s) \dot x_{\alpha}(s')$ or in $\dot x_{\alpha}(s) x_{\alpha}(s')$. Indeed, all these monomials necessarily contain at least one factor of  $\dot x_{\alpha}$ since the gauge field along the loop has the index contracted with one $\dot x_{\alpha}$. The possible factor of $x_{\alpha}$ arises from the dependence of Schwinger functions on the coordinates.
We now specialize to twistor Wilson loops. 
In the limit $\theta \rightarrow \infty$ of the non-commutative gauge theory $O(4)$ invariance is recovered, because the theory becomes the large-$N$ limit of the commutative theory,
that obviously is $O(4)$ invariant, say in the Feynman gauge. 
Therefore, all the monomials just mentioned vanish when evaluated on the Lagrangian submanifold which
a twistor Wilson loop lies on, because they are of the form $\dot z(s) \dot{\bar z}(s')- \dot z(s) \dot{\bar z}(s')=0$ or $z(s) \dot{ \bar z}(s')-z(s) \dot{\bar z}(s')=0$. Thus the effective propagators that connect
a Feynman graph at any order to a twistor Wilson loop vanish. The only factors that may spoil the triviality occur if singularities due to denominators of Feynman diagrams
arise, since $ x_{\alpha}(s)  x_{\alpha}(s')$ vanishes too on the Lagrangian submanifold for the same reason. To cure this, we analytically continue twistor Wilson loops
from Euclidean to Minkowski or to ultra-hyperbolic space-time, in order to get the $i \epsilon$ prescription $z_+(s)  z_-(s')-z_+(s)  z_-(s')+i \epsilon=i\epsilon$
in the denominators. \par
At the lowest perturbative order the triviality of the v.e.v. of twistor Wilson loops is simply the cancellation between gauge propagators due to the factor of $i$ in the covariant derivative
in the Feynman gauge in the large-$\theta$ limit:
\bea
&&<Tr_{\cal N}  \big( \int_{L_{ww}}(\hat A_z+\lambda \hat D_u) dz+(\hat A_{\bar z}+ \lambda^{-1}\hat D_{\bar u}) d \bar z \int_{L_{ww}}(\hat A_z+\lambda \hat D_u) dz+(\hat A_{\bar z}+ \lambda^{-1} \hat D_{\bar u}) d \bar z\big)> \nonumber \\
&& \sim 2 \int_{L_{ww}} dz\int_{L_{ww}}d \bar z (<Tr_{\cal N} (\hat A_z  \hat A_{\bar z})>+i^2 <Tr_{\cal N} (\hat A_u  \hat A_{\bar u})>)=0 
\eea

\section{Change to the $ASD$ variables and contribution of the Jacobian and zero modes to the beta function} \label{s5}

In \cite{MB3} and \cite{3MB} we introduced the change to the $ASD$ variables in pure $YM$ theory, and more generally in any gauge theory that extends pure $YM$. In $n=1$ $SUSY$ $YM$ in the light-cone gauge this change of variables is known as the Nicolai map. For a detailed exposition with historical background see \cite{BP}.
In \cite{BP} it is also proved the perturbative equivalence to $YM$ in the usual variables by direct evaluation of the perturbative one-loop effective action in the $ASD$ variables. The $YM$ partition function is:
\bea \label{zeta}
Z=\int  \exp \big(-\frac{16\pi^2NQ}{2g^2}-\frac{N}{4g^2} \int tr_f (F_{\alpha\beta}^-)^2d^4 x\big)\delta A
\eea
where $Q$ denotes now the second Chern class, we have used the well known identity $tr_f(F_{\alpha\beta}^2)=tr_f(F_{\alpha\beta}^-)^2/2+tr_f(F_{\alpha\beta} \,  ^{*}\!{F}_{\alpha\beta})$, and the trace is in the fundamental representation. We change variables in the gauge-fixed theory from the connection to the $ASD$ curvature, introducing in the functional integral the appropriate resolution of the identity \cite{MB3}:
\bea \label{ri}
1=\int \delta(F_{\alpha\beta}^-(A)-\mu_{\alpha\beta}^-) \delta\mu_{\alpha\beta}^-
\eea
In the gauge-fixed theory this is a well-defined change of variables: The $3$ independent components of the gauge connection in the gauge-fixed theory (the $4$ components of the gauge connection minus the gauge-fixing condition) are mapped to the $3$ independent components of the $ASD$ curvature (the $6$ components of the curvature minus the $3$ $ASD$ conditions). Thus:
\bea
Z=\int  \exp \big(-\frac{16\pi^2NQ}{g^2}-\frac{N}{4g^2}\int tr_f(\mu_{\alpha\beta}^-)^2d^4 x\big)\delta(F_{\alpha\beta}^--\mu_{\alpha\beta}^-) \delta \mu_{\alpha\beta}^- \delta A
\eea
Exchanging the order of integration we can perform the integral on the gauge connection because of the delta functional, in order to get formally the Jacobian of the change of variables.
The partition function becomes:
\bea \label{3}
Z&&=\int  \exp \big(-\frac{16\pi^2NQ}{g^2}-\frac{N}{4g^2}\int tr_f (\mu_{\alpha\beta}^-)^2d^4 x\big)
Det^{-1/2}(-\Delta_A \delta_{\alpha \beta}+ D_\alpha D_\beta -i ad F_{\alpha \beta}^+ ) \, \delta \mu^-  \nonumber \\
\eea
The determinant in Eq.(\ref{3}) does not exist unless the gauge is fixed. This is most conveniently done in a background Feynman gauge.
As a consequence the gauge-fixed partition function in the $ASD$ variables is:
\bea
Z&&=\int  \exp \big(-\frac{16\pi^2NQ}{g^2}-\frac{N}{4g^2}\int tr_f (\mu_{\alpha\beta}^-)^2d^4 x\big) 
 Det^{-1/2}(-\Delta_A \delta_{\alpha \beta} - i ad F_{\alpha \beta}^+ ) Det(-\Delta_A) \delta \mu^- \nonumber \\
\eea
As reported in the computation below, the Jacobian of the change of variables and the $FP$ determinant contribute to the first coefficient of the beta function the term $-\gamma_0=-\frac{5}{3}\frac{1}{(4 \pi)^2}$ \cite{3MB} (see also for a detailed computation section 3.3, 3.4 and 3.1 in \cite{GGI} and \cite{BP}), that does not reproduce the complete one-loop beta function of $YM$. This is most easily understood noticing that the differential operator in the Jacobian differs by the corresponding object in the one-loop contribution to the perturbative one-particle irreducible effective action just for the spin term: $- i ad F_{\alpha \beta}^+ $ instead of $ - i 2 ad F_{\alpha \beta}$.
However, by standard arguments, if the Jacobian develops zero modes the partition function gets additional divergent counterterms due to the Pauli-Villars regulator of the zero modes:
\bea
Z&&=\int  \exp \big(-\frac{16\pi^2NQ}{g^2}-\frac{N}{4g^2}\int tr_f (\mu_{\alpha\beta}^-)^2d^4 x \big) 
Det^{-1/2}(-\Delta_{A} \delta_{\alpha \beta}-i ad \mu_{\alpha \beta}^{- }) Det(-\Delta_{A}) \Lambda^{n_b [\mu^-] } \omega^{\frac{n_b[\mu^-]}{2}} \, \delta \mu^-\nonumber \\
\eea
where $\omega$ is a Kahler form on the moduli space that parametrizes the $n_b [\mu^-]$ zero modes, induced by a Kahler form on the connections (see section \ref{s7}), and we have conveniently substituted the operator occurring in the Jacobian with one that is isospectral to the non-zero modes of the original operator, employing standard spinor identities \cite{Rev} (see section 3.3 in \cite{GGI} and \cite{BP}). We show in section \ref{s8} that the moduli of surface operators in a neighborhood of the critical points furnish
the zero modes necessary to reproduce the correct beta function in the $TFT$ (see section \ref{s9}). Explicitly, the contribution of the Jacobian and of the $FP$ determinant to minus the exponential of the effective action,
to quadratic order in $\mu^-$ in the coordinate representation, in the gauge $\partial_{\alpha}A_{\alpha}=0$ for the background field, is:
\bea \label{div}
&& Det^{-1/2}(-\Delta_A \delta_{\alpha \beta} - i ad F_{\alpha \beta}^- ) Det(-\Delta_A) 
 =Det^{-1} (-\Delta_A) Det^{-1/2} (1- i (-\Delta_A )^{-1} ad F^-_{\alpha \beta}) \nonumber \\
&&\sim \exp \big(\frac{1}{2}Tr((-\Delta)^{-1} 2iA_\alpha \partial_\alpha (-\Delta)^{-1} 2iA_\alpha \partial_\alpha ) \big)  \exp\big(-\frac{1}{4}Tr((-\Delta)^{-1}adF^-_{\alpha \beta}(-\Delta)^{-1}adF^-_{\beta \alpha})\big) \nonumber \\
&&=\exp \big(\frac{N}{(4 \pi^2)^2}\int \int \frac{1}{6}\Delta A^a_\alpha (x) \frac{1}{(x-y)^4} A^a_\alpha(y) + \frac{1}{4} F^{a-}_{\alpha \beta}(x)  \frac{1}{(x-y)^4} F^{-a}_{\alpha \beta}(y)d^4x d^4y) \nonumber \\
&& \sim \exp \big(\frac{N}{(4 \pi)^2} \int \frac{1}{3}\log\frac{\Lambda}{M} \Delta A^a_\alpha (x)  A^a_\alpha(x) + \frac{2}{4} \log\frac{\Lambda}{M} F^{a-}_{\alpha \beta}(x)  F^{-a}_{\alpha \beta}(x)d^4x  ) 
\eea
where we have employed repeatedly the identity: $4 \frac{z_{\alpha}z_{\beta}}{z^8}=\frac{2}{3 z^6} \delta_{\alpha \beta}+\partial_{\alpha}\partial_{\beta}\frac{1}{6 z^4}$. The contribution to the beta function $-\frac{5}{3}\frac{1}{(4 \pi)^2}$ follows noticing that the first and second term in the last line are respectively $-\frac{1}{3}\frac{1}{(4 \pi)^2}$ and  $+2\frac{1}{(4 \pi)^2}$ times the quadratic part
of the classical action in the aforementioned background gauge.

\section{Holomorphic loop equation} \label{s6}

The equations of $ASD$ type in the resolution of identity,
$F_{01}-F_{23}=\mu^-_{01} ,
F_{02}-F_{31}=\mu^-_{02} ,
F_{03}-F_{12}=\mu^-_{03} $,
can be rewritten in the form of a Hitchin system, taking into account the central extension that occurs in the non-commutative case (we skip the superscript $\hat \,$ and we write $\partial_{A}  \bar D$ for $[\partial_{A},  \bar D]$, to simplify the notation):
\bea \label{4}
-i F_A+[D,\bar D] -\theta^{-1}1&&=\mu^0=\frac{1}{2}\mu^-_{01} \nonumber \\
-i\partial_{A}  \bar D&&= n=\frac{1}{4}(\mu^-_{02}+i\mu^-_{03}) \nonumber \\
-i\bar \partial_A D&&=\bar n=\frac{1}{4}(\mu^-_{02}-i\mu^-_{03}) 
\eea
or equivalently in terms of the non-Hermitian twistor 
connection with parameter $\rho$,
$B_{\rho}=A+\rho D+ \rho^{-1} \bar D=(A_z+ \rho D_u) dz+(A_{\bar z}+ \rho^{-1} D_{\bar u}) d \bar z$:
\bea \label{5}
-i F_{B_{\rho}} -\theta^{-1}1&&= \mu_{\rho}=\mu^0+\rho^{-1} n- \rho \bar n  \nonumber \\
-i\partial_{A}  \bar D&&= n  \nonumber \\
-i\bar \partial_A D&&=\bar n
\eea
The resolution of identity in the functional integral
then reads:
\bea
1=\int \delta n \delta  \bar n \int_{C_{\rho}} \delta \mu_{\rho} \delta(-i F_{B_{\rho}} - \mu_{\rho}-\theta^{-1}1) \delta(-i\partial_{A} \bar D- n) \delta(-i\bar \partial_A D- \bar n) 
\eea
where the measure $\delta \mu_{\rho}$ along the path $C_{\rho}$ is over the non-Hermitian path with fixed $n$ and $\bar n$ and varying $\mu^0$. The resolution
of identity is independent, as $\rho$ varies, on the complex path of integration $C_{\rho}$, because Eq.(\ref{5}) is equivalent to Eq.(\ref{4}).
We now specialize for simplicity to the case $\rho=1$ for the twistor connection $B \equiv B_{1}$ and curvature $\mu \equiv \mu_1$, since the general case does not add anything essentially new.  The partition function reads: 
\bea \label{positive}
Z&&=\int \delta n \delta \bar n \int_{C_{1}}\delta  \mu'  \frac{\delta  \mu } { \delta  \mu' }  \exp(-\frac{N 8 \pi^2 }{g^2} Q-\frac{N 4}{g^2}  \int Tr_f( \mu \bar \mu)  +Tr_f(n + \bar n)^2 d^4x) \nonumber\\
&& \delta(-i F_{B} - \mu-\theta^{-1}1) \delta(-i\partial_{A} \bar D- n) \delta(-i\bar \partial_A D- \bar n) 
\delta A \delta \bar A \delta D \delta \bar D
\eea
where we have further changed variables to the holomorphic gauge $B_{\bar z}=0$, defined by a gauge transformation in the complexification of the gauge group, for the $RHS$ of the holomorphic loop equation to be field independent. Now the v.e.v. is taken with respect to the measure:
\bea \label{holomorphic}
<\cdots>&&= Z^{-1} \int \delta n \delta \bar n \int_{C_{1}} \delta  \mu'  \cdots \exp(-\frac{N 8 \pi^2 }{g^2} Q-\frac{N 4}{g^2}  \int Tr_f( \mu-n+\bar n)^2  +4Tr_f(n \bar n) d^4x) \nonumber\\
&& \delta(-i F_{B} - \mu-\theta^{-1}1) \delta(-i\partial_{A} \bar D- n) \delta(-i\bar \partial_A D- \bar n) \frac{\delta  \mu } { \delta  \mu' }  
\delta A \delta \bar A \delta D \delta \bar D
\eea
where in the last equation we have rewritten the classical action as a holomorphic functional of $\mu$. 
The holomorphic loop equation is obtained following the Makeenko-Migdal technique \cite{MM,MM1}, as an identity that expresses the fact that the functional integral of a functional derivative vanishes:
\bea
\int  Tr\frac{\delta }{\delta \mu'(z,\bar z)} (e^{-\Gamma}\Psi(B'; L_{zz})) \delta \mu'=0
\eea
The new holomorphic loop equation for twistor loops follows:
\bea
&&<Tr(\frac{\delta \Gamma}{\delta \mu'(z,\bar z)}\Psi(B'; L_{zz}))> 
=\frac{1}{ \pi } \int_{L_{zz}} \frac{ dw}{z-w} <Tr\Psi(B'; L_{zw})> <Tr\Psi(B'; L_{wz})> \nonumber \\
\eea
where $\Psi(B'; L_{zz})$ is the holonomy of $B$ in the gauge $B'_{\bar z}=0$.
The Cauchy kernel arises as the kernel of the operator $\bar \partial^{-1}$, that occurs by functionally differentiating $\Psi(B'; L_{zz})$ with respect to $\mu'$, employing $i \bar \partial B_z'=\mu'$ (see Eq.(\ref{flat0})).
Going back to a unitary gauge, since everything is gauge invariant, we can simply skip the superscript $'$ to get Eq.(\ref{hol0}).

\section{Integrating on surface operators in the $TFT$} \label{s7}

On the physics side, because of translational invariance, critical points of the effective action occur at each space-time point. 
Therefore, the physics of $YM$ theory requires that the lattice of punctures on a local wedge in the $TFT$ defined in section \ref{s2} becomes dense in the continuum limit. 
This is a new version of lattice $YM$ theory, for which the continuum limit is obtained, as in standard Wilson formulation, by renormalization. 
Thus we represent the resolution of identity in the map to the $ASD$ variables on the punctured disks as a functional integral on infinite-dimensional parabolic bundles,
as suggested long ago \cite{MB2,MB3}:
\bea
1= \int \delta(-i[\hat D_{\alpha},\hat D_{\beta}]^--\sum_p \hat \mu^{-}_{\alpha \beta}(p)  \delta^{(2)} (z-z_p) - \theta^{-1}_{\alpha \beta} \hat 1 ) \prod_{p} \delta \hat \mu^{-}_{\alpha \beta}(p) 
\eea
In this notation all the dependence on the non-commutative coordinates is absorbed into the infinite dimensional nature of the connection on non-commutative space-time (see section \ref{s4}).
Therefore, the base of the infinite-dimensional parabolic bundles is the two-dimensional surface $R^2$, labelled by the commutative coordinates $(z, \bar z)$. 
This amounts to substitute the continuous field $\hat \mu^{-}_{\alpha \beta}(z, \bar z)$ with the lattice field $\hat \mu^{-}_{\alpha \beta}(p)$
by the resolution: 
$\hat \mu^{-}_{\alpha \beta}(z, \bar z)=\sum_p \hat \mu^{-}_{\alpha \beta}(p)\delta^{(2)} (z-z_p)$,
that is dense in the sense of distributions, since for any smooth test function of compact support
$N^{-1}_D\sum_p f(z_p ,\bar z_p)  \hat \mu^{-}_{\alpha \beta}(p) \rightarrow \int f(z,\bar z) \hat \mu^{-}_{\alpha \beta}(z, \bar z) d^2z$.
Codimension-two singularities of this kind have been introduced in \cite{MB2,MB3} in pure $YM$ theory as an "elliptic fibration of parabolic bundles", for the purpose of getting control over the large-$N$ limit exploiting the integrability of Hitchin fibration. In \cite{W2} they have been introduced in $n=4$ $SUSY$ $YM$ in relation to the geometric Langlands correspondence under the name of "surface operators", and this is now the name universally used in the physics literature. 
Formally, the $3$ constraints of $ASD$ type are the Hermitian and the complex moment maps for the Hamiltonian action of the infinite-dimensional unitary gauge group on
the commutative plane $R^2$ (or its compactification to a sphere, or locally on a disk):
\bea \label{6}
-i F_{\hat A}+[\hat D,\hat {\bar D}] -\theta^{-1}1&=& \sum_p \hat \mu^0_p \delta^{(2)} (z-z_p)  \nonumber \\
-i\partial_{\hat A} \hat {\bar D}&=&\sum_p  \hat n_p  \delta^{(2)} (z-z_p)  \ \nonumber \\
-i\bar \partial_{\hat A} \hat D&=&\sum_p \hat {\bar n}_p \delta^{(2)} (z-z_p)
\eea
with respect to the $3$ symplectic forms \cite{MB2,MB3,3MB} \footnote{We use the same labels $(I,J,K)$ of the symplectic forms as for the finite dimensional case in \cite{W2}, but we multiply the symplectic forms by a factor of
$2 \pi \theta$, that is naturally associated to the trace of the non-commutative gauge theory by the function/operator correspondence (see Eq.(\ref{nctr}) and for more details section 4 in \cite{GGI}).}:
\bea
\omega_I&&= \theta \int d^2 z  tr_f \hat Tr (\delta {\hat A}_{z} \wedge \delta {\hat A}_{\bar z}+ \delta {\hat D}_u \wedge \delta {\hat D}_{\bar u}) \nonumber \\
\omega_J-i\omega_K&&=\frac{\theta}{ i}\int d^2 z  tr_f \hat Tr (\delta {\hat A}_{z} \wedge \delta {\hat D}_{\bar u} ) \nonumber \\
\omega_J+i\omega_K&&=\frac{\theta }{ i}\int d^2 z  tr_f \hat Tr (\delta {\hat A}_{\bar z} \wedge \delta {\hat D}_{u}) 
\eea
as it follows immediately from the interpretation as (infinite-dimensional) Hitchin systems \cite{H,HKL}. In order to reduce to finite dimensional bundles, a technical requirement needed to get control on the moduli space of surface operators and of the associated zero modes in the functional integral on $ASD$ variables in section \ref{s5}, we compactify the non-commutative plane $R^2_{\theta}$ on a non-commutative torus of large area $\hat L^2$.
As recalled in section \ref{s4}, the corresponding non-commutative $U(N)$ gauge theory enjoys, for rational values of the dimensionless non-commutative parameter, $2 \pi \theta \hat L^{-2}= \frac{\hat M}{\hat N}$, 
Morita equivalence \cite{Szabo2,AG,Gu} to a theory on a commutative torus of area  $\hat L^2 \hat N^{-2}$, with gauge group $U(N \times \hat N)$, with the same 't Hooft coupling constant $g$, and 
with twisted boundary conditions. 
Thus in the Morita equivalent theory the $3$ operators $ \hat \mu^{-}_{\alpha \beta}(p)$ are reduced in fact to large finite-dimensional matrices. Besides, after taking the thermodynamic limit on the torus, the effect of the twisted boundary conditions disappears. As explained in section \ref{s2}, in the finite dimensional case the matrices $\hat \mu^{-}_{\alpha \beta}(p)$ all commute at each puncture, as a consequence of the local model of Hitchin equations \cite{S5,S6}. Thus there is a unitary gauge in which the $3$ Hermitian $ASD$ lattice fields can be diagonalized at the same time at each lattice point.
As recalled in section \ref{s2}, this local Abelianization implies the suppression of the local large-$N$ fluctuations around the critical points associated to the punctures, due to the reduction to the $O(3N)$ eigenvalues as opposed to the $O(3N^2)$ fluctuations of the original smooth $ASD$ field. 
However, the gauge connection is still non-Abelian: The missing degrees of freedom occur as moduli. 
Thus we come to the remarkable conclusion that the critical points of the effective action of the $TFT$ on the punctured surface can be interpreted as gauge orbits of connections
parametrized by hyper-Kahler moduli spaces, that arise as the quotient of the manifold defined by Eq.(\ref{6}) for the action of the unitary gauge group \cite{H,HKL}.
There is a symplectic form associated to the twistor connection:
\bea
\omega_{\rho}
&&=\theta \int d^2 z  tr_f \hat Tr (\delta {\hat B}_{\rho z} \wedge \delta {\hat B}_{\rho \bar z}) \nonumber \\
&&=\omega_I-i\rho (\omega_J+i\omega_K)+i\rho^{-1}(\omega_J-i\omega_K)
\eea
For $\rho=1$ it will be employed as an ingredient of the holomorphic/anti-holomorphic fusion in section \ref{s9}. Indeed, $\omega \equiv \omega_{1}$ depends holomorphically on $\mu$ via the connection $B$.
We employ also a modification $\omega'$ of $\omega$, defined over a punctured surface rather than over its compactification obtained adding the singular divisor. The relation between the two forms is (Eq.(3.30) in \cite{Malkin}):
\bea
\frac{1}{2 \pi \theta} \omega=\omega'+\sum_p  Tr (\mu_p(\delta g_p g_p^{-1})^2 )
\eea
where the terms in the sum over $p$ represent Kirillov forms on the adjoints orbits at $p$. $\omega'$ depends only on the holonomy of the connection (Eq.(3.13) in \cite{Malkin}). \par
Despite the parabolic singularity of surface operators, the topological term in $YM$ action Eq.({\ref{zeta}) has a well-defined mathematical extension to parabolic bundles, as parabolic Chern class. With the notation in \cite{W2}:
\bea
\frac{1}{ 16 \pi^2} \int d^4x F_{\alpha\beta}\, ^{*}\!{F}_{\alpha\beta}=Q+\sum_p tr_f(\alpha_p m_p)+\frac{1}{2} \sum_p  D_p \cap D_p tr_f(\alpha_p^2)
\eea
where $Q$ is the usual second Chern class of the $U(N)$ bundle without the parabolic structure, $\alpha_p$ is the vector of the parabolic weights at the point $p$, i.e.
the vector of the eigenvalues of $F^-_{01}$ divided by $2 \pi$ modulo $1$ in the fundamental representation, $m_p$ the magnetic flux through the surface $D_p$ of the singular divisor $p \times D_p$ of the surface operator and $D_p \cap D_p$ the index of self-intersection of the surface $D_p$. \par
Yet, the term in the action Eq.({\ref{zeta}) that involves the $ASD$ field
does not extend smoothly to surface operators \cite{3MB}.
As a consequence the classical $YM$ action is quadratically divergent at the singular divisor $p \times D_p$ of a surface operator,
with a divergence proportional
to the area of the singular locus of the surface operator.
We need a way to handle this classical divergence. 
For a non-commutative codimension-two surface the $YM$ action of the corresponding regularized non-commutative
gauge theory (see Eq.(\ref{NC})) is rescaled by a power of the inverse cutoff, that cancels precisely \cite{3MB} the quadratic
divergence that occurs evaluating the classical $YM$ action on surface operators. 
This allows us to define a new kind of semi-classical computation \cite{3MB} for which the classical $YM$ action is finite on parabolic bundles.
Since in our case the non-commutative $EK$ reduction is only partial (see section \ref{s4}), the action reads:
\bea
\frac{N}{2 g^2}  \hat N (\frac{2 \pi}{\Lambda})^2 \int d^2x tr_N Tr_{\hat N} (i [\hat \partial_{\alpha}+i \hat A_{\alpha},\hat \partial_{\beta}+i \hat A_{\beta}]+ \theta^{-1}_{\alpha \beta} 1)^2
\eea
where the trace $Tr_{\hat N}$ is over a subspace of dimension $\hat N$,
with  $\hat N (\frac{2 \pi}{\Lambda})^2=2 \pi \theta$ in the large $\hat N, \theta, \Lambda $ limit.
The contribution of each parabolic singularity in the action reads:
\bea
&&\int d^2x \delta^{(2)}(x-x_p)^2 
= \delta^{(2)}(0) \int d^2x \delta^{(2)}(x-x_p) 
=  (\frac{\Lambda}{2 \pi})^2
\eea
and its divergence is cancelled by its inverse, in the regularized $EK$ non-commutative action. In the Morita-equivalent theory, regularized in the $UV$ by the cutoff $\Lambda$ on a torus of size $\frac{\hat L}{\hat N}$: 
\bea
\frac{N}{2 g^2}  \hat N  \int d^4x tr_N Tr_{\hat N} (i [ \partial_{\alpha}+i  A_{\alpha}, \partial_{\beta}+i  A_{\beta}])^2
\eea
the contribution of each parabolic singularity of a surface operator is:
\bea \label{comm}
&&\int d^4x \delta^{(2)}(x-x_p)^2 
= \delta^{(2)}(0) (\frac{\hat L}{\hat N})^2 \int d^2x \delta^{(2)}(x-x_p)
=  (\frac{\Lambda}{2 \pi})^2 (\frac{\hat L}{\hat N})^2
= \frac{N_2}{\hat N^2}
\eea
where $(\frac{L}{\hat N})^2$ is the area of the singular divisor of the surface operator and the last equality is the definition of $N_2$. Thus the two regularized actions are normalized in the same way on surface operators provided
$\hat N^2=N_2$. \par
Hence the regularized partially-non-commutative $EK$ action, and its Morita equivalent version as well, are related to the action of the regularized commutative gauge theory on a torus of size $\hat L$ by the factor of $N_2^{-1}$, precisely the same factor that arises in the quenched version of $EK$ reduction \cite{Rt}. 

\section{Hyper-Kahler and Lagrangian moduli of surface operators, behavior around the parabolic divisor, and local Abelianization}  \label{s8}

In order to compute the beta function of the $TFT$, it is necessary to evaluate the dimension of the space of zero modes in a neighborhood of the critical points. This dimension coincides with the dimension of the moduli space of surface operators by the same standard argument as for instantons. The proper definition of the moduli space \cite{S5,S6} involves $L^2$ spaces for a conformally flat cusp metric adapted to the parabolic divisor. The conformal modification of the metric involves
at quantum level a conformal anomaly (see section \ref{s9}).
To study the moduli space it is convenient to compactify the $(z, \bar z)$ plane on a sphere.
The moduli space has three different equivalent descriptions that are usefully employed in this paper. There is a vast mathematics \cite{S1,S2,KM,Konno1,Konno2,S5,S6,Moc} and physics literature \cite{W2} on parabolic Hitchin bundles \footnote{These references are by no means a complete list.}. Thus we summarize briefly the 
essential results \cite{S1,S2,S5,S6}. \par
The first description of the moduli space is of differential geometric nature, as a Hitchin system and hyper-Kahler quotient. This is the description that occurs in the resolution of identity that defines the change to the $ASD$ variables for our special choice of a basis dense in function space, as we just discussed in section \ref{s7}.
In the hyper-Kahler description the structure group of the bundles involved is compact, in our case $U(N)$ or $SU(N)$. Thus we refer to a gauge fixing in this framework as to a unitary gauge.
Eq.(\ref{6}) reduces exactly to Hitchin equations for a connection constant \footnote{This suffices to compute the moduli in the translational-invariant vacua of the $TFT$.} on the, now commutative by Morita equivalence, ($u, \bar u$) plane (we disregard the constant $u(1)$ central extension that splits):
\bea \label{H}
-i F_A-[A_u, A_{\bar u}]&&=\sum_p \mu^0_p \delta^{(2)}(z-z_p)\nonumber \\
\partial_{A} A_{\bar u}&&= \sum_p n_p \delta^{(2)}(z-z_p)\nonumber \\
\bar \partial_A  A_{u}&&=\sum_p \bar n_p \delta^{(2)}(z-z_p)
\eea
Because of the delta function at $p$, in general the gauge connection has a pole singularity. The triple $(\mu^0_p, n_p,\bar n_p)$ determines the coefficients of the leading behavior of the gauge connection in a unitary gauge around the pole. The local model arises by restricting to such leading behavior \cite{S5,S6} and implies that the triple $(\mu^0_p, n_p,\bar n_p)$ is mutually commutative.  \par
This is a quite remarkable fact, referred to in this paper as local Abelianization (see section \ref{s2}) \footnote{A proof in the physicists style of the result in \cite{S5,S6} about the commutativity of the triple $(\mu^0_p, n_p,\bar n_p)$ can be found in \cite{W3} p.6 and p.31.}.
Hence there is a unitary gauge in which all the coefficients of the delta functions can be diagonalized at the same time, that is the ultimate reason that allows us the explicit computation in section \ref{s11} in the large-$N \hat N$ limit. \par
As for the behavior of the connection around the parabolic points, let us consider first the semi-simple case, for which by definition the eigenvalues  $e^{2i \mu^0_p}$ are all different (and different from $1$) \footnote{This implies that the parabolic structure is completely non-trivial \cite{S6}.} or the eigenvalues of $n_p$
are all different.
In this case a study of the local model \cite{S6} implies that the singular behavior of the connection $A_z$ and of the Higgs field $A_u$ in a unitary gauge around the parabolic divisor is a pure pole plus smooth terms. \par
In this description it is clear which are the local degrees of freedom of the $ASD$ field, but it is less clear what the moduli are for fixed eigenvalues, but for the fact that the moduli space is hyper-Kahler. \par
The structure of the moduli becomes clearer in the second description, that is of holomorphic nature. It arises by a meromorphic connection in a holomorphic gauge. 
Indeed, Hitchin equations imply the flatness equation:
\bea \label{flat}
-i F(B)&&=\sum \mu_p \delta^{(2)}(z-z_p) \nonumber \\
F(B)&&=\partial_z B_{\bar z}-\partial_{\bar z } B_{ z}+i[B_z, B_{\bar z}]
\eea
for the non-Hermitian connection:
\bea
B_z=A_z+i A_u \nonumber\\
B_{\bar z}=A_{\bar z}+i A_{\bar u} 
\eea
The moduli space arises as the Kahler quotient of the space of solutions of the flatness equation Eq.(\ref{flat}) for the action of the complexification of the gauge group. Because of a well known result \cite{H,HKL} it coincides with the hyper-Kahler quotient $\mathcal{H}$ of the three equations Eq.(\ref{H}) for the action of the compact gauge group \cite{Konno2,Konno1,S5,S6}.
The structure of the moduli space is particularly transparent in a holomorphic gauge:
\bea \label{flat0}
B'_{\bar z}=0
\eea
In this gauge $B'_z$ is a meromorphic connection:
\bea
i \partial_{\bar z }B'_{ z}=\sum_p \mu'_p \delta^{(2)}(z-z_p)
\eea
with residue at $p$ determined by $\mu'_p$, that is conjugate to $\mu_p$ by a gauge transformation in the complexification of the gauge group.
This description is the most transparent
to understand the moduli space, because the local moduli \footnote{By local moduli we mean moduli associated to the holonomy around a point.} are labelled by the adjoint orbits in the complexification of the gauge group $\mu_p'=G_p \lambda_p G_p^{-1}$. 
The holomorphic description arises in the holomorphic loop equation (see section \ref{s2} and \ref{s6}).  
It implies also that Hitchin equations are associated to local systems \cite{S1,S2}, i.e. to fiber bundles with locally-constant transition functions.  
A local system on a complex curve is the same as a representation of the fundamental group of a Riemann surface with punctures \cite{S1,S2}. \par
This is the topological description of the moduli, and it is also the easiest to understand globally.
Indeed, the residues of the meromorphic connection $B'_z$ determine its holonomy around $p$:
\bea
M_p&=& Pe^{{i \int_{L_p}} B'_z dz} = e^{2 i \mu_p'}
\eea
The global moduli space on a punctured sphere is therefore the quotient of the algebraic variety:
\bea
\prod_p M_p=1
\eea
by the adjoint action of the complexification of the global gauge group (this description has been employed in section \ref{s1}). \par
Now we come to the non-semi-simple case \cite{S5,S6}. In this case some eigenvalues of $n_p$ and of its Hermitian conjugate  $\bar n_p$ are degenerate, and the parabolic structure as well. Consequently, it follows from the local model that the Higgs field $A_u$ in a neighborhood of the parabolic divisor cannot be diagonalized, but can be put in Jordan form in a unitary gauge. Moreover, $A_u$ has not anymore a pole singularity along directions in color that are degenerate, but only a milder one, a pole divided by a power of a logarithm, determined by the off-diagonal parameters in the Jordan form of the holonomy \cite{S5,S6}. \par
The complex dimension of the local moduli space is the dimension of the adjoint orbit for $GL(C,N)$ associated to the holonomy, that is given by \cite{Konno2}:
\bea \label{HK}
 \dim_{C}(\lambda)=N^2-\sum_i m_i^2
 \eea
where $m_i$ are the multiplicities of the eigenvalues of the holonomy. \par
We are interested in special solutions of Hitchin equations for connections with $Z_N$ holonomy, that occur at the critical points in the $TFT$. These connections have no local moduli, since the adjoint orbit of the center is the center.
They satisfy Hitchin equations Eq.(\ref{H}) with $\mu^0_p=\tilde \mu_p$, $n_p=\bar n_p=0$,
and $e^{2i \tilde \mu_p}\in Z_N$. Therefore:
\bea 
\label{82}
2 \tilde \mu_p=diag(\underbrace{2\pi(k-N)/N}_{k},\underbrace{2\pi k/N}_{N-k})
\eea
Indeed, for a surface operator of $Z_N$ holonomy of magnetic charge $k$ around the point $p$, i.e. such that $M_p= e^{\frac{i 2 \pi k}{N}}$,
$N-k$ eigenvalues, $2\tilde \mu_p$, of the $ASD$ curvature  at $p$, $F^-_{01}=2 \tilde \mu_p \delta^{(2)}(z-z_p)$, are equal to
$\frac{2 \pi k}{N}$ and $k$ eigenvalues are equal to $\frac{2 \pi (k-N)}{N}$, for the curvature to be traceless and to give rise
to the holonomy $M_p= e^{\frac{i 2 \pi k}{N}}$ \footnote{The curvature in not uniquely determined by the holonomy, since parabolic bundles admit extensions over the punctures such that the eigenvalues of the $ASD$ curvature differ by shifts of $2 \pi$. Our choice is in some sense minimal. This feature together with many others, known in the mathematical literature, is reviewed in \cite{W2}. }.
These equations are invariant for the following $U(1)$ action
$A_u \rightarrow e^{i\phi} A_u $, $A_{\bar u} \rightarrow e^{-i\phi} A_{\bar u}$.
Since there are no moduli, this $U(1)$ must act by gauge transformations:
\bea
g _{\phi}A_u g _{\phi}^{-1} &&= e^{i\phi} A_u \nonumber \\
g _{\phi} A_{\bar u}g _{\phi}^{-1} &&= e^{-i\phi} A_{\bar u}  \nonumber \\
g _{\phi}A_z g _{\phi}^{-1} &&=  A_z \nonumber \\
g _{\phi} A_{\bar z}g _{\phi}^{-1}&& =  A_{\bar z}    
\eea
More generally, the set of moduli fixed by this $U(1)$ action is the Lagrangian submanifold of the hyper-Kahler moduli space \cite{Konno1} for which the Higgs field $A_u$ is nilpotent \cite{Konno2}. The connections valued in $Z_N$, that occur at the critical points, sit on the tip of the Lagrangian cone. \par
The Lagrangian cone is of particular interest, since as we will show in section \ref{s9}, only moduli of gauge orbits in the Lagrangian cone can reproduce the $YM$ Wilsonian beta function.
There are fundamentally two interesting types of orbits in the Lagrangian cone of the hyper-Kahler moduli space \cite{SL}:
The orbits with unitary holonomy, for which the Higgs field vanishes identically;
The orbits that correspond to more general Hodge bundles, for which the holonomy is valued in a real version of the complexification of the gauge group \cite{SL}. 
In the first case the holonomy can always be diagonalized, despite the eigenvalues may not be all different. 
In the second case the holonomy cannot be diagonalized, but can be set in Jordan form. \par
The first component of the Lagrangian cone corresponds to $A_u=0$, and gives rise to unitary representations of the fundamental group.
The complex dimension of an adjoint orbit at a parabolic point of a unitary bundle of rank $N$ is given by \cite{Konno1,Konno2}: 
\bea \label{flag}
\dim_{C}(\lambda)=\frac{1}{2} (N^2-\sum_i m_i^2)
\eea
that is one-half the dimension in the generic case (see Eq.(\ref{HK})), as it must be for a Lagrangian submanifold.
Yet, we observe that, as in Nekrasov localization \cite{Nek} \footnote{Nekrasov localization occurs in $n=2$ $SUSY$ $YM$ by cohomology (see section \ref{s1}). In this case the cohomology ring localizes on instantons, whose moduli contribute to the beta function. However, the integral on the moduli as well is
localized on fixed points (for a torus action) that do not carry families of zero modes. Thus renormalization must be performed before exploiting this second localization.}, it is necessary to evaluate and renormalize the functional measure in a neighborhood of the critical points and thereafter sitting on the critical points.
Since the critical points have no moduli and thus no zero modes, the inverse order, first sitting on the critical points and then renormalizing the functional measure, would not lead to the correct result. \par
The following slight deformation of the eigenvalues, keeping constant their multiplicity in order not to affect the the glueball potential and spectrum, that depends only on the degeneracy of the eigenvalues at the critical points (see section \ref{s10}), gives rise to a non-trivial adjoint orbit for the holonomy:
\bea 
\label{vorticesZN}
2 \lambda =diag(\underbrace{2\pi(k-N)/N+\epsilon}_{k},\underbrace{2\pi k/N-\epsilon k/(N-k)}_{N-k})
\eea
Thus the real dimension of the orbit is:
\bea \label{flag1}
\dim_{R}(\lambda)= N^2-k^2-(N-k)^2= 2k(N-k)
\eea
Hence the real dimension is even but not a multiple of $4$ in general, as it must be for a Lagrangian submanifold of a hyper-Kahler manifold, whose real dimension is always a multiple of $4$. \par
There is a more intrinsic characterization of the Lagrangian neighborhood of the critical points.
Instead of deforming slightly the eigenvalues, we may deform the holonomy along nilpotent directions. Again, this deformation does not modify the glueball potential and glueball spectrum, that depend only on the eigenvalues of $\mu$.\par
Therefore, the eigenvalues of the $ASD$ field are given exactly by Eq.(\ref{82}),
but now we allow the Higgs field $A_u$ to have a nilpotent residue. In the Lagrangian cone not only the Higgs field has a nilpotent residue, but it is nilpotent itself \cite{S6}.
These are bundles of Hodge type \cite{SL} for which the twistor connection has a holonomy that cannot be diagonalized, but it can be set in Jordan canonical form. \par
In the Lagrangian cone the local moduli at a point are parametrized by orbits of Jordan canonical form for a real version of the complexification of the compact gauge group \cite{SL}.  
Thus in our case, for which the diagonal part of the holonomy is in $Z_N$, the local holonomy is unipotent.
Since the diagonal part of the holonomy is central, to compute the dimension of the orbit we need to consider only the nilpotent part.
Any such matrix is conjugate by the Jordan theorem to a direct sum of $k$ blocks of dimension $d_i$ such that $\sum^k_{i=1} d_i=N$, where $N$ is the total rank. Each block has $d_i$ zero eigenvalues on the diagonal and it is upper triangular with all $1$
on the super-diagonal. Thus a nilpotent orbit is associated to a partition of $N$ given by the sequence of the dimensions of Jordan blocks in decreasing order \cite{Nil1,R}. This defines a Young tableaux whose rows have length $d_i$.
To this Young tableaux it is associated a dual tableaux with the rows and columns exchanged and a dual partition $\hat d_i$, that will play a role momentarily. \par
To get the same contribution to the beta function as in the unitary case, we need to construct orbits of nilpotent Jordan matrices with precisely the same real dimension as in Eq.(\ref{flag1}). The general formula for the complex dimension of the adjoint orbit in $GL(C,N)$ of a nilpotent is \cite{Nil1}: 
\bea \label{part}
&&\dim_C(O_N)= N^2-N-2 \sum_i (i-1)d_i
\eea
that can be conveniently rewritten in terms of the dual partition \cite{R}:
\bea \label{Nil}
&&\dim_C(O_{N})=N^2-\sum_i \hat d^2_i
\eea
Choosing $\hat d_1=N-k$, $\hat d_2=k$, we get a real dimension that is the double of the dimension in Eq.(\ref{flag1}). But restricting to the Lagrangian cone amounts to computing the dimension of the orbit for a real version of the complexification of the gauge group. Hence for such an orbit we get from Eq.(\ref{Nil}) the real dimension $2k(N-k)$.
The nilpotent orbit corresponding to the dual Young tableaux
is the direct sum of $k$ nilpotent Jordan
blocks of dimension $2$ and of $N-2k$ Jordan blocks of dimension 1. \par
As a check, the same dimension follows from Eq.(\ref{part}):
\bea
\dim_C(O_{N})&=&N^2-N-2 \sum^k_{i=1}(i-1)2-2\sum^{N-k}_{i=k+1}(i-1) \nonumber \\
 &=&N^2-N +2(2k+N-2k)-2\sum^k_{i=1}i-2\sum^{N-k}_{i=k+1}i \nonumber \\
 &=&N^2-N +2N-(N-k+1)(N-k)-(k+1)k  \nonumber \\
&=&N^2+N-(N-k+1)(N-k)-(k+1)k  \nonumber \\
&=& 2k(N-k)
\eea

\section{Effective action of the $TFT$, Wilsonian and canonical beta function} \label{s9}

The partition function of the $TFT$ reduces by the local Abelianization, i.e. by the choice of gauge that diagonalizes at the same time the three components of the residue of the $ASD$ fields at each puncture, to an integral over the eigenvalues of $(\mu_p, n_p, \bar n_p)$, that by an abuse of notation we indicate by the same symbols when no confusion can arise. Moreover, the gauge connection can be gauge fixed away from the punctures by the usual background Feynman gauge (that is understood but not explicitly displayed in Eq.(\ref{TFT})), and thus the integral on the gauge connection induces an integral on the moduli of surface operators, that give rise to the zero modes of the Jacobian of the change of variables. For generic hyper-Kahler orbits the partition function of the $TFT$ reads:
\bea \label{TFT}
 Z=|\int_{\mathcal{H}}  \delta A   \delta \bar{A} \delta  D   \delta { \bar D} \, \delta(-i F_{ B} - \sum_p  \mu_p  \delta^{(2)}_p) 
 \delta(- i \partial_{ A}{\bar D}- \sum_p  {n}_p  \delta^{(2)}_p)\delta(- i\bar \partial_{A}  D - \sum_p  {\bar n}_p  \delta^{(2)}_p)  \nonumber \\ 
\exp{(-\frac{4N \hat N}{g_W^2}  \sum_p  tr_N    Tr_{\hat N } ((\mu_p -n_p+\bar n_p)^2+ 4 n_p {\bar n}_p))}
\frac{\Delta(n_p) \Delta( \bar n_p) }{\Delta(\mu_p)} (\Lambda \sqrt \theta)^{n_b} \omega'^{\frac{n_b}{2}} 
\prod_p \delta \mu_p \wedge  \delta n_p \wedge \delta \bar n_p |^2\nonumber \\
\eea
Some observations are in order. \par
The partition function is explicitly factorized in its holomorphic/anti-holomorphic form, because of the factorization of Wilson loops in the adjoint representation into the fundamental
and conjugate representations at the leading $\frac{1}{N}$ order. \par
We have normalized the classical and quantum effective action according to $EK$ non-commutative reduction (see end of section \ref{s7}). In ordinary $YM$ theory on commutative space-time the classical and quantum effective action would get an overall factor of $N_2$ according to Eq.(\ref{comm}). \par
The functional integral on the connection involves now only the non-zero modes, while the measure on the zero modes is represented by $\omega'^{\frac{n_b}{2}}$.
The superscript in $\omega'$ refers to the version of $\omega$ defined on the punctured surface, as discussed in section \ref{s7}. $\omega'$ depends on the curvature in a holomorphic gauge $\mu'_p$ only through the holonomy, a fact that has importance computationally (see section \ref{s10}). Substituting $\omega'$ to $\omega$ is allowed, since excluding the singular divisor is equivalent to omitting in $\omega$
the sum of Kirillov forms of the adjoint orbits on the singular divisor \cite{Malkin}, because the volume form on these orbits is already taken into account by the product measure on the adjoint orbits $\prod_p \delta \mu_p'$
that occurs by the resolution of identity. As a consequence: 
\bea
 \omega^{\frac{n_b[ \mu']}{2}} \wedge \prod_p \delta \mu'_p = \omega'^{\frac{n_b[ \mu']}{2}} \wedge \prod_p  \delta \mu'_p
\eea
$\Delta(\lambda)=\prod_{i>j}(\lambda_i-\lambda_j)$ is the Vandermonde determinant of the eigenvalues of $\lambda$.\par
The Vandermonde determinant that occurs in the numerator in Eq.(\ref{TFT}) is due to gauge fixing $n_p$ in triangular form, by the action of the unitary gauge group in a unitary
gauge. As a consequence $n_p$ is  automatically diagonal on the locus for which Hitchin equations admit solutions (see section \ref{s8}). \par
The Vandermonde determinant in the denominator arises as follows.
Gauge fixing in a unitary gauge $ \mu_p= g_p( \lambda_ p + \text{ nilpotent})g_p^{-1}$ in triangular form, that is automatically diagonal because of the local Abelianization, i.e. $\text{nilpotent}=0$ for a solution of Hitchin equations, produces one power of the Vandermonde determinant in the measure $\prod_p \delta \mu_p |_{\text{ Hitchin locus}}$ restricted to solution of Hitchin equation in a unitary gauge. \par
The change of variables to a holomorphic gauge $ \mu_p'=  G_p \lambda_p G_p^{-1}$
produces two powers of the Vandermonde determinant in the measure $\prod_p \delta \mu'_p$ in a holomorphic gauge. Hence we get the Jacobian to a holomorphic gauge as the Radon-Nykodim derivative of the measure in a unitary gauge with respect to the measure in a holomorphic gauge:
\bea
\wedge  \omega'  \prod_p \frac{\wedge  \delta \mu_p|_{\text{ Hitchin locus}}}{\wedge \delta \mu'_p}= \frac{ \wedge \omega' }{\prod_p \wedge \delta G_p} \prod_p \frac{\Delta(\mu_p) \wedge \delta \lambda_p}{\Delta(\mu_p)^2 \wedge \delta \lambda_p} = Det(\omega')^{\frac{1}{2}} \prod_p \Delta(\mu_p)^{-1}
\eea
where $\wedge$ means here the maximal power, and the dependence on the moduli $G_p$ of the adjoint orbits is included in $\omega'$. \par
The occurrence of the inverse of Vandermonde determinant of $\mu$ as the Jacobian to a holomorphic gauge is the most important technical point of this paper, because its logarithm furnishes the glueball potential in the effective action (see Eq.(\ref{hol}) and section \ref{s10}). \par
At large-$N \hat N$ it is not restrictive to assume that only one set of eigenvalues or a discrete sum of them actually contribute to the partition function. 
The eigenvalues $ \tilde \mu_p$ at critical points are fixed requiring that the global gauge group be
unbroken. It follows that twistor Wilson loops in the fundamental representation have
$Z_N$ holonomy at critical points, and thus $n_p=\bar n_p=0$. Hence the induced measure at the critical points, up to constant factors, is:
\bea
Z&&=\big| \int_{\mathcal{L}}   \delta A  \delta \bar{A} \delta  D   \delta   { \bar D} 
\delta(-i F_{ B} - \sum_p  \tilde \mu_p  \delta^{(2)}_p) \delta( \partial_{ A}{\bar D})\delta(\bar \partial_{A}  D) 
 e^{-\frac{4N \hat N}{g_W^2}  \sum_p  tr_N    Tr_{\hat N } ( \tilde \mu_p^2)} \Delta(\tilde \mu_p)^{-1} (\Lambda \sqrt \theta)^{n_b} \omega^{\frac{n_b}{2}} \big|^2  \nonumber \\ 
\eea
Now we compute the Wilsonian beta function in a Lagrangian neighborhood $\mathcal L$ of the critical points (see section \ref{s9}) in three steps. \par
Firstly, we evaluate the classical action of surface operators of $Z_N$ holonomy. The Morita equivalent commutative \footnote{We rescale the area of the torus in the Morita equivalent theory from $\frac{L^2}{\hat N^2}$ to $\hat L^2$ in order to perform the thermodynamic limit uniformly in
$\hat N$.} theory has gauge group $U(N \times \hat N)$ and the center of $SU(N)$ is embedded diagonally in $U(N \times \hat N)$ as $e^{\frac{i 2 \pi k}{N}} 1_{\hat N}$.
The trace of the square of the eigenvalues of the $ASD$ curvature in the fundamental representation is thus:
\bea \label{cond}
tr_N(4\tilde \mu^2)=(N-k) (\frac{2 \pi k}{N})^2 + k (\frac{2 \pi (k-N)}{N})^2 
=(2 \pi)^2 \frac{k(N-k)}{N}
\eea
with $\tilde \mu \equiv \tilde \mu_p$, where we have disregarded the contribution of the tensor product with $1_{\hat N}$, since it produces just an overall factor everywhere. It is convenient either to sum the contributions of the fundamental and conjugate representation at a point, or
to sum the contributions of the pairwise-identified punctures associated to a double point (see section \ref{s2}). Since the curvature at the critical points is charge conjugate for pairwise-identified punctures, the result is the same.
This is compatible also with the counting of moduli and zero modes associated to double points. They coincide with moduli associated with either one of the pairwise-identified punctures, that is the same as counting only complex zero modes associated to the fundamental representation and complex-conjugate zero modes associated to the conjugate representation. \par
Thus the regularized non-commutative $EK$ action, including the fundamental and conjugate representation (see section \ref{s4} and \ref{s7}), is: $S_{EK}= \frac{N \hat N (4 \pi)^2}{ g^2}  \frac{Q}{N_2} + \frac{ \hat N^2}{ g^2}  \sum_p 2(2 \pi)^2 k(N-k)$,
while the corresponding  commutative the $YM$ action on a commutative torus of size $\hat L$ is: $S_{YM}= \frac{N \hat N (4 \pi)^2}{ g^2}  Q + \frac{ \hat N^2}{ g^2} N_2 \sum_p 2(2 \pi)^2 k(N-k)$. \par
It follows that the contribution of any finite second Chern class $Q$ to the classical and quantum effective action is irrelevant with respect to the one of the $ASD$ curvature at the
parabolic singularities. This fact allows us to ignore issues related to the possible global non-triviality of the bundles and to the counting of the global moduli \cite{KM}, and to concentrate only on the local
counterterms associated to the sum over the points of the parabolic divisor. \par
Secondly, there are logarithmic divergences from
the Jacobian of the change of variables and the $FP$ determinant.
They have already been computed in Eq.(\ref{div}), and we can adapt our previous calculation to the case 
of a lattice of surface operators. They contribute a renormalization factor $Z^{-1}(\frac{\Lambda}{M})=1- g^2_W\frac{1}{(4\pi)^2} \frac{10}{3}
\log (\frac{\Lambda}{M})$ for the Wilsonian coupling:
\bea \label{J}
\frac{(4\pi)^2 k(N-k)}{2 g^2_W(M)}
&& = (4 \pi)^2 k(N-k)( \frac{1}{2 g^2_W(\Lambda)}- \frac{1}{(4\pi)^2} \frac{5}{3}
\log (\frac{\Lambda}{M})) \nonumber \\
&& =(4 \pi)^2 k(N-k) \frac{Z^{-1}(\frac{\Lambda}{M})}{2 g^2_W(\Lambda)}
\eea
For further use it is convenient to compute explicitly, by means of Eq.(\ref{div}) and of the local model of the solutions of Hitchin equations (see section \ref{s8}), the complete quadratic counterterm in the effective action evaluated on a lattice of surface operators and not only its divergent part:
\bea \label{fluc}
- \frac {5}{ 3 (4 \pi^2)^2 } \sum_{ p, p' }  \int d^2u d^2v  \frac{N 4 Tr(\tilde \mu_p \tilde  \mu_{p'})}{(|z_p-z_{p'}|^2+|u-v|^2)^2}
\eea
where the sum over $p, p'$ runs over the planar lattice of the parabolic divisors of the surface operators. \par
To avoid quadratic divergences, in order to satisfy the usual power counting
as in the background-field computation of the beta function (see section 3.1 in \cite{GGI}),
we restrict the sum to $p \neq p'$. Although quadratic divergences in the commutative Morita equivalent theory correspond to finite counterterms
for the $EK$ reduced action, we avoid divergences at coinciding points in higher orders of the loop expansion for the conventional power counting to hold.
On the translational invariant condensate Eq.(\ref{fluc}) reduces to:
\bea
&&- \frac {5}{ 3(4 \pi^2)^2 } \sum_{ p \neq p' }  \int d^2u d^2v  \frac{N 4 Tr(\tilde \mu^2) }{(|z_p-z_{p'}|^2+|u-v|^2)^2} \nonumber \\
&&\rightarrow - \frac {5}{3 (4 \pi^2)^2 }   \int d^2u a^{-2} d^2z a^{-2} \int d^2w d^2v  \frac{N 4 Tr(\tilde \mu^2) }{(|z-w|^2+|u-v|^2)^2} \nonumber \\
&&\sim - N_2^2   N \frac {5}{ 3(4 \pi)^2 } Tr(8\tilde \mu^2) \log \frac{\Lambda}{M} 
= - N_2 \sum_p   N \frac {5}{ 3(4 \pi)^2 } Tr(8\tilde \mu^2) \log \frac{\Lambda}{M}
\eea
with $x=(z, \bar z, u, \bar u)$ and $y=(w, \bar w, v, \bar v)$,
that is logarithmically divergent, both at the ultraviolet and at the infrared in the thermodynamic limit, with $a$ a lattice scale  related to the cutoff by $a^{-1}=\frac{\Lambda}{2 \pi}$ and with $M$ an infrared renormalization scale. \par
One factor of $N_2$ is the sum on lattice points, the other factor is the "phase-space area" $N_2=(\frac{\Lambda}{2 \pi})^2 \hat L^2$ of one surface operator Eq.(\ref{comm}). \par
Thirdly, there is the contribution of zero modes. It follows from Eq.(\ref{J}) that, in order to get the correct one-loop beta function, the contribution of the zero modes to the renormalization of the action should be $-2k(N-k) \log (\frac{\Lambda}{M})$, including the fundamental and conjugate representation.
The sign is consistent with Pauli-Villars regularization of zero modes, yet the absolute value of the coefficient of the logarithm 
is in general an even integer but not a multiple of $4$, as it would be implied by the
hyper-Kahler reduction \footnote{A hyper-Kahler manifold has necessarily a real dimension that is a multiple of $4$.}. \par
Thus the neighborhood of the critical points cannot be generic. We have seen in section \ref{s8} that the critical points sit inside the Lagrangian cone of the moduli space
for which $A_u$ is nilpotent. Since in a Lagrangian neighborhood of the critical points the dimension of the moduli space is generically one-half the dimension of a hyper-Kahler neighborhood, the correct beta function may arise. 
We have identified in section \ref{s8} the Lagrangian neighborhoods of the critical points that lead to the correct Wilsonian beta function. Including the zero modes, the renormalization of the Wilsonian coupling at both the $UV$ and $IR$ divisors, provided the subtraction scale $M$ is chosen to be the same, reads:
\bea \label{eff}
\frac{(4\pi)^2 k(N-k)}{2 g^2_W(M)}
&& = (4 \pi)^2 k(N-k)( \frac{1}{2 g^2_W(\Lambda)}- \frac{1}{(4\pi)^2} (\frac{5}{3}+2)
\log (\frac{\Lambda}{M})) 
\eea
that implies that the Wilsonian beta function is one-loop exact Eq.(\ref{beta}). 
Moreover, choosing the subtraction scale at the $IR$ close to the infrared Landau singularity $\Lambda_{W}$ of the Wilsonian coupling,
and the subtraction scale $M$ at the $UV$ close to the cutoff $\Lambda$, we get for the local part of the effective action (including the contribution of the tensor product with $1_{\hat N}$):
\bea
&&Z= e^{-  \sum_p  tr_N    Tr_{\hat N }\frac{(4\pi)^2 \hat N^2 k(N-k) }{2 g^2_W(M)} } \omega'^{\sum_p \hat N^2 k(N-k)   } \big| _{UV} \nonumber \\ 
&& 
 e^{-  \sum_p  tr_N    Tr_{\hat N }\frac{(4\pi)^2 \hat N^2 k(N-k)Z^{-1}(\frac{M}{\Lambda_W}) }{2 g^2_W(M)} }  \omega'^{\sum_p \hat N^2 k(N-k)   } \big|_{IR} 
\eea
where we have taken into account the contribution of the $IR$ divisor as well. The fields at the $UV$ and at the $IR$ are not canonically normalized
in the Wilsonian effective action. \par
In order to obtain the canonical beta function, as in the standard computation in $n=1$ $SUSY$ $YM$ (see \cite{Sh} and section 3.2 in \cite{3MB}), we rescale the gauge connection $A$ in the resolution of identity Eq.(\ref{ri}) at the $UV$ divisor by a factor of $g(M)$ and at the $IR$ divisor by a factor of $g(M) Z(\frac{M}{\Lambda_{W}})^{\frac{1}{2}}$ \footnote{We rescale at the $IR$ by $g(M) Z(\frac{M}{\Lambda_{W}})^{\frac{1}{2}}$ rather than by $g_W(M) Z(\frac{M}{\Lambda_{W}})^{\frac{1}{2}}$. The difference between the two choices is just a change of scheme that affects only the higher-order coefficients of the beta function but not its universal part. The reason for the first choice will be apparent in section \ref{s12}, because it allows the existence of a scheme-independent zero in the beta function at the point where the magnetic condensate becomes unstable in massless Veneziano limit of $QCD$.}. The rescaling of the connection induces rescaling factors in the symplectic form on the moduli, i.e. the zero modes at the $UV$ and $IR$, and in the partition function including the fundamental and conjugate representation:
\bea
\omega'^{\sum_p \hat N^2 k(N-k)}  \rightarrow  g(M)^{\sum_p \hat N^2 2k(N-k)} \omega'^{\sum_p \hat N^2 k(N-k)}  \nonumber \\
\omega'^{\sum_p \hat N^2 k(N-k)}  \rightarrow  (g(M) Z^{\frac{1}{2}}(\frac{M}{\Lambda_W}))^{\sum_p \hat N^2 2k(N-k)} \omega'^{\sum_p \hat N^2 k(N-k)}
\eea
Exponentiating these factors, they combine with the action at the $UV$ divisor (the action at the $IR$ is vanishing small compared to the action at the $UV$ because of the Landau infrared singularity of the Wilsonian coupling, or alternatively because $Z^{-1}(\frac{M}{\Lambda_W})$ vanishes in the large-$M$ limit by Eq.(\ref{an})). Thus defining $g(M)$ at the $UV$ divisor by:
\bea \label{be}
\frac{1}{2g_W^2(M)}=\frac{1}{2g^2(M)}+\frac{2}{(4\pi)^2}\log g^2(M)+\frac{1}{(4\pi)^2}\log Z(\frac{M}{\Lambda_W}) \nonumber \\
\eea
the effective action is canonically normalized at the $UV$ (at the $IR$ it was already by the aforementioned rescaling). Besides, the canonical effective action at the $UV$ is made finite provided the action density is rescaled by a factor of $g^2(M)$, that corresponds to its anomalous dimension.
Taking the derivative of Eq.(\ref{be}) with respect to $\log M$ and using the fact that $g_W$ is one-loop exact, the canonical beta function Eq.(\ref{alfa}) follows. \par
Eq.(\ref{be}) can be rewritten as an equality between the Wilsonian and canonical $RG$-invariant scales for 't Hooft coupling:
\bea \label{RGW}
M e^{ -\frac{1}{2 \beta_0 g_W^2}} = M e^{ -\frac{1}{2 \beta_0 g^2}} (g^4 Z)^{-\frac{1}{(4 \pi)^2 \beta_0}} 
\eea
From Eqs.(\ref{be},\ref{z},\ref{g},\ref{z1},\ref{an}) it follows with leading and next-to-leading logarithmic accuracy for large-$M$:
\bea
\beta_0 \log M \sim \frac{1}{2g^2(M)}-\frac{1}{ (4\pi)^2} 2 \log \log M +\frac{1}{ (4\pi)^2}  \frac{10}{22} \log \log M
\eea
Comparing with Eq.(\ref{sc}), we get that the first-two coefficients of the canonical beta function of the $TFT$ coincide with the perturbative ones.
The one-loop exactness of $Z$ implies:
\bea \label{an}
\frac{\partial \log Z}{\partial \log \Lambda} =\frac{ \frac{1}{(4\pi)^2} \frac{10}{3} g_W^2}
{1-g_W^2 \frac{1}{(4\pi)^2} \frac{10}{3}
\log (\frac{\Lambda}{\sigma M })}
\eea
where now we have included the contribution of the conformal anomaly, due to the singular rescaling by a factor of $e^{\sigma}$ of the metric, that is adapted to the cusps (see section \ref{s8}). This is equivalent to rescaling the subtraction point $M$ by the factor of $\sigma$,
to give a finite but arbitrary result for the higher-order contributions to the anomalous dimension. \par
Assuming that the anomalous dimension is independent
on the subtraction point, as required by general principles of the $RG$, the $RG$ trajectory must be followed along the line $c'=-\frac{1}{(4\pi)^2} \frac{10}{3}
 \log (\frac{\Lambda}{\sigma M})=const$. It is precisely
the contribution of the conformal anomaly that allows the anomalous dimension to be a function of the coupling
only, according to the $RG$. \par
To describe the continuum limit, we introduce the density of surface operators
$\rho=\sum _{p'} \delta^{(2)}(z-z_{p'})$,
normalized in such a way that $\int d^2z \sum _{p'} \delta^{(2)}(z-z_{p'})= N'_2$
is the number of lattice points for which the holonomy is non-trivial at the scale for which the density is $\rho$. 
This allows $\rho$ to scale non-trivially with the $RG$. \par
At the leading $\frac{1}{N}$ order in the holomorphic loop equation and in Eq.(\ref{TFT}) we have regarded the $\mu$ and $\bar \mu$ integrations as independent. Indeed, since twistor Wilson loops are holomorphic functionals of $\mu$, we can interpret the localization due to the holomorphic loop equation as a non-$SUSY$ analog of the localization of the holomorphic ring \cite{V,Kawai,Kawai1} in $n=1$ $SUSY$ gauge theories on holomorphic matrix models \cite{Laz}. But at the next to leading order, we can integrate on a path for which $\bar \mu$ is the adjoint of $\mu$, by means of the following version of holomorphic/anti-holomorphic fusion \cite{CV}: The effective action (with the standard normalization of the commutative gauge theory) in the continuum limit is implied by Eq.(\ref{TFT}) considered as a functional of $(\mu,\bar \mu,\nu)$:
\bea \label{hol}
\Gamma&=&\frac{4N \hat N}{g_W^2} \int d^2u d^2z   \rho^2   tr_N    Tr_{\hat N } (\mu\bar \mu+ \nu^2)+\int d^2u d^2z  \rho^2 (\log \Delta(\mu) 
- \log|\Delta(\nu+ \mu-\bar \mu)|^2 )\nonumber \\
&-& \log Det^{-1/2}(-\Delta_{A} \delta_{\alpha \beta}-i ad \mu_{\alpha \beta}^- ) Det(-\Delta_{A})  -  \int d^2u d^2z   \rho^2 n_b [\mu']  \log \Lambda- \int d^2u \rho \log Det(\omega')^{\frac{1}{2}}  \nonumber \\
&+& \text{complex conjugate}
\eea
where $\nu=n+\bar n$.
This form of the effective action is of the utmost importance, because it shows that the coefficient $\rho^2$ of the glueball potential is in fact $RG$ invariant, as it should be.
Indeed, the local divergent part
of the Wilsonian effective action for a condensate of surface operators of area $\hat L^2$ in the hypercube of volume $\hat L^4$ with holonomy in $Z_N  \times Z_{\bar N}$ of magnetic charge $(k,-k)$ with surface density $\rho$ reads:
\bea
\Gamma_k
&&= -  \hat N^2 \hat L^4  k(N-k)(4 \pi)^2  \rho^2  \frac{\beta_0}{4} \log \frac{N (\Lambda e^{-\frac{1}{2 \beta_0 g_W^2}})^4  }{\hat N k(N-k) (4 \pi)^2  \rho^2 }  
\eea
where in the logarithm we have conveniently chosen the infrared subtraction point $M$ at a scale \footnote{Other choices of the subtraction scale are possible in order to match physical requirements. But the choice that we have made is the most convenient computationally. While the glueball spectrum in units of the scheme-dependent $RG$-invariant scale $\Lambda_{W}$ and the universal asymptotic behavior of correlators turn out to be independent on our specific choice, the scheme dependent objects are in fact affected by the choice.
Thus the scheme can be completed fixed only imposing physical conditions and computing carefully finite parts in the given scheme. We will not pursue further this direction in this paper.} of the order of the action density $M^4 = \frac{\hat N}{N} k(N-k)(4 \pi)^2 \rho^2$, as in the analog in the $TFT$ of Veneziano-Yankielowicz  $n=1$ $SUSY$ $YM$ effective action \cite{VY}:
\bea
\Gamma&=& \hat N N\int d^4x (\frac{1}{2 g_W^2} Tr(F^2) - \frac{\beta_0}{4} Tr(F^2) \log\frac{\Lambda^4}{Tr(F^2)}) \nonumber \\
&=& - \hat N N \int d^4x \frac{\beta_0}{4} \log \frac{(\Lambda e^{-\frac{1}{2 \beta_0 g_W^2}})^4  }{Tr(F^2) }  
\eea
As a consequence in the sector labelled by $k$ the minimum of the density $\rho_k$ satisfies:
\bea \label{con}
\frac{\delta \Gamma_k}{\delta \rho}|_{\rho=\rho_k} &&=  -  \hat N^2  \hat L^4    k(N-k)(4 \pi)^2  \frac{\beta_0}{2}   
( \rho  \log \frac{(\Lambda e^{-\frac{1}{2 \beta_0 g_W^2}})^4  }{  \frac{\hat N}{N}k(N-k)(4 \pi)^2  \rho^2 }   - \rho) =0\nonumber \\
\eea
For $k \neq 0$, $\rho_{k}^2  = ( \frac{\hat N}{N} k(N-k)) ^{-1}    \Lambda_W^4 $ and
$ \Lambda_W^4 = \frac{1}{(4 \pi)^2 e} (\Lambda e^{-\frac{1}{2 \beta_0 g_W^2}})^4$, since
$\log \frac{(\Lambda e^{-\frac{1}{2 \beta_0 g_W^2}})^4  }{\frac{\hat N}{N} k (N-k)   (4 \pi)^2  \rho_k^2 } =1$. \par
The value $\Gamma_k(\rho_k)$
of the renormalized effective action is large and negative and equal, for every $k$ but $k=0$, to:
\bea \label{con1}
\Gamma_k(\rho_k)=\Gamma_1(\rho_1) 
=  -  \frac{\beta_0}{4} (4 \pi)^2 \hat N N \hat L^4  \,  \Lambda_W^4
\eea
The trivial solution with magnetic charge $k=0$ is excluded, since it has a greater action, actually zero. \par
Therefore, the $TFT$ admits an infinite number of degenerate vacua labelled by $k=1, 2, \cdots$ for which the magnetic charge condenses. 
Since for large $\mathcal N = N \hat N $ the subtraction scale grows with $\mathcal N$, the density $\rho_k^2=( \frac{\hat N}{N} k (N-k) )^{-1}    \Lambda_W^4$ vanishes in the large-$\mathcal N$ limit.
This has remarkable consequences discussed later. 

\section{Mass gap and kinetic term} \label{s10}

The mass matrix in the holomorphic/anti-holomorphic sector Eq.(\ref{hol}) is the second derivative of the logarithm of the modulus of Vandermonde determinant: 
\bea
\label{glueballM}
M^2_{ij}&&=\frac{\partial^2}{\partial\mu_i\partial\bar{\mu}_j}\log|\Delta(\mu)|^2 
=2 \pi\sum_{\beta\neq i}\delta^{(2)}(\mu_i-\mu_\beta)\delta_{ij}-\delta^{(2)}(\mu_i-\mu_j)(1_{ij}-\delta_{ij}) \nonumber \\
\eea
where $1_{ij}=1$ for all $i,j$, and $\delta^{(2)}(\mu_i-\mu_j)$ is a distribution in color space supported on degenerate eigenvalues only. 
In fact, the glueball potential in the effective action generates attraction between the eigenvalues of $\mu$ since it arises from the inverse of the Vandermonde determinant of the eigenvalues in the partition function restricted to surface operators. 
We show below Eq.(\ref{A}) that
the other contributions to the glueball potential in Eq.(\ref{hol}) are irrelevant for the glueball masses, in the only channel for which a non-trivial glueball kinetic term arises. \par
Besides, the term $\log |Det(\omega')|$ in the effective action contributes to the mass matrix at most a $u(1)$ term  that decouples in the large-$\mathcal N$ limit, since it depends on the eigenvalues only through the holonomy (see section \ref{s7}), that is central at the critical points. \par
We now specialize to $Z_N$ surface operators (see Eq.(\ref{82})) with multiplicity $\hat N$ due to the tensor product with $1_{\hat N}$ in the untwisted sector of the Morita equivalent theory. It is not restrictive to require $k=1,\cdots,\frac{N}{2}$ for $N$ even, since our formulae are symmetric
for the exchange $k \rightarrow N-k$. 
For the the first block $i,j=1,\ldots,k$, Eq.(\ref{glueballM}) implies
$M^2_{ij}=2 \pi\delta^{(2)}(0)((\hat N k-1)\delta_{ij}-(1_{ij}-\delta_{ij}))= 2 \pi\delta^{(2)}(0)(\hat N k\delta_{ij}-1_{ij})$,
with $\delta^{(2)}(0)= \frac{  N \hat N} {(2 \pi)^2 } $ the delta function at zero in color space, regularized according to the color analog of Eq.(\ref{comm}).
Since the diagonal terms scale as $\hat N $, the non-diagonal corrections to the mass term are negligible in the large-$\hat N$ limit,
and in this limit $M^2_{ij}$ becomes diagonal without zero eigenvalues (for $\hat N=1$ the theory has a massless eigenvalue in addition to the trivial diagonal
$u(1)$ that decouples):
$M^2_{ij}=    \frac{  N \hat N} {2 \pi }  \hat N k\delta_{ij}$. \par
Similarly, for the second block $i,j=k+1,\ldots,\frac{N}{2}$, we get
$M^2_{ij}=2 \pi\delta^{(2)}(0)(\hat N (N-k)\delta_{ij}-1_{ij})$,
that implies that the glueball masses are at the cutoff scale for large $N$ in this block. \par
Thus the glueball mass term in the effective action of the commutative theory is:
\bea \label{mass}
 \frac{  k  N \hat N^2} {2 \pi } \rho_k^2   \int d^2z d^2u \,  tr_k Tr_{\hat N}(\delta \mu \delta \bar \mu) = \frac{ N'_2\Lambda_W^2  N^2  \hat N^2 (2 \pi)^2}{\hat N  (N-k)2 \pi}  \int d^2u  \,  tr_k Tr_{\hat N}(\delta \mu \delta \bar \mu) \nonumber \\
\eea
where we have set by definition $N'_2= \frac{1}{(2\pi)^2} \int d^2z \Lambda_W^2=  \frac{1}{(2\pi)^2} \hat L^2 \Lambda_W^2$, with $\hat L^2$ the area of the commutative disk with coordinates $(z, \bar z)$ on which the surface operators are supported on, that is chosen to coincide with the area of the rescaled torus in the Morita equivalent theory by the diagonal immersion. \par
Thus the mass gap in the $TFT$ originates from a holomorphic anomaly in the glueball potential. Besides, the Hessian of the glueball potential is a positive matrix in Euclidean space, corresponding to attraction between the eigenvalues in the potential. Hence this is the stable manifold of the critical points. It arises from critical points with $\hat N k$ degenerate eigenvalues, i.e. from configurations that are $U(\hat N k)$ invariant, so that the group $U(\infty)$ is unbroken in the large-$\hat N$ limit at the critical points, as it should be. \par
The classical action cannot furnish the kinetic term for glueballs, since it is ultralocal once expressed in terms of the eigenvalues of the $ASD$ curvature. Therefore, the kinetic term must be generated by radiative corrections around surface operators that arise from the Jacobian of the change to the $ASD$ variables. This is the case for fluctuations of Lagrangian-embedded surface operators analytically continued to ultra-hyperbolic space-time, because the effective action is naturally defined on a real Lagrangian surface. \par
It arises as the analytically continued Lagrangian surface $(u \rightarrow - i u_+, \bar u \rightarrow  - i u_-)$ that originates from the analytically continued twistor Wilson loops:
\bea \label{A}
Tr_{\cal N} \Psi(\hat B_{\rho};L_{ww}) \rightarrow Tr_{ \cal{N}} P \exp i \int_{L_{ww}}(\hat A_{z+}- i \rho \hat D_{u_+}) dz_++(\hat A_{ z_-} - i \rho^{-1} \hat D_{ u_{-}}) d z_-  \nonumber \\
\eea
By Eq.(\ref{supp}) the support is now
$( z,  \bar z, u,  {\bar u}) \rightarrow (z_+, z_-, \rho  z_+, \rho^{-1} z_-)$.
Again, for computational simplicity we set $\rho=1$ in the following. Because of the analytic continuation, the twistor connection and its curvature $\mu$ become Hermitian $\mu \rightarrow \mu_{+-}$. 
Hence analytically continued twistor Wilson loops are supported on a physical wedge, as opposed to the topological ones. \par
The second derivative of the glueball potential in the holomorphic or anti-holomorphic sector separately are proportional to the holomorphic or anti-holomorphic first derivative of the analog in color of the distribution in Eq.(\ref{prin}). The derivative of the principal part is subleading in $\frac{1}{\hat N}$ at the critical points since it gets contributions only by different eigenvalues because of the definition of the principal part, the derivative of the delta function vanishes at zero, and in any case it is cancelled by the complex conjugate, because of the analytic continuation that involves Hermitian fluctuations. \par
Hence $\mu_{+-}$ couples only to the mass term that we actually computed, since the further contributions to the potential in the effective action couple to $\nu$ and $\mu-\bar \mu$ (see Eq.(\ref{hol})), that are zero at the critical points and thus they may contribute only to the $u(1)$ part of the fluctuations. Besides, only one polarization of $\mu_{\alpha \beta}$ propagates in the $TFT$ analytically continued to the physical wedge. From now on we write $\mu$ for $\mu_{+-}$ when no confusion can arise. \par
We compute the kinetic term on the analytically continued Lagrangian surface, starting for convenience with its Euclidean version, that is more natural for the effective action. We choose the surface $(z=u,\bar z= \bar u)$ diagonally immersed in $R^4$. Since we have defined a lattice in the $(z, \bar z)$ plane,
this defines a lattice also in the $(u, \bar u)$ plane by the diagonal map $(z_p=u_p,\bar z_p= \bar u_p)$. This lattice in the $(u, \bar u)$ plane has a set of dual plaquettes such
that the $(u, \bar u)$ plane is the union of the plaquettes. We define the function $z_p(u, \bar u)=u$ with domain the interior of the plaquette dual to $p$, and analogously for the complex conjugate. \par
We also define a lattice fluctuating field $\delta \mu_p(u,\bar u)$, supported on the plaquette dual to $p$ and locally constant as $(u, \bar u)$ vary in the support, that is zero outside its support.
Thus we suppose that in addition to the translational invariant background of surface operators there are locally-defined fluctuating surface operators. Both the background and the fluctuations are diagonal matrices in color space by the local Abelianization (see section \ref{s2} and \ref{s8})
$-i F_{B} = \sum_p  \tilde \mu  \delta^{(2)} (z-z_p)+ \sum_p \delta \mu_p(u,\bar u) \delta^{(2)} (z-z_p(u, \bar u))$. \par
We examine the loop expansion of the effective action. This is in fact a one-loop expansion, because only the logarithm of functional determinants occurs in the effective action, with multiple insertions
of trees, due to the inversion of the change to the $ASD$ variables. This complicated expansion carries a multiple insertion of the background field and of the fluctuating field supported on the lattice of surface operators. The divergent parts, that contain the background field with $Z_N$ holonomy, determine the beta function, and have been already investigated. \par
We are now interested in the finite parts to second order, that contain the quadratic part of the fluctuating field. 
We may limit ourselves to the second order, within the leading large-$\hat N$ accuracy. Indeed, every term of the loop expansion contains a trace in the adjoint representation, and thus it is proportional to $\hat N^2$, that of course diverges for large $\hat N$. \par
However, the loop expansion is in fact an expansion in powers of the density $\rho_k$ of surface operators.
But since the density scales as $ \hat N^{-\frac{1}{2}}$, only the leading quadratic term survives the large-$\hat N$ limit. 
Thus at the leading large-$\hat N$ order the effective action for the fluctuations of surface operators in the commutative theory on the
diagonal Lagrangian submanifold, with $x=(z, \bar z, u, \bar u)$ and $y=(w, \bar w, v, \bar v)$, is:
\bea \label{2}
&&-\frac {5 \hat N N 4}{3 (4 \pi^2)^2 } \sum_{ p \neq p' }  \int d^2u d^2v  \frac{ tr_N Tr_{\hat N} (\delta \mu_p(u, \bar u) \delta  \bar \mu_{p'}(v, \bar v) ) } {(|z_p(u, \bar u)-z_{p'}(v, \bar v)|^2+|u-v|^2)^2} \nonumber \\
&&=-\frac {20 \hat N  N}{3 (4 \pi^2)^2 } \int d^2z d^2w  \, \rho_k^2   \int d^2u d^2v\frac{ tr_N Tr_{\hat N} (\delta \mu(u, \bar u) \delta \bar  \mu(v, \bar v) ) } {(|u-v|^2+|u-v|^2)^2} \nonumber \\
&&\rightarrow -\frac {20  {N'}^2_2 (2 \pi)^4 N^2 }{ 3 (4 \pi^2)^2  k(N-k)}   \int du_+ du_- dv_+ dv_- \frac{ tr_N Tr_{\hat N} (\delta \mu(u_+,  u_-) \delta  \mu(v_+,  v_-) ) } {2 (u_+-v_+ +i \epsilon)^2(u_- - v_-+i \epsilon)^2}   
\nonumber \\
&&= \frac {10  {N'}^2_2 (2 \pi)^2 N^2 }{ 3 k(N-k)}    \int du_+ du_-  tr_N Tr_{\hat N} (\delta \mu(u_+,  u_-) \, \partial_+\partial_- \delta  \mu(u_+,  u_-) )  
\eea
Some comments are in order. \par
The first line is the counterterm quadratic in the $ASD$ field that arises in Eq.(\ref{fluc}).\par
The second line expresses the first one in terms of the density of surface operators $\rho_k$,
that reduces to the third line 
by analytic continuation to Minkowski signature of the Lagrangian surface \footnote{It is necessary to assume that $\delta \mu(u_+,  u_-)$ be the boundary value 
of a holomorphic function on the upper-half plane for each of the independent variables $(u_+,  u_-)$, with suitable properties at infinity. }. \par
The result depends crucially on a number of ingredients of the $TFT$: The fluctuations occur as surface operators, the support of fluctuations is immersed as a diagonal Lagrangian submanifold in Euclidean space-time, and the analytic continuation to the Minkowskian Lagrangian submanifold is performed. \par
Summing the kinetic and the mass term in Eq.(\ref{mass}) and dividing by the factor of $\hat N^2$ that occurs in the Morita-equivalent theory according to Eq.(\ref{comm}) \footnote{This is necessary because of the extended nature of surfaces operators in order to match the natural $O(1)$ normalization of the $ASD$ correlator (see section \ref{s11}), but being just a rescaling of the effective action it does not affect the glueball spectrum.}, we get the effective action for the fluctuations in the Morita-equivalent theory in each $k$-sector:
\bea \label{Wil}
\Gamma_k(\delta \mu, \delta  \mu) = 2 \pi  \frac{ N^2 N_2'^2 }{\hat N^2\delta (N-k)} \big(  \int du_+ du_-  \frac {\alpha'}{ k}  tr_kTr_{\hat N} (\delta \mu \, \partial_+\partial_- \delta  \mu )  +   \int du_+ du_-  \,  \Lambda_W^2  tr_k Tr_{\hat N}(\delta \mu \delta  \mu)  \big) \nonumber \\
\eea
with $\alpha'= 4\pi (4 \pi)^2 \gamma_0 \delta   =\frac { 20 \pi }{ 3 } \delta  $ and $\delta =\frac{N'_2}{\hat N}$. The precise value of $\delta$ will be determined in the next section.
We construct here a family of schemes in which $N'_2= \delta \hat N$. \par
By definition,
$N'_2= (\frac{\Lambda_W}{2 \pi})^2 \hat L^2$, where $\hat L^2$ is the area of the commutative disk and of the commutative torus as well (see section \ref{s4}). Now it must be $ N'_2 \sim  \hat N$ and $\hat N \sim N$, for the effective action to describe a theory weakly coupled for large $\hat N$, with propagating degrees of freedom, and mass gap $m^2_{gap}= \alpha'^{-1}  \Lambda^2_W$. 
Indeed, this is the same as writing
$\Lambda_{W}^2= \frac{\delta \hat N (2 \pi)^2}{\hat L^2}$,
i.e. requiring that the large-$\hat N$ limit and the large-$\hat L$ limit be taken in this prescribed $RG$-invariant way. \par
This relation is implied by the following family of schemes that defines the $RG$ flow of the vacua of the $TFT$:$  \frac{\delta \hat M}{\hat N}=\frac{ \Lambda_W^2}{ \Lambda^2} = (e^{-\frac{1}{2 \beta_0 g_W^2}})^2$,
i.e. $\frac{1}{ \beta_0 g_W^2} =  \log \frac{ \hat N }{\delta \hat M}$,
that is compatible both with $AF$ and with the large-$\theta$ limit, as it follows that $2 \pi \theta=  \Lambda_W^{-2} \delta \hat M$. \par
It implies a fine tuning of the continuum limit with $\hat N$ and of the large-$\theta$ limit with $\hat M$, since a number of space-time  degrees of freedom of the non-commutative gauge theory by which the $TFT$ is originally defined (see section \ref{s4}) are converted into color degrees of freedom by a certain choice of the map that defines Morita equivalence, in order to provide the vacua of the $TFT$ on a commutative torus. Then the size of the commutative torus is rescaled by large factors, in such a way that the fluctuations are computed in the commutative theory
in the thermodynamic limit, in order for $ \Lambda_W^2$ and $\rho$ to be largely separated from the $UV$ cutoff $\Lambda^2=\frac{\hat N}{\delta \hat M}\Lambda^2_{W}$, from the infrared cutoff $\hat L^{-2}$, and at the same time for $\theta$ to diverge. \par
Indeed, from Eguchi-Kawai regularization (see Eq.(\ref{cutoff})) 
$\hat N (\frac{2 \pi}{\Lambda})^2=2 \pi \theta$ and from Morita equivalence
$2 \pi \theta = \hat L^2 \frac{\hat M}{\hat N}$.
Therefore, 
$\hat N=(\frac{\Lambda}{2 \pi})^2 \hat L^2 \frac{\hat M}{\hat N}$ and $ (\frac{\Lambda}{2 \pi})^2 = \frac{\delta \hat N}{\hat L^2} \frac{\hat N}{\delta \hat M} =(\frac{\Lambda_{W}}{  2 \pi })^2 \frac{\hat N}{\delta \hat M}$,
that implies $\Lambda_{W}^2=\frac{\delta \hat N (2 \pi)^2}{\hat L^2}$. Besides, $\hat M$ cannot grow too fast with $\hat N$, but it cannot grow too slowly, since $\theta$ must diverge in order for the non-commutative gauge theory be equivalent to the large-$N$ limit on commutative space-time. However, we are plenty of choices for $\hat M$, for example $\hat M=O(\sqrt {\hat N})$\footnote{This is possibly related to the lattice formulation of  twisted $EK$ reduction for which it has been suggested \cite{GA1} that the twist $r$, that is related to $\hat M$ (see section \ref{s4}), has to grow with $\hat N$ in order to define a non-trivial $AF$ continuum limit \cite{GA2} for such a lattice model. In our case the main, but crucial, difference is that  fluctuations are actually computed in commutative space-time in the thermodynamic limit, but the $AF$ $RG$ flow at the saddle-points, that defines the vacua of the $TFT$, is related to the non-commutative space-time, because the technical definition of twistor Wilson loops involves necessarily such non-commutativity.}. Finally, the Wilsonian effective action in this scheme reads:
\bea
\Gamma_k(\delta \mu, \delta  \mu) = 2\pi  \frac{ \delta N^2}{N-k} \int du_+ du_-  \big(\frac {\alpha' }{  k}  tr_k Tr_{\hat N} (\delta \mu \, \partial_+\partial_- \delta  \mu )
 +  \Lambda_W^2  tr_k Tr_{\hat N}(\delta \mu \delta  \mu) \big) \nonumber \\
\eea
Hence $YM$ theory has a mass gap, and in fact an infinite tower of glueball states, for fluctuations around the topological sector, provided the $AF$ continuum limit is taken along any sequence of bare 't Hooft Wilsonian couplings $\frac{1}{ \beta_0 g_W^2} =  \log \frac{ \hat N }{\delta \hat M}$, in any scheme satisfying $\hat N, \hat M \rightarrow \infty$ with $\frac{\hat M}{\hat N} \rightarrow 0$. After fixing $\delta$  in next section, the mass gap $m^2_{gap}= \alpha'^{-1} \Lambda^2_W$ is scheme independent in units of the scheme dependent $RG$-invariant scale $\Lambda_W$ in this family of schemes.

\section{$ASD$ glueball propagator in the $TFT$} \label{s11}

Now we compute in the $TFT$ the two-point correlator of the surface operator that corresponds to the local operator $tr_{\mathcal{N}} (\mu^{-2}_{\alpha \beta})(x)$
in $YM$. In fact, there are two possible definitions in the $TFT$, that correspond to the correlators in Eq.({\ref{corr}) and Eq.({\ref{corrM}). \par
We start with the Euclidean version
$ 4tr_N Tr_{\hat N}(\mu(x) \bar \mu(x) + \nu(x)^2)$.
The corresponding surface operator is
$ 4 \sum_p \delta^{(2)}(0) \delta^{(2)}(z-z_p) tr_N Tr_{\hat N}(\mu_p \bar \mu_p+\nu_p^2)$,
that is a divergent composite operator supported on lattice points. \par
This suggests two possibilities. \par
In the first version we substitute the overall divergent factor by the corresponding power of the $RG$-invariant scale, and we compute simply the correlator of $ 4 \Lambda_W^4 tr_N Tr_{\hat N}(\mu_p \bar \mu_p+\nu_p^2)$
at different points. \par
In the second version we introduce the density $\rho$, to define a smoothing of the surface operator in the same fashion as for the action density 
$4 \rho^2 tr_N Tr_{\hat N}((\mu \bar \mu)+\nu^2)$. \par
It is very instructive to compute the two-point correlator in both cases, since the $OPE$ of the local $ASD$ operator in $YM$ theory (see section \ref{s3}) is matched differently by the two versions in the $TFT$.\par
Indeed, for the second definition the condensate is well defined in the $TFT$, contrary to perturbation theory, but the universal part of the first coefficient of the $OPE$ vanishes. \par
For the first definition the condensate is not necessarily well defined, i.e. it may be divergent, but this divergence does not affect the correlator, because the universal part of the coefficient of the condensate in the $OPE$ vanishes in the $TFT$. \par
Remarkably, these statement have exact analogs in perturbative $QCD$ as recalled in section \ref{s1} and as described in detail in section \ref{s3}, because of relations and cancellations  between the first-two coefficients functions
in the $OPE$ of the scalar and the pseudoscalar operator. \par
In both cases, to make contact with $YM$ perturbation theory we pass to a canonical scheme in which our surface operator includes
the renormalization factor, proportional to $g^2$ (see section \ref{s9}), due to its anomalous dimension. Besides, in the $TFT$ we expand around the saddle points $\mu=\tilde \mu_k$, $\tilde \nu_k=0$ in each $k$ sector.\par
In the first case we evaluate at the leading $\frac{1}{\mathcal N}$ order in ultra-hyperbolic signature:
\bea \label{can}
&&  g^4 \Lambda_W^8  \int d^4x e^{i(p_+x_-+ p_-x_+)} 
 <tr_N Tr_{ \hat N}(\mu \bar \mu+\nu^2)(x_+, x_-) tr_N Tr_{\hat N}(\mu \bar \mu+\nu^2)(0,0)>_{conn}\nonumber \\ 
&& \rightarrow  4 g_k^4  \Lambda_W^8  \int d^4x    e^{i(p_+x_-+ p_-x_+)} 
 <tr_N Tr_{\hat N}(\delta\mu_k \tilde \mu_k)(x_+, x_-) tr_N Tr_{\hat N}(\tilde \mu_k \delta  \mu_k)(0,0)>_{conn}\nonumber \\
\eea
since fluctuations of $\nu$ do not contribute, being coupled to a vanishing condensate. This correlator couples to scalars and pseudoscalars. The correlator of $tr_N Tr_{\hat N}(\mu(x)^2-\bar \mu(x)^2)$, that would couple purely
to spin-$2$ glueballs, vanishes on the physical wedge because of the Hermitian condition on $\mu$, thus confirming the aforementioned interpretation of the spectrum. \par
In the sector with degeneracy $N-k$, the mass squared is $O(N)$ and 
the corresponding propagator decouples in the large-$N$ limit.
Hence the interesting sector has degeneracy $k=1,2 \cdots N/2$, described by the Wilsonian effective action Eq.(\ref{Wil}).
Summing and averaging over each $k$ sector, we get at the leading large $\frac{1}{N \hat N}$ order:
\bea
&& \frac{2}{N} \sum_{k=1}^{\infty} 4 g_k^4   \Lambda_W^8 (2 \pi)^2 \int d^4x   e^{i(p_+x_-+ p_-x_+)}   <tr_k Tr_{ \hat N}(\delta \mu (\frac{k-N}{N}))(x_+, x_-) tr_k Tr_{\hat N}((\frac{k-N}{N})\delta  \mu)(0,0)>_{conn}  \nonumber \\
&&= \frac{2}{N} \frac{4 (2 \pi)^2  L^2 \Lambda^2_W}{ 4 \pi}     \sum_{k=1}^{\infty} \frac{ k  \hat N  (N-k)^3}{ N^4 \delta}  \frac{k g_k^4  \Lambda^6_W}{ - \alpha'  p_+ p_-+ k \Lambda_{W}^2} \nonumber \\
&&= 4 (2 \pi)^3 (\frac{  \hat N }{N})^2   \sum_{k=1}^{\infty}  \frac{g_k^4 k^2 \Lambda_{W}^6 }{ - \alpha'  p_+ p_-+ k \Lambda_{W}^2} +O(\frac{1}{N}) \eea
where the overall extra factor of $\frac{2}{N}$ arises by the average over critical points.
Moreover, using the identity:
\bea
&&  \sum_{k=1}^{\infty} \frac{ \Lambda_W^4 k^2  g_k^4  \Lambda_W^2}{ - \alpha'  p_+ p_-+ k \Lambda_{W}^2}  
=  \sum_{k=1}^{\infty} \frac{( k  \Lambda_W^2+ \alpha'  p_+ p_-)(k  \Lambda_{W}^2- \alpha'  p_+ p_-)+(- \alpha'  p_+ p_-)^2  ) g_k^4  \Lambda_W^2}{ - \alpha'  p_+ p_-+ k \Lambda_{W}^2 }   \nonumber \\
&&=  (- \alpha' p_+ p_-)^2  \sum_{k=1}^{\infty} \frac{g_k^4  \Lambda_W^2}{ - \alpha'  p_+ p_-+ k \Lambda_W^2} + \sum_{k=1}^{\infty}  g_k^4  (k  \Lambda_W^4+ \alpha'  p_+ p_- \Lambda_W^2)
\eea
and subtracting the physically irrelevant contact terms \footnote{Remarkably, an infinite constant contact term was discovered in the coefficient function of the $OPE$ at two loop in \cite{Che3}.}, we get the physical part of the correlator:
\bea
&& \int d^4x  e^{i(p_+x_-+ p_-x_+)}  < g^2  \Lambda_W^2 tr_N Tr_{ \hat N}(\mu \bar \mu+\nu^2)(x_+, x_-)  g^2 \Lambda_W^2  tr_N Tr_{\hat N}(\mu \bar \mu+\nu^2)(0,0)>_{phys} \nonumber \\
&&=4(2 \pi)^3  (\frac{\hat N}{N})^2  (- \alpha' p_+ p_-)^2  \sum_{k=1}^{\infty} \frac{g_k^4  \Lambda_W^2}{ - \alpha'  p_+ p_-+ k \Lambda_W^2} 
\eea
The overall normalization of the correlator is ambiguous because the first definition arises by the regularization of an infinite quantity by a multiplicative factor. We can redefine the surface operator in order to match the
normalization of the $RG$-improved perturbative result.
Thus with the first definition of the surface operator the two-point correlator matches asymptotically the $RG$-improved $ASD$ correlator in $QCD$ in perturbation theory up to an overall constant (see section \ref{s3}). \par
We show momentarily that the second definition $4 \rho^2    tr_N Tr_{\hat N}(\mu^2)$ matches asymptotically the correlator in Minkowski space-time. 
We evaluate at the leading $\frac{1}{\mathcal N}$ order:
\bea \label{can0}
&&16  g^4 \rho^4    \int d^4x e^{i(p_+x_-+ p_-x_+)} 
 <tr_N Tr_{ \hat N}(\mu^2)(x_+, x_-) tr_N Tr_{\hat N}(\mu^2)(0,0)>_{conn}\nonumber \\ 
&& \rightarrow  64 g_k^4   \rho_k^4    \int d^4x    e^{i(p_+x_-+ p_-x_+)} 
 <tr_N Tr_{\hat N}(\delta\mu_k \tilde \mu_k)(x_+, x_-) tr_N Tr_{\hat N}(\tilde \mu_k \delta  \mu_k)(0,0)>_{conn}\nonumber \\
\eea
Summing and averaging over each $k$ sector, we get at the leading large $\frac{1}{N \hat N}$ order:
\bea \label{can1}
&& 16 \frac{2}{N} \sum_{k=1}^{\infty} 4 g_k^4   \rho_k^4  (2 \pi)^2 \int d^4x   e^{i(p_+x_-+ p_-x_+)}   <tr_k Tr_{ \hat N}(\delta \mu (\frac{k-N}{N}))(x_+, x_-) tr_k Tr_{\hat N}((\frac{k-N}{N})\delta  \mu)(0,0)>_{conn}  \nonumber \\
&&= 16 \frac{2}{N} \frac{4 (2 \pi)^2   L^2 \Lambda^2_W}{4 \pi}     \sum_{k=1}^{\infty}  \frac{ k  \hat N  (N-k)^3}{ N^4 \delta} (\frac{N}{\hat N k (N-k)})^2  \frac{k g_k^4 (k-N)^2 \Lambda^6_W}{ - \alpha'  p_+ p_-+ k \Lambda_{W}^2} \nonumber \\
&&=64 (2 \pi)^3  \frac{1}{ N^2}  \sum_{k=1}^{\infty}  \frac{g_k^4  \Lambda_{W}^6}{ - \alpha'  p_+ p_-+ k \Lambda_{W}^2} + O(\frac{1}{N^3}) \nonumber \\
&&=64 \frac{ 8 \pi^3  }{ 8 \pi^2 { N}^2}  8 \pi^2 \Lambda_{W}^4 \sum_{k=1}^{\infty}  \frac{g_k^4  \Lambda_{\overline W}^2}{ -  p_+ p_-+ k \Lambda_{\overline W}^2} \nonumber \\
&&=\frac{64 \pi \hat N}{N}  < \frac{1}{N \hat N} \mathcal{O}_{ASD}> \sum_{k=1}^{\infty}  \frac{g_k^4  \Lambda_{\overline W}^2}{ -  p_+ p_-+ k \Lambda_{\overline W}^2}
\eea
We may fix the ratio $\frac{16 \pi \hat N  }{N}=  \beta_0 $ by matching the overall normalization of the perturbative result in Minkowski space-time. \par
The only ambiguity that remains in the $TFT$ is the choice of the constant $\delta$,
that occurs in the relation between $\Lambda_W$ and the mass gap: $ \Lambda^2_W=\alpha' \Lambda^2_{\overline W}$, i.e. that defines the glueball condensate in terms of the mass gap (see below). We proceed as follows. \par
Firstly, we completely fix the canonical scheme in the beta function, requiring that the canonical coupling be defined at the scale $\Lambda_W$ at which the Wilsonian coupling diverges. Since the canonical flow is defined in terms of the Wilsonian coupling,
it can extend at most until it reaches $\Lambda_W$. \par
Hence we require that the canonical scheme extends maximally. Now the canonical flow ends either at a zero in the numerator of Eq.(\ref{alfa}) or at the fixed point in the denominator. But a zero in the numerator would correspond to a conformal fixed point. Thus the canonical flow ends at the fixed point $g^2=\frac{4}{(4 \pi)^2}$, and $c'$ if fixed by the condition that the Wilsonian and canonical $RG$-invariant scales coincide for $g^2_W=\infty$ and $g^2=\frac{4}{(4 \pi)^2}$ (see Eq.(\ref{RGW})). \par
Secondly, we fix $\delta$ \emph{ imposing} the equality that expresses the $NSVZ$ low-energy theorem (see section 2.5 in \cite{MBN} and references therein): The $RHS$ evaluated in Wilsonian scheme must be equal to the $LHS$ evaluated in a canonical scheme specified by $\delta$, with $g_k \equiv g(k \Lambda^2_{\overline W})=g(k \alpha'^{-1} \Lambda^{2}_ W)$ in Eq.(\ref{can1}). This means that in the residues of the physical $ASD$ correlator the running coupling must be evaluated on shell in the scheme in which the $NSVZ$ theorem holds:
\bea \label{10}
\int < \mathcal N\Tr{F^-}^2(x)\mathcal N \Tr{F^-}^2(0)>_{phys} \,d^4x 
=\frac{8}{\beta_0} <\mathcal N\Tr{F^-}^2(0)>
\eea
where $<\mathcal N \Tr{F^-}^2(x)> \rightarrow 4 < \mathcal N \rho^2 tr_N Tr_{\hat N}(\mu \bar \mu+\nu^2)>=   \hat N N 8 \pi^2 \Lambda_{W}^4$ in the $TFT$ in the Wilsonian scheme because of Eq.(\ref{cond}) and Eq.(\ref{con}),
but we will not perform the explicit computation in this paper. \par
This procedure is consistent because the Minkowskian $ASD$ correlator in the $TFT$ occurs deprived of contact terms, that therefore cannot mix with the contact term that occurs in the $RHS$ of $NSVZ$ low-energy theorem (at zero momentum). Thus the $TFT$ solves the difficulty pointed out in \cite{Che3} to give a meaning to the $NSVZ$ low-energy theorem in $QCD$, because of the infinite contact terms arising instead in Eq.(\ref{contact}). To mention their words (see p. 12 ibidem ): "The two-loop part is new and has a feature that did not occur in lower orders,
namely, a divergent contact term. Its appearance clearly demonstrates that non-logarithmic
perturbative contributions to $C_1$ are not well defined in $QCD$, a fact seemingly ignored by the
the $QCD$ sum rules practitioners."
 
\section{Beta function in Veneziano large-$N$ limit of $QCD$, conformal window and quark-mass anomalous dimension} \label{s12}

The exact $n=1$ $SUSY$ $NSVZ$ beta function, that has an extension that includes $N_f$ $SUSY$ matter fields, say in the fundamental representation \cite{NSVZ}:
\begin{align}\label{beta}
&\frac{\partial g}{\partial \log M}
=\frac{- \frac{1}{(4\pi)^2}(3-\frac{N_f}{N})
g^3-\frac{\gamma_m(g)}{(4\pi)^2}g^3\frac{N_f}{N}}{1-\frac{2}{(4\pi)^2}g^2} \nonumber \\
&\frac{\partial\log Z_m}{\partial\log M} =-\gamma_m(g) 
\end{align}
has been be employed to determine the mass anomalous dimensions $\gamma_m=-1$ of the $SUSY$ matter fields at the lower edge of the conformal window \cite{Sei2}, by means of independent information furnished by the chiral ring on the critical value of $\frac{N_f}{N}=\frac{3}{2}$ at which the conformal transition occurs. \par
We proceed analogously in our non-$SUSY$ case firstly
extending in the $TFT$ the exact large-$N$ beta function from pure $YM$ to $QCD$ with renormalized quark mass $m$ in the massless large-$\theta$ (i.e. large-$N$) limit  $m \sqrt \theta=const , \theta \rightarrow \infty$
by employing the quark-mass anomalous dimension $\gamma_m$ computed in $QCD$ perturbation theory \cite{Chean}, finding a spectacular agreement in Veneziano limit $ N \rightarrow \infty, \frac{N_f}{N}=const $:
 \begin{align} \label{betaV}
&\frac{\partial g}{\partial \log M}
=\frac{-\beta_0' g^3+\frac{1}{(4\pi)^2}g^3(\frac{\partial \log Z'}{\partial \log M}-\gamma_m(g)\frac{N_f}{N}) }{1-\frac{4}{(4\pi)^2}g^2} \nonumber \\
&=\frac{(-\beta_0+\frac{2}{3(4\pi)^2}\frac{N_f}{N}) g^3+\frac{1}{(4\pi)^2}g^5(2\gamma_0-\frac{4}{3(4\pi)^2}\frac{N_f}{N} +\frac{9}{3(4\pi)^2}\frac{N_f}{N}(1-\frac{1}{N^2})+ \cdots)}{1-\frac{4}{(4\pi)^2}g^2} \nonumber \\
&=- (\beta_0-\frac{2}{3(4\pi)^2}\frac{N_f}{N}) g^3-(\beta_1- \frac{1}{(4\pi)^4} (\frac{13}{3}- \frac{3}{N^2}) \frac{N_f}{N}) g^5+\cdots \nonumber \\
&\frac{\partial g_W}{\partial \log M}=-\beta_0' g_W^3 \\ 
&\frac{\partial\log Z'}{\partial\log M} =\frac{2\gamma_0' g_W^2}{1+c' g_W^2}=2\gamma_0' g^2 +\cdots  \nonumber \\
&\gamma_0'=\gamma_0-\frac{2}{3(4\pi)^2}\frac{N_f}{N} \nonumber \\
&\frac{\partial\log Z_m}{\partial\log M} =-\gamma_m(g)= \frac{9}{3(4\pi)^2}\frac{N^2-1}{N^2} g^2+\cdots   \nonumber \\
& \gamma_0=\frac{5}{3(4\pi)^2} \nonumber \\
& \beta_0'=\beta_0- \frac{2}{3(4\pi)^2}\frac{N_f}{N}\nonumber \\
&\beta_0=\frac{11}{3(4\pi)^2} \nonumber \\
&\beta_1=\frac{34}{3(4\pi)^4}
\end{align} 
with the universal part of the $QCD$ beta function in perturbation theory: 
\bea
\frac{\partial g}{\partial \log M}=- (\beta_0-\frac{2}{3(4\pi)^2}\frac{N_f}{N}) g^3-  (\beta_1- \frac{1}{(4\pi)^4}(\frac{13}{3}-  \frac{1}{N^2}) \frac{N_f}{N})  g^5+\cdots \nonumber \\
\eea
up to terms on the order of $\frac{1}{N^2}$ in the second coefficient. \par
Eq.(\ref{betaV}) follows observing that in the aforementioned large-$N$ massless limit the contribution of the fermion determinant $Det(i \slashed D-m_0)$, with $m_0=Z_m m$ the bare quark mass \cite{Chean}, to the partition function in the $TFT$ evaluated around the critical points, is:
\bea
Det'(i \slashed D) ( Z_m m \sqrt \theta)^{n_f}
\eea
with $Det'(i \slashed D)$ the determinant of the non-zero modes of the Dirac operator $i \slashed D-m_0$ for $m_0=0$, the other factor being the contribution of $Det(i \slashed D-m_0)$ restricted to the $n_f$ zero modes of the Dirac operator $i \slashed D$, with $n_f=\frac{N_f}{2N} n_b$, where $n_b$ is the the number of gluon zero modes in the background of surface operators. \par
This can be compared with the $n=1$ $SUSY$ case, in which the contribution of the renormalization factor $Z_m$ \cite{Sh} in the \emph{kinetic} term of $N_f$ \emph{massless} $SUSY$ matter fields in the fundamental representation to the partition function evaluated around the instanton background leads to the $NSVZ$ beta function Eq.(\ref{beta}).
In both the $SUSY$ case and $QCD$ we obtain the effective action in the local approximation by means of the inclusion of the fermion determinant evaluated at the saddle point with the $Z_m$ factor of the fermion fields computed in perturbation theory, that certainly does not contain all the non-local contributions, but only the local ones that originate the fermion-mass anomalous dimension. \par
Contrary to the $SUSY$ case, the $TFT$ in $QCD$ cannot be defined directly in the massless case, because of the fine tuning $m \sqrt \theta=const , \theta \rightarrow \infty$ at large-$\theta$, i.e.
at large-$N$, that it is absolutely necessary to avoid that the quark zero modes in the massless limit give a (wrong) contribution to the first coefficient of the beta function.\par
It is easy to check that the effect of $Det'(i \slashed D)$ 
is to shift the coefficient of the kinetic term (obtained from Eq.(\ref{2})) in the glueball effective action Eq.(\ref{Wil}) to the value:
\bea \label{alpha'}
\alpha'= (4 \pi)^3 \delta \gamma'_0
\eea
because $Det'(i \slashed D)$ has the same structure as the determinant of the non-zero modes of the Jacobian to the $ASD$ variables by the standard spinor identities mentioned in section \ref{s5}, and therefore it contributes 
to the kinetic term of the glueball propagator in Eq.(\ref{fluc}) a counterterm with the same structure but with overall coefficient $+\frac{N_f}{N}\frac{2}{3}$ instead of $-\frac{5}{3}$. \par
The lower edge of the conformal window is determined as the value of $\frac{N_f}{N}$ at which fluctuations around the magnetic condensate become unstable in the kinetic term for the glueball propagator:
\bea
\gamma_0'=\gamma_0-\frac{2}{3(4\pi)^2}\frac{N_f}{N}=0
\eea
thus signaling a phase transition to a non-confining phase identified with a conformal Coulomb phase.
If the phase transition is driven by the existence of an infrared Banks-Zaks fixed point \cite{BZ} in the beta function, at the value $\frac{N_f}{N}=\frac{5}{2}$ the exact beta function must develop a zero, that determines the value of the quark-mass anomalous dimension at the critical coupling $g^*$:
 \begin{align}
&\frac{\partial g}{\partial \log M}
= \frac{(-\frac{6}{3(4\pi)^2} -\frac{1}{(4\pi)^2}\frac{N_f}{N} \gamma_m(g^{*}))g^{*3}}{1-\frac{4}{(4\pi)^2}g^{*2}}=0 \nonumber \\
&-\frac{\partial\log Z_m}{\partial\log M} =\gamma_m(g^{*})= -\frac{4}{5}  
\end{align}
This value is well inside the unitarity bound, and a previous estimate for small volumes coincides with it exactly \cite{Unsal}. Remarkably, the zero of the beta function may exist universally, despite the scheme dependence of our beta function in Eq.(\ref{betaV}), because at the unstable point for the magnetic condensate the scheme dependence in the bracket in the numerator of our beta function disappears as well. The coincidence is not accidental, since both the kinetic term in the glueball effective action
and the $Z'$ contribution in the beta function originate by the same counterterm in the effective action of the $TFT$. \par
It follows also that in the local approximation for the effective action in the confining phase the glueball spectrum of the $TFT$ in 't Hooft and in massless Veneziano limit are the same, and the structure of the $ASD$ correlator as well, modulo the appropriate change of the renormalization scheme that defines the mass gap $\Lambda_{\overline W}$ in Eq.(\ref{corr}) and Eq.(\ref{corrM}). \par
Morover, the glueball condensate 
 $< \Tr{F^-}^2(x)> \rightarrow <  4 \rho^2 tr_N Tr_{\hat N}(\mu \bar \mu+\nu^2)>=  \alpha'^2 8 \pi^2 \Lambda_{\overline W}^4$ in units of the mass gap, that is kept constant by construction in the confining phase while $\frac{N_f}{N}$ varies, vanishes identically \footnote{In fact, it vanishes unless the zero of $\gamma_0'$ in Eq.(\ref{alpha'}) is cancelled by $\delta$ as determined by the low-energy theorem (see section \ref{s11}).} in the $TFT$ at the point $\gamma'_0=0$, thus confirming the occurrence of the conformal phase.

\thispagestyle{empty}
\appendix
\section{Phenomenology} \label{App}

As a heuristic aside to make contact with reality, we discuss here the actual experimental meson and tentative glueball spectrum in relation to our large-$N$ results. \par
We expect that the slope of the scalar meson trajectories $2 \Lambda^{-2}_{\overline W}$ at large-$N$ be the double of the slope of the scalar glueball trajectories $\Lambda^{-2}_{\overline W}$, because of the large-$N$ factorization of physical Wilson loops in the adjoint representation into the product of the fundamental and conjugate representation, and of the standard string picture. Besides, the exact linearity of the joint scalar and pseudoscalar glueball spectrum of positive charge conjugation implied by the $TFT$ leads us to conjecture \cite{MBS} that also the meson spectrum be exactly linear in the large-$N$ massless 't Hooft limit of $QCD$, both conjectures being in good agreement 
with the lattice measurements \cite{MBS} (see Fig.\ref{fig:regge1} and Fig.\ref{fig:regge2}). From the theory point of view this conjecture is related \cite{MBS} to the existence of a Topological String Theory for glueballs and mesons dual to the $TFT$ of this paper.  \par
A priori there is no reason by which the aforementioned exactly-linear large-$N$ glueball spectrum should match the spectrum of the real world with $1 \%$ accuracy. Neither there is any a-priori reason for the linearity to extend to the meson trajectories with the same accuracy. 
Nevertheless, heuristically we have compared the result of our large-$N$ theoretical computation $\frac{m_{0^{++*}}}{m_{0^{++}}}=\sqrt 2=1.414 \cdots$, in addition to the large-$N$ lattice results \cite{MBS} (see Fig.\ref{fig:regge1}), with the value implied by the Particle Data Group (2014) $\frac{m_{f_0(2100)}}{m_{f_0(1500)}}=1.397(008)$ on the basis of the identification of the lower-mass glueballs $f_0(1500)$ and $f_0(2100)$ suggested indirectly by the experimental findings of the $BES$ collaboration, according to the criteria stated in their final summary \cite{BES}.
The remarkable agreement of our tentatively-identified low-lying scalar glueball spectrum with the large-$N$ prediction (both from the theory in this paper and from the lattice \cite{T0,T}) is suggestive of smallness in full-fledged $QCD$ of next-to-leading $\frac{1}{N}$ contributions. Small $\frac{1}{N}$ corrections seem to occur also for all the meson leading Regge trajectories
in the real world as demonstrated in Fig.\ref{fig:Real}. 
\begin{figure}[t] 
\centering
\includegraphics[width=.78\textwidth]{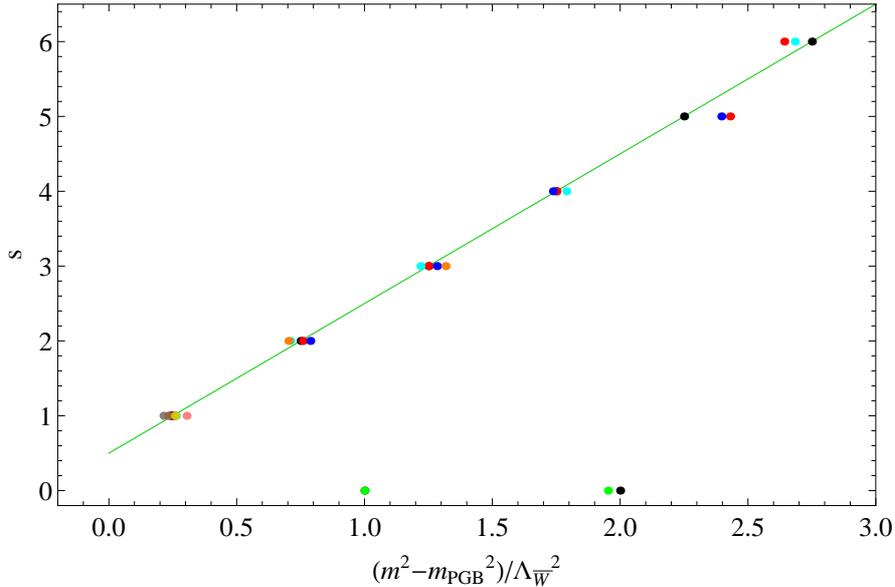}\\
\caption{Experimental scalar glueballs (green dots), and meson leading Regge trajectories (green line) named by the spin-$1$ meson for any flavor, $\rho$ red, $K^*(892)$ blue, $\omega$ cyan, $\phi$ orange, $D^*$ purple,
$B^*$ grey, $B^*_s$ brown, $D^*_s$ yellow, $J / \psi$ pink, extracted by the Particle Data Group (2014). The experimental masses squared of the candidate glueballs $f_0(1505)$ and $f_0(2100)$ (green points) are compared with Eq.(\ref{s}): $m^2_k= k \Lambda^2_{\overline W}$ (black points on the $s=0$ axis).
The experimental data for mesons are compared with the theoretical formula (black points on the green line): $m^2_s - m^2_{PGB}=\frac{1}{2} \Lambda^2_{\overline W}(s-\frac{1}{2})$, that arises conjecturally in a large-$N$ Topological String Theory \cite{MBS} dual to the $TFT$ of this paper. $s$ is the spin along the leading Regge trajectory, $m_{PGB}$ is the mass of the pseudo-Goldstone boson for any flavor (e.g. $\pi^0$ for the $\rho$ trajectory, $K^0$ for the $K^*$ trajectory and so on; for the $\phi$ trajectory $m_{PGB}^2=\frac{3}{4} m^2_{\eta}+\frac{1}{4} m^2_{\eta'}$, because of $\eta$-$\eta'$ mixing), $\Lambda_{\overline W}=1505 \, \text{MeV}$ is the mass of the lowest-mass glueball, that sets a universal slope even for the real mesons, and not only for their large-$N$ counterparts in Fig.\ref{fig:regge1} and Fig.\ref{fig:regge2}. The actual spectrum agrees with the theoretical formulae for the lower states and any flavor with an accuracy on the order of $1 \%$ or better for the masses in units of $\Lambda_{\overline W}$, quite surprisingly since in principle $\frac{1}{N}$ corrections are expected to occur, while for larger masses a few-percent deviations start to arise. This is rather opposite to the naive expectation that trajectories become linear only asymptotically, and it has no qualitative and quantitative explanation in conventional string models, including those based on the $AdS$ String/Gauge Theory correspondence. We suggest \cite{MBS} that the (semi-)integral nature of the spectrum reflects the existence of a Topological String Theory dual to the $TFT$ of this paper, with quite small effective $\frac{1}{N}$ corrections for the meson leading Regge trajectories and the glueballs.}
\label{fig:Real}
\end{figure}
Indeed, we have discovered, analyzing the meson spectrum of the Particle Data Group (2014), that all the low-lying meson masses squared of the leading Regge trajectories for any flavor in the real world, after subtracting the mass squared of the corresponding pseudo-Goldstone boson (PGB) due to the explicit breaking of the chiral symmetry, are (semi-)integer valued in units of $\frac{1}{2} \Lambda^2_{\overline W}$ to a surprising accuracy for the masses in units of $\Lambda_{\overline W}$ on the order of $1 \%$ or better, according to the theoretical formula (conjecturally large-$N$ exact in the dual Topological String Theory) $m^2_s - m^2_{PGB}=\frac{1}{2} \Lambda^2_{\overline W}
(s-\frac{1}{2})$, very much as (or even better than) their large-$N$ lattice counterparts (compare Fig.\ref{fig:Real} with Fig.\ref{fig:regge1} and Fig.\ref{fig:regge2}).
\begin{figure}[t] 
\centering
\includegraphics[width=.78\textwidth]{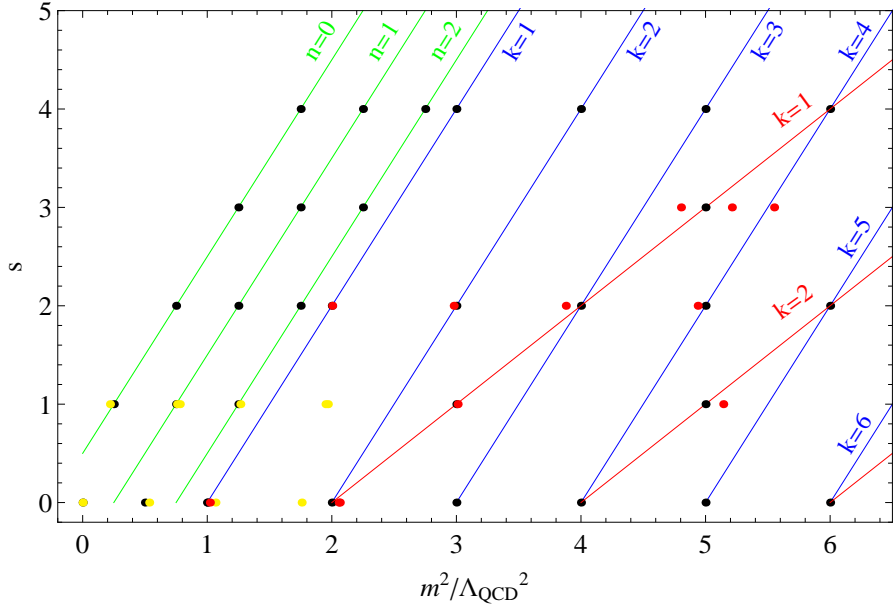}\\
\caption{Glueball (blue and red lines) and meson (green lines) Regge trajectories in massless large-$N$ quenched $QCD$ from lattice gauge theory \cite{T0,T,L0,L}, mesons in yellow, glueballs in red. The black points represent the following spectrum.
For even-spin glueballs $m^2_{s,k} = \Lambda^2_{QCD}(k+\frac{s}{2})$, for odd-spin glueballs $m^2_{s,k} = 2\Lambda^2_{QCD}(k+\frac{s}{2})$, $k=1, 2, \cdots$. For mesons $m^2_{s,n} = \frac{1}{2}\Lambda^2_{QCD}(n+s-\frac{1}{2})$, $s=1,2, \cdots$ (trajectories displayed) and $m^2_{s,n} = \frac{1}{2}\Lambda^2_{QCD}(n+s)$, $s=0,1, \cdots$ (trajectories not displayed), $n=0,1,\cdots$. The plot is based on \cite{MBS}. The case $s=0$ for the glueballs is Eq.(\ref{s}) in this paper. The other formulae are conjectured to occur in a Topological String Theory dual to the $TFT$ of this paper \cite{MBS}.}
\label{fig:regge2}
\end{figure}
We stress that our observation, that for the low-lying states of the meson leading Regge trajectories the masses squared shifted by the masses squared of the pseudo-Goldstone bosons are linear 
and (semi-)integer valued in units of the universal scale $\frac{1}{2}\Lambda^{2}_{\overline W}$ fixed by the mass squared of the lowest-lying scalar glueball $f_0(1500)$ $\Lambda^{2}_{\overline W}=(1505 \,\text{ MeV})^{2}$ to a high accuracy, was unsuspected so far, and in fact it contrasts with the standard string-inspired folklore that the trajectories are only asymptotically linear for large masses. Moreover, deviations from exact linearity of the experimental meson leading trajectories seem to occur more for higher rather than for lower masses (see Fig.\ref{fig:Real}). Besides, we stress that the aforementioned universality and (semi-)integral nature of the leading meson Regge trajectories (see Fig.\ref{fig:Real}) has not been manifest in previous linear fits of the meson trajectories \cite{Regge},
because of the lack of the theoretical prejudice arising from the $TFT$ (this has to do with the fact that in usual plots both slopes and intercepts are fitted independently for each trajectory, while in our case, given the masses of the pseudo-Goldstone bosons and of the lowest-mass glueball, we have no parameter). \par
We have suggested \cite{MBS} that the (semi-)integral nature of the masses squared (see Fig.\ref{fig:regge2}) is in fact consequence of the existence of a Topological String Theory for glueballs and mesons dual to the $TFT$ for glueballs of this paper, in our interpretation apparently with very small effective large-$N$ corrections: Part of the large-$N$ corrections for the mesons are effectively reabsorbed in the mass of the pseudo-Goldstone, conjecturally part of the large-$N$ corrections for the glueballs might be dynamically suppressed by their mixing with mesons in full-fledged $QCD$, because of the vast degeneracy with meson spectra occurring at large-$N$ due to the (semi-)integral nature of the spectrum in universal units. \par
In the specific case \cite{BES} the $f_0(1500)$ meson has been for long suspected to be a glueball and the $f_0(2100)$ meson has been considered as a possible glueball candidate (in the real world and in Veneziano limit of $QCD$  glueballs mix with flavor-singlet mesons of similar mass). The aforementioned facts together with the mass formula for the scalar and pseudoscalar glueballs of this paper and the relation $m_{f_0(1505)}^2=4(m^2_{\rho(775)}-m^2_{\pi(135)})$ verified to a surprising accuracy, since in principle it would need $\frac{1}{N}$ corrections, and expected \cite{MBS} by the aforementioned doubling at large-$N$ of the slope of meson Regge trajectories with respect to glueball trajectories (see formulae below Fig.\ref{fig:regge2}), would imply the identification of the lowest-mass scalar glueball with $f_0(1500)$ and of the first-excited scalar glueball with $f_0(2100)$.  At the same time the spin-$2$ meson $f_2(2150)$ could be identified with the first spin-$2$ glueball on the basis of the mass formula for even-spin glueballs $m^2_{s,k} = \Lambda^2_{QCD}(k+\frac{s}{2})$ reported below Fig.\ref{fig:regge2}.  Another possibility for the scalar excited glueball is $f_0(2200)$, that would be essentially degenerate with a candidate first pseudoscalar glueball $\eta(2225)$ and a candidate first spin-$2$ glueball $f_J(2200)$ according to our plot in Fig.\ref{fig:regge1} of the lattice results \cite{T0,T}, yet it would be compatible with $5\%$ accuracy with our large-$N$ mass formula.
A conclusive statement will be possible only if $\frac{1}{N}$ corrections for the masses and for the decay rates become available. \par

\end{document}